\documentclass[superscriptaddress,amsmath,onecolumn,amssymb]{revtex4}

\usepackage{xr}

\usepackage{subfigure}
\usepackage{amsmath}
\usepackage{pgfplots}
\usepackage{dsfont}
\pgfplotsset{compat=1.5}
\usepgfplotslibrary{groupplots}

\usepackage{color}

\usepackage{xcolor}
\usepackage{multirow}
\usepackage{placeins}
\usepackage{soul}
\usepackage{color,xcolor}
\usepackage[colorlinks,linkcolor=red,anchorcolor=blue,citecolor=blue,urlcolor=blue]{hyperref}
\usepackage{cleveref}

\usepackage{subfigure}
\usepackage{graphicx}
\usepackage{amsmath}

\usepackage{pgfplots}
\pgfplotsset{compat=1.5}
\usepgfplotslibrary{groupplots}

\usepackage{color}
\usepackage{algpseudocode}
\usepackage{algorithm}

\begin{document}

\title{Sample Caching Markov Chain Monte Carlo Approach to Boson Sampling Simulation}

\author{Yong Liu}
\affiliation{Institute for Quantum Information \& State Key Laboratory of High Performance Computing, College of Computer, National University of Defense Technology, Changsha 410073, China}

\author{Min Xiong}
\affiliation{Institute for Quantum Information \& State Key Laboratory of High Performance Computing, College of Computer, National University of Defense Technology, Changsha 410073, China}

\author{Chunqing Wu}
\affiliation{Institute for Quantum Information \& State Key Laboratory of High Performance Computing, College of Computer, National University of Defense Technology, Changsha 410073, China}

\author{Dongyang Wang}
\affiliation{Institute for Quantum Information \& State Key Laboratory of High Performance Computing, College of Computer, National University of Defense Technology, Changsha 410073, China}

\author{Yingwen Liu}
\affiliation{Institute for Quantum Information \& State Key Laboratory of High Performance Computing, College of Computer, National University of Defense Technology, Changsha 410073, China}

\author{Jiangfang Ding}
\affiliation{Institute for Quantum Information \& State Key Laboratory of High Performance Computing, College of Computer, National University of Defense Technology, Changsha 410073, China}

\author{Anqi Huang}
\affiliation{Institute for Quantum Information \& State Key Laboratory of High Performance Computing, College of Computer, National University of Defense Technology, Changsha 410073, China}

\author{Xiang Fu}
\affiliation{Institute for Quantum Information \& State Key Laboratory of High Performance Computing, College of Computer, National University of Defense Technology, Changsha 410073, China}

\author{Xiaogang Qiang}
\affiliation{Institute for Quantum Information \& State Key Laboratory of High Performance Computing, College of Computer, National University of Defense Technology, Changsha 410073, China}
\affiliation{National Innovation Institute of Defense Technology, AMS, 100071 Beijing, China}

\author{Ping Xu}
\affiliation{Institute for Quantum Information \& State Key Laboratory of High Performance Computing, College of Computer, National University of Defense Technology, Changsha 410073, China}

\author{Mingtang Deng}
\affiliation{Institute for Quantum Information \& State Key Laboratory of High Performance Computing, College of Computer, National University of Defense Technology, Changsha 410073, China}

\author{Xuejun Yang}
\affiliation{Institute for Quantum Information \& State Key Laboratory of High Performance Computing, College of Computer, National University of Defense Technology, Changsha 410073, China}

\author{Junjie Wu}
\email{junjiewu@nudt.edu.cn}
\affiliation{Institute for Quantum Information \& State Key Laboratory of High Performance Computing, College of Computer, National University of Defense Technology, Changsha 410073, China}

\begin{abstract}
  Boson sampling is a promising candidate for quantum supremacy. It requires to sample from a complicated distribution, and is trusted to be intractable on classical computers. Among the various classical sampling methods, the Markov chain Monte Carlo method is an important approach to the simulation and validation of boson sampling. This method however suffers from the severe sample loss issue caused by the autocorrelation of the sample sequence. Addressing this, we propose the Sample Caching Markov chain Monte Carlo method that eliminates the correlations among the samples, and prevents the sample loss at the meantime, allowing more efficient simulation of boson sampling. Moreover, our method can be used as a general sampling framework that can benefit a wide range of sampling tasks, and is particularly suitable for applications where a large number of samples are taken.
\end{abstract}

\date{\today}

\maketitle

\section{Introduction}

Demonstrating quantum supremacy is a milestone in quantum computing, representing that quantum devices can outperform the fastest classical hardware on some task~\cite{Preskill2012,Arute2019,Spagnolo2018,Nielsen2002}.
Evaluating the demonstration of quantum supremacy should base on the intense competition between the two sides: the developing quantum devices on some selective tasks, and the classical computers running the benchmark for the corresponding tasks to explore the supremacy threshold.
The chosen task should be suitable for near-term implementation as well as able to be significantly accelerated by quantum computing.

Such a task is boson sampling~\cite{Aaronson2011}.
On the one hand, its linear optical implementation merely requires identical bosons (typically photons), linear transformation, and passive detection.
On the other hand, its classical simulation involves computing permanents of Gaussian complex matrices~\cite{Scheel2008}, which is likely to be classically intractable, even in approximation cases~\cite{Valiant1979}.
The classical hardness of boson sampling attracts enormous efforts to build large-scale physical devices to beat classical computers, and remarkable achievements have been made~\cite{Spring2013,Broome2013,Tillmann2013,Spagnolo2014,Lund2014,Bentivegna2015,Seshadreesan2015,Latmiral2016,Zhong2018, Paesani2019,Wang2019}. It is promising to show quantum advantage via boson sampling.

On the classical side, various sampling approaches have been raised for simulating boson sampling. The direct computation of the whole distribution of boson sampling is restricted within small scales because of the exponential growth of the state space. This stimulates a variety of methods for sampling from an exponentially large state space by calculating a few number of probabilities, such as the Clifford-Clifford (C\&C) sampling method~\cite{Clifford2018}, the rejection sampling method~\cite{Villalonga2019} and the Metropolised independent sampling (MIS) method~\cite{Neville2017, Liu1996}. Among the various sampling approaches, the MIS method allows to generate an effective sample by evaluating only a constant number of probabilities. More importantly, it can serve as a general sampling framework for far-ranging applications.

\begin{figure*}[!t]
  \centering
  \includegraphics[width=0.95\textwidth]{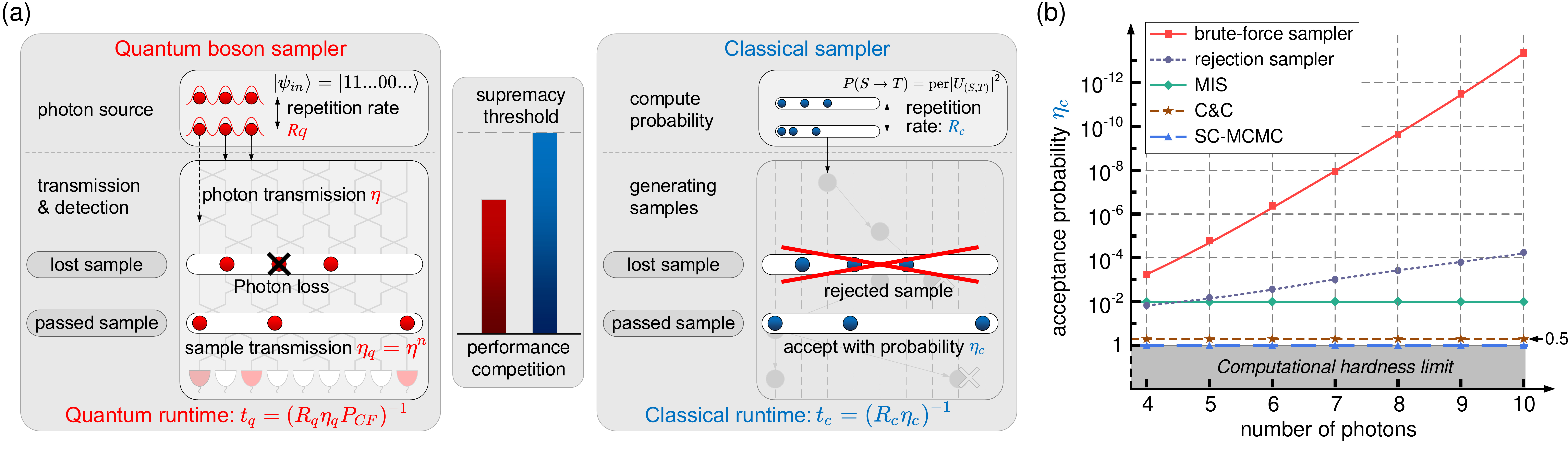}\\
  \caption{\footnotesize {Demonstrating quantum supremacy is a competition between quantum computing and classical computing.}
  (a) Sampling schemes of the quantum boson sampler and the classical sampler, where sample loss is a fatal issue in both of the two samplers. The quantum runtime of a boson sampler is mainly determined by the repetition rate $R_q$ of the multi-photon source, the transmission probability for a single photon $\eta$ (the sample transmission probability $\eta_q=\eta^n$), and the probability for a collision-free event $P_{CF}$. The classical runtime consists of the repetition rate of generating a candidate sample $R_{c}$ that is decided by the time for calculating a permanent, and the probability for keeping a sample $\eta_c$.
  (b) The average acceptance probability of the mainstream general sampling methods when a large number of samples are taken. The {\em computational hardness limit} represents the computational complexity of the core problems. In boson sampling, it is the computation of a permanent, which corresponds to $\eta_c=100\%$. The C\&C sampler has an equivalent cost of two permanents calculations for one effective sample, corresponding to an equivalent acceptance probability of 0.5.
  }\label{Fig:topic}
\end{figure*}

However, similar with the optical implementation of boson sampling, the sample loss is also a serious problem of classical samplers, as shown in Fig.~\ref{Fig:topic}{\color{red}{(a)}}. Current sampling methods generate massive candidate samples with each involving the evaluation of a probability, but only a small fraction of candidates are kept as effective samples, as shown in Fig.~\ref{Fig:topic}{\color{red}{(b)}}. For the MIS method, it has to discard massive candidates in order to tackle the autocorrelation issue inherited from the Markov chain Monte Carlo (MCMC) method. To this point, we define the {\em computational hardness limit} of the sampling tasks as the complexity of computing one probability. As for boson sampling, it is the complexity of computing one permanent~\cite{Aaronson2011}, and to reach this limit, it has to eliminate the overhead of the sampling methods. In this work, we propose a method that could reach this limit.

Our method, namely Sample Caching-Markov chain Monte Carlo (SC-MCMC), makes contributions on two sides.
(1) On the classical side, the SC-MCMC sampler provides a general sampling framework which allows to generate one effective sample from averagely one candidate sample. SC-MCMC deals with the autocorrelation issue of MCMC following a similar manner to MIS, but caches the discarded candidate samples for further reuse. In this way, the SC-MCMC sampler avoids the sample loss, and allows much more efficient sampling than a MIS sampler. Particularly, the SC-MCMC method can be used for quantum-enhanced applications based on boson sampling such as simulating the molecular vibronic spectra~\cite{Huh2015}, finding dense subgraphs~\cite{Arrazola2018} or approximate optimization~\cite{Arrazola2018_2}. Moreover, Our method can also be directly applied on other quantum supremacy candidates, such as the Gaussian boson sampling~\cite{Hamilton2017}, the IQP circuit sampling~\cite{Bremner2016,Shepherd2009} and random quantum circuit (RQC) sampling~\cite{Boixo2018,GuoLiu2019}, especially when a lot of samples are taken for validating the device~\cite{Hangleiter2019}.
(2) On the quantum side, our results suggest that the hardness of demonstrating quantum supremacy by boson sampling is increasing with the development of classical simulator. Based on the results of the simulation, we emphasize the importance of avoiding the photon loss in the optical implementation of boson sampling, showing that an improvement of 5\% for the loss rate would spare hundreds of photons.

\section{Boson sampling and Markov chain Monte Carlo}
The task of boson sampling is to sample the output distributions of an $m$-mode interferometer network with $n$ identical photons as input.
Because of the interferences of photons, the probability of a certain output pattern of photons is related to the permanent of the transformation matrix decided by the interferometer network.
Specifically, the output patterns are post-selected within the ``collision-free'' regime where photons are no-bunching in each output port.
The value of $m$ is often chosen to be $n^2$ to meet the requirement for the hardness proof of boson sampling, leading to that the number of all the possible output patterns grows exponentially with $n$.
In summary, the task of classically simulating boson sampling is to generate samples from the probability distribution over the $\left(_n^m\right)$ output patterns.

Instead of calculating the probabilities of the whole distribution, a feasible method is Metropolis-Hastings algorithm, a kind of MCMC method~\cite{Chib1995}.
The sampling is processed by constructing a Markov chain, with the state space corresponding to the set of the $\left(_n^m\right)$ possible output patterns of boson sampling, and the probabilities of all the patterns as the stationary probabilities.
To expand the chain, a state $s'$ is chosen as the candidate state (current state is $s$) following a easy-to-sample symmetric proposal probability distribution, and this candidate state would be accepted and added on the chain with probability
\begin{equation}
P_{\rm accept}=\min\left(1,p(s')/p(s)\right),
\end{equation}
where $p(s)$ is the boson sampling probability of state $s$, or be rejected with the rest probability.
If the candidate is rejected, the current state would duplicate on the end of the chain.
Each time a node is added in the chain, the pattern corresponding to the added node is outputted as a candidate sample.
In this way, it seems ideal that one may just need to evaluating one probability for one sample.

\begin{figure}[t]
  \centering
  \includegraphics[width=0.95\textwidth]{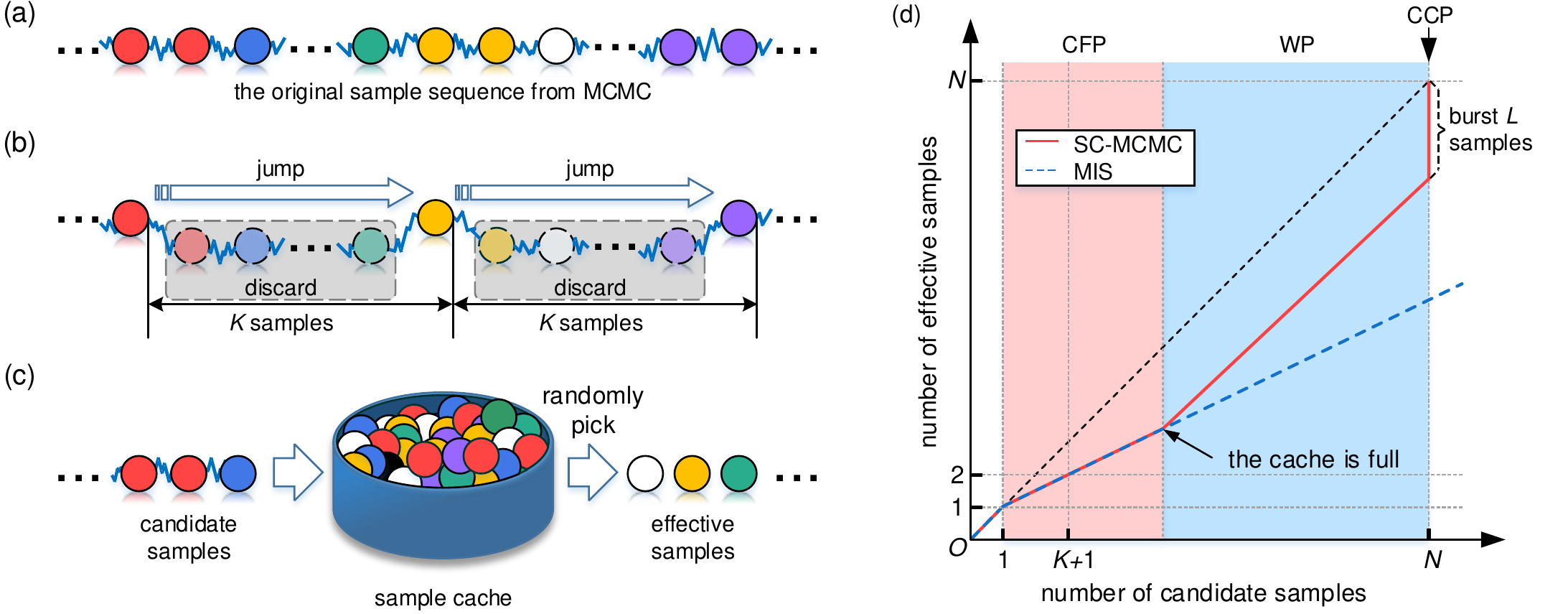}\\
  \caption{\footnotesize {Scheme of jump sampling and sample caching.} (a) The original sequence generated by MCMC sampler, where the samples are strongly correlated. (b) The jump sampling method. The samples within the leap are all discarded. The remained samples are approximately uncorrelated. (c) The sample caching method. Samples are filled in a cache first, and are afterwards randomly outputted from the cache. The output samples are approximately independent. (d) The three phases of SC-MCMC. After outputting the first sample, SC-MCMC enters the cache filling phase (CFP), in which it performs similar to MIS. After the cache is full, SC-MCMC enters the working phase (WP) and starts to output samples by every evaluation of a probability. Finally, in the cache cleaning phase (CCP), there can be a burst of $L$ samples that can be the compensation for the delay in CFP. Averagely, only one probability is evaluated for one effective sample in the SC-MCMC sampler.}\label{Fig:scheme}
\end{figure}

Unfortunately, the generated samples may be erroneous because the generated sample sequence suffers from severe autocorrelation, as shown in Fig.~\ref{Fig:scheme}{\color{red}{(a)}}.
The first-order autocorrelation of a sequence is often estimated using Durbin-Watson statistic~\cite{Durbin1971}, while in this paper, we follow a more straightforward way to estimate the autocorrelation by
\begin{equation}\label{EQ:1autoc}
r_1=\frac{\sum_i(x_i-\bar{x})(x_{i+1}-\bar{x})}{\sum_i(x_i-\bar{x})^2},
\end{equation}
where $x_i$ is the value of the $i^{th}$ sample, $\bar{x}$ is the mean of the samples. Each $x_i$ is assigned a value as the order of patterns after been sorted.
The value of $r_1$ should be in $[-1,1]$, and it reflects the autocorrelation in the sequence.
The closer $|r_1|$ is to 1, the stronger the samples are correlated, and the sign of $r_1$ indicates if the samples are positively or negatively correlated.

To overcome the autocorrelation of the sample sequence, in MIS, a method called ``jump sampling'' (or ``thinning procedure'') is applied to obtain independent samples, as shown in Fig.~\ref{Fig:scheme}{\color{red}{(b)}}.
In jump sampling, an effective sample is kept in every $K$ candidates with the rest $K-1$ candidates discarded.
The remained samples are approximately independent.
The value of $K$ is determined by checking the autocorrelation of the obtained sequence.
In MIS, the value of $K$ is 100, which is claimed to be sufficient for simulating boson sampling of more than 30 photons~\cite{Neville2017}. However, this suggests that in MIS the ratio for keeping a candidate sample is $\eta_c=\frac{1}{K}=1\%$. Other methods such as the delayed rejection~\cite{Au2001,Tierney1999,Zuev2011} could be used to reduce the correlations among the samples at the cost of calculating more probabilities for one sample.
If the autocorrelation issue could be tackled without abandoning any samples, the sampling process could be accelerated by 100 times.

\section{The sample caching method}
The main scheme of the SC-MCMC protocol is shown in Fig.~\ref{Fig:scheme}{\color{red}{(c)}}. It mainly contains two parts: one is a MCMC sampler providing candidate samples, with which the second part combined is a procedure that we call ``sample caching''. In Fig.~\ref{Fig:scheme}{\color{red}{(d)}} we show the working processes of the SC-MCMC sampler, which can be divided into three phases:
\begin{enumerate}
  \item The cache filling phase (CFP). In this phase, the SC-MCMC sampler follows the similar protocol to the MIS sampler: It outputs an effective sample in every $K$ candidates, where $K$ is the jump step used in the MIS sampler. The difference is that the unselected candidates are used to fill a sample cache till the cache is full. The sample cache can be completely filled after generating $\lceil L/(K - 1)\rceil+L$ candidate samples, where $L$ is the size of the cache. In this stage it has to calculate $K$ probabilities for one effective sample;
  \item The working phase (WP), in which the cache is full. For each time a candidate sample is generated by the MCMC process, randomly output a sample from the cache first and then store this new candidate into the cache. This procedure is repeated till the MCMC process generates $N$ candidates. Therefore in WP, the SC-MCMC approach drives efficient sample generation that an effective sample can be generated by calculating only one probability;
  \item The cache cleaning phase (CCP). In this phase, the samples stored in the cache are outputted in random order without calculating any probabilities, and therefore there can be a burst of sample generation. This phase can be regarded as a compensation for the initial overhead of CFP, and result in that averagely the evaluation of every probability can support the generation of one effective sample.
\end{enumerate}
Nevertheless, the first candidate sample can be directly outputted as an effective sample since there is no other sample that can be correlated with. As reflected from Fig.~\ref{Fig:scheme}{\color{red}{(d)}}, if only a small number of samples are taken (i.e. $N<\lceil L/(K-1)\rceil+L$), the sampling process may stop in CFP, in this case the SC-MCMC sampler is actually executing the MIS protocol. Generally, in the final CCP, cleaning the cache provides a burst of sample generation. However, this step can be dangerous when the value of $N$ is not large enough because the burst output could result in sample correlation, and whether a sample can be kept is determined by how much it is correlated with the sample sequence. While if a large number of samples are taken, SC-MCMC can be $K$ times as fast as MIS.

It is worth noting that the generation of candidate samples discussed here is slightly different from that in~\cite{Neville2017} where the proposal probability distribution is chosen as the distribution of distinguishable particles corresponding to the boson sampling instance. This choice can reduce the correlations between samples, but introduces extra computational cost of a real permanent. Here we choose the uniform distribution as the proposal distribution so that only one permanent is required to calculate for every candidate sample, while the autocorrelation can be eliminated by the sample cache without abandoning any candidates (as we will show in the following discussion). In this way, compared with the C\&C sampler which has an approximately equivalent cost of calculating two $n\times n$ permanents for one effective sample~\cite{Clifford2018}, the SC-MCMC sampler can modestly obtain a 2-fold improvement over the C\&C sampler when a large number of samples are taken.

Similar to MIS, SC-MCMC is also case independent. Besides boson sampling, SC-MCMC can be directly used as a general sampling framework for any sampling tasks as long as we can evaluate the probabilities for given events, for example the Gaussian boson sampling~\cite{Hamilton2017}, IQP circuit sampling~\cite{Bremner2016,Shepherd2009} and random quantum circuit sampling~\cite{Boixo2018,GuoLiu2019}. Meanwhile, SC-MCMC is particularly suitable for tasks where a large number of samples are taken.

\begin{figure}[!b]
  \centering
  \begin{tabular}{cc}
  \includegraphics[width=0.45\textwidth]{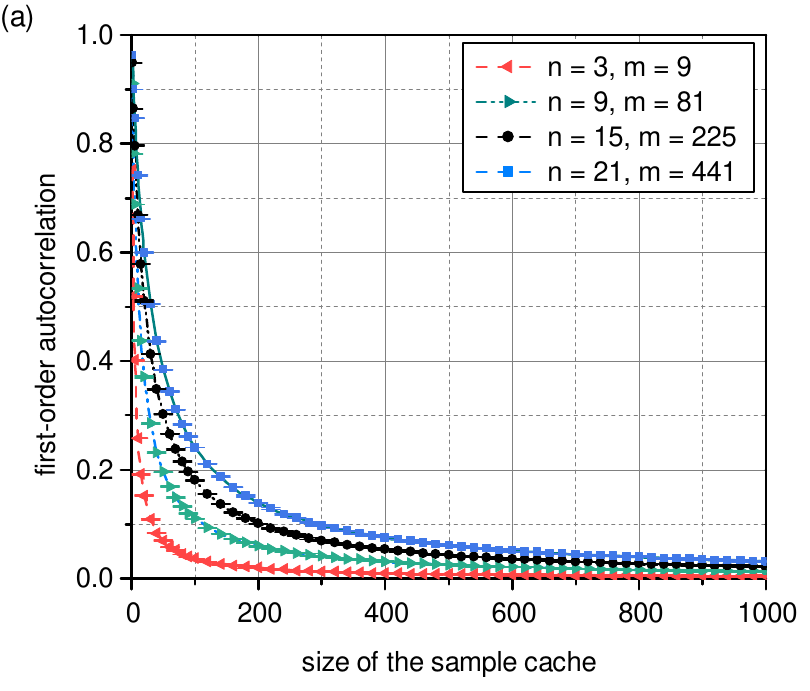}&
  \includegraphics[width=0.45\textwidth]{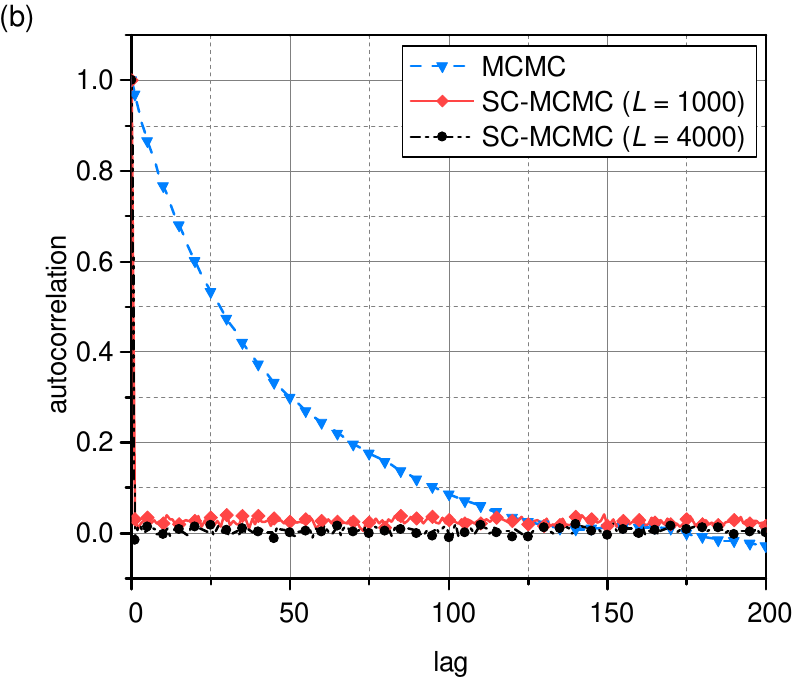}\\
  \end{tabular}
  \caption{\footnotesize {(Color online) The results of simulating boson sampling with the SC-MCMC method. (a) The first-order autocorrelation against the size of the sample cache.} The autocorrelation is quite insignificant when the size of cache reaches 1,000. The number of samples taken is 1,000,000 for each boson sampling instance with $n$ photons and $m$ modes. For each configuration we ran 5 times for average, and the error bars correspond to one standard deviation. (b) The autocorrelation at different orders obtained through simulating boson sampling with 30 photons in 900 modes using original MCMC approach or the SC-MCMC approach with a cache of size $L$. The blue dashed line with triangle is the result from the original MCMC method, the red solid line with rhombus indicates the autocorrelation of the sample sequence generated by SC-MCMC with $L=1,000$ and the black dash-dotted line with square is obtained from SC-MCMC with $L=4,000$. For each simulation 10,000 samples are taken.}
  \label{Fig:r1vsCacheSize}
\end{figure}

Essentially, the correctness of SC-MCMC is guaranteed by the MCMC process. Some methods can be applied to validate the sampling results~\cite{Aaronson2014,Bentivegna2014,Agresti2019}, but a more straightforward way is to compare the frequency graph with the probability distribution. We found an empirical result that the number of samples may has to be in the order of $100\cdot\left(_n^m\right)$ to construct a frequency graph with a similarity of 99\% with the probability distribution.

More importantly, the sample caching process eliminates the autocorrelations within the sample sequence without losing any candidate samples.
Under the asymptotic condition where the size of the cache is large enough, the number of samples stored in the cache follows the probability of the samples, and the uniformly random choice makes it independent among the draws.
Practically, the correlations among the samples can be eliminated with a sample cache in a limited size. In
Fig.{\ref{Fig:r1vsCacheSize}\color{red}{(a)}} we plot the first-order autocorrelation against the size of sample cache for different scales of boson sampling. In Fig.{\ref{Fig:r1vsCacheSize}\color{red}{(b)}} we plot the autocorrelation of the sample sequences obtained from different approaches in simulating an instance of boson sampling with 30 photons and 900 modes, where the ``lag'' indicates the orders of the autocorrelations. In the sample sequence from the original MCMC approach, we can still find remained correlation between two samples even in lag 100 owing to the improper choice of the proposal probability distribution (see supplementary for details~\cite{supplementary}). In comparison, the SC-MCMC sampler with $L=1,000$ can reduce the autocorrelation at every order to a quite low level, and a cache with $L=4,000$ can approximately generate independent samples, even though the proposal distributions in the SC-MCMC and the MCMC samplers are the same.

To understand how sample cache works, we analyze the jump sampling method first.
For the MCMC sampler with state space $S=\{s_1,\dots,s_N\}$, the transition probabilities can be described by a matrix $P=\{p_{ij}\}_{N\times N}$ with $p_{ij}$ equals the probability to transit from state $s_i$ to $s_j$ in one step.
The $k$-step transition probability matrix is $P^{(k)}=P^k$, and $p_{ij}^{(k)}$ is the probability to transit from $s_i$ to $s_j$ in $k$ steps.
It satisfies that $\lim\limits_{k\to\infty}p_{ij}^{(k)}=p(s_j)$ for arbitrary $i$ and $j$, where $p(s_j)$ is the probability of state $s_j$~\cite{Liu1996,Gagniuc2017,Wan2016}.
Thus a sample is independent from another that is infinite steps away.
Actually, in a certain number of steps (e.g. $K$ steps), the $K$-step transition probability $p_{ij}^{(K)}$ approximately equals to $p(s_j)$ for arbitrary $i$ and $j$, which means that a sample is approximately independent from the samples with a distance of more than $K$ steps. The value of $K$ depends on how fast the Markov chain converges~\cite{Schmitt2001}.
Empirically, we can choose a sufficiently large value.
If the samples between the $K$ steps are discarded, the correlations among the remained samples are negligible.

In SC-MCMC, the result can be regarded as a reordered sequence from the original one.
The probability that the two adjacent samples are at a distance of $k$ $(k\geq 1)$ steps in the original sequence is
\begin{equation}
p(k,L)=\left(\frac{L-1}{L}\right)^{k - 1}\cdot\frac{1}{L},
\end{equation}
where $L$ is the size of the sample cache.
Then we can obtain the probability that two adjacent samples are still correlated (the distance between the two samples in the original sequence does not exceed $K$, the corresponding jumping step in jump sampling) by
\begin{equation}\label{EQ:PCORR}
P_{\rm cr}\equiv\sum_{k=1}^Kp(k,L)=1-\left(\frac{L-1}{L}\right)^K\approx \frac{K}{L}.
\end{equation}
Clearly, $\lim\limits_{L\to\infty}P_{\rm cr} = 0$, and the first-order autocorrelation is thus eliminated if the sample cache is sufficiently large.
The autocorrelations of higher orders are eliminated in the same way~\cite{supplementary}.

The next question is that practically what size should the cache be?
The exact size of the sample cache required varies case-by-case, however we can choose a sufficiently big cache.
This answer is easy to obtain from Eq.(\ref{EQ:PCORR}) by ensuring $P_{\rm cr}<\varepsilon$, then $L>\frac{K}{\varepsilon}$.
Setting $K=200$ and $\varepsilon=0.05$, resulting in $L=4,000$, is sufficient for the case of $n=30$ and $m=900$.
It is supposed to be sufficient for boson sampling with larger scale, and the size of cache could be further enlarged if needed.
Most significant of all, no sample is lost no matter what size the cache is, which suggests that our method reaches a time complexity of simulating boson sampling to $O(n2^n)$.

\section{Numerical simulation}

\begin{table}[!t]
  \caption{\footnotesize {The results of sample caching and jump sampling}. The results of the two methods are compared for simulating boson sampling instances with different number of photons (p.) and modes (m.). $r_o$ is the first-order autocorrelation of the original sample sequence generated by MCMC. $r^{(200)}$ is the first-order autocorrelation of the sequence obtained by jump sampling with the jumping step of 200. $r_{500}$, $r_{1000}$, $r_{2000}$ and $r_{4000}$ are the first-order autocorrelation of the sequences obtained through sample caching with the sizes of caches to be $500$, $1000$, $2000$ and $4000$ respectively.}
  \label{TAB:COMPSCJS}
  \begin{tabular}{p{0.3cm}p{2.0cm}p{1.5cm}p{1.5cm}p{1.5cm}p{1.5cm}p{1.5cm}p{1.5cm}}
    \hline
    \hline
      &     Scale & $r_o$ & $r^{(200)}$ & $r_{500}$ & $r_{1000}$ & $r_{2000}$ & $r_{4000}$ \\
    \hline
      &   9p.81m. & 0.9115 &  0.0102  & 0.0261 & 0.0119 & 0.0067 &  0.0035 \\
      & 15p.225m. & 0.9490 &  0.0128  & 0.0451 & 0.0205 & 0.0102 &  0.0057 \\
      & 21p.441m. & 0.9633 &  0.0030  & 0.0622 & 0.0311 & 0.0162 &  0.0081 \\
      & 25p.625m. & 0.9673 & -0.0183  & 0.0642 & 0.0328 & 0.0167 &  0.0089 \\
      & 30p.900m. & 0.9702 & -0.0832  & 0.0684 & 0.0309 & 0.0095 & -0.0162 \\
    \hline
    \hline
  \end{tabular}
\end{table}

\begin{figure}[!b]
  \centering
  \includegraphics[width=0.55\textwidth]{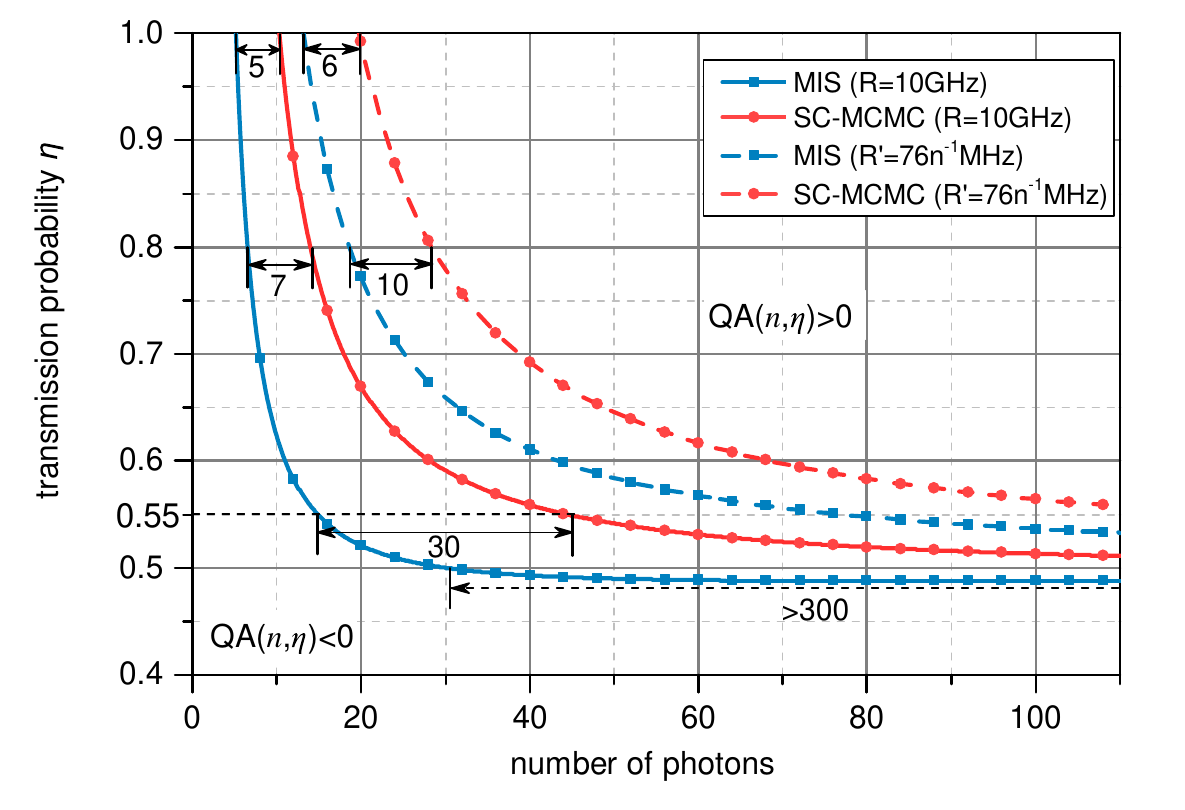}\\
  \caption{\footnotesize {(Color online) The performance comparison between the quantum Boson sampling device and the classical computer.} The blue solid (dashed) line represents $QA(n,\eta)=0$ with the classical runtime estimated by MIS as $t_c=3\cdot n^22^n\times10^{-13}$ with the $n$-photon repetition rate of the photon source $R_q = 10{\text {GHz}}$ ($R'_q=76n^{-1} {\text {MHz}}$, the leading parameter of proposed photon source~\cite{Wang2019}, which is obtain from a $76$MHz quantum dot source), and the red lines are the results obtained from SC-MCMC. The increment of photon number is vast when $\eta<0.5$, and it will reduce by more than 300 when $\eta$ is improved by only 5\%.}
  \label{FIG:PerformanceComparison}
\end{figure}

We demonstrate our method for boson sampling simulation by implementing the parallelized Glynn's algorithm to exactly compute the permanents of arbitrary square matrices~\cite{Glynn2010}.
The numerical simulations were done on 64 nodes of Tianhe-2 supercomputer~\cite{Liao2014,Liao2014_2} or another local cluster, as shown in Tab.~\ref{TAB:COMPSCJS}.
The sampling rate reached 1.01Hz on Tianhe-2 nodes when simulating boson sampling with 30 photons, and the 32 cluster nodes afforded the simulation for 21 photons with a sampling rate of 18.09Hz.
The percentage of the time used for calculating permanents exceeds 99\% when $n$ is sufficiently large.

Combined with the performance estimation on Tianhe-2 supercomputer\cite{Wu2018}, the average time estimated (in seconds) for obtaining a $n$-photon candidate sample is $T(n)=1.9925\cdot n^22^n\times 10^{-15}$ if we are allowed to simulate the generation of a large number of samples. This scaling result indicates that the average time estimated for a 50-photon sample can be reduced from about 10 days to within 100 minutes. To compare the performance of the simulator with the physical setups, we use the function defined in ref.~\cite{Neville2017}
\begin{equation}
QA(n, \eta)=\log\left(t_c/t_q\right),
\end{equation}
where $t_c$ is the estimated classical runtime for a boson sampling instance with $n$ photons and $m=n^2$ modes.
$t_q\propto (R_q\eta^{n})^{-1}$ is the quantum runtime for the corresponding instance in which $R_q$ is the $n$-photon repetition rate of the photon source, and $\eta$ is the transmission probability for a photon~\cite{supplementary}.
$QA > 0$ suggests that the performance of a quantum boson sampling setup surpasses that of Tianhe-2 supercomputer.

In Fig.~\ref{FIG:PerformanceComparison} we plot four lines corresponding to $QA(n,\eta)=0$ under different combinations of classical approaches and physical setups. Associated with the transmission probability realized in recent experiments ($<0.4$~\cite{Spring2013,Broome2013,Tillmann2013,Spagnolo2014,Bentivegna2015,Zhong2018,Wang2019}), the increment of photon number required to reach a corresponding performance is vast. For example, when $\eta=0.5$ and $R_q=10{\text {GHz}}$ (probably beyond the reach of near-term experiments) the increment exceeds 300 photons, while it reduces to 30 when $\eta$ is improved to 0.55. Further, for a network with given shape, there exist the minimum request for $\eta$, even when we have sufficiently many photons~\cite{supplementary}. Our results suggest that currently, enhancing the transmission probability can greatly reduce the required photon number for quantum advantage.

\section{Conclusion}
In summary, we proposed a classical sampling approach which can be used to simulate boson sampling. Our method tackles the autocorrelation issue which leads to the sample loss in MIS, and can be 100 times as fast as MIS if a large number of samples are taken. For cases where only a small number of samples are required, our method can perform as well as MIS. Therefore our method enables more efficient simulation of boson sampling, especially in the cases where boson sampling is applied for quantum-enhanced problems. Above all, our method is a general sampling framework and could support customized optimization for specific tasks to gain further advanced efficiency.

The progress in classical simulators can be challenges to physical implementations, which however could provide some instructive feedback on physical experiments. The comparison between classical computing and quantum boson sampling suggests that currently, reducing photon loss may approach closer to quantum advantage, and is perhaps more important than merely enlarging the scale of boson sampling devices.

\begin{acknowledgments}
We gratefully acknowledge the help from China Greatwall Technology in Changsha, and the National Supercomputer Center in Guangzhou. We appreciate the helpful discussion with other members of QUANTA group. J. W. acknowledges the support from the National Natural Science Foundation of China under Grant 61632021. P. X. acknowledges support from National Natural Science Foundation of China under Grants No. 11621091 and No. 11690031. X. Q. acknowledges support from National Natural Science Foundation of China under Grants No. 11804389.
\end{acknowledgments}

\bibliographystyle{unsrt}

\section{acknowledgments}
We gratefully appreciate the kindly support from China Great wall Technology in Changsha. We also appreciate the helpful discussion with other members of QUANTA team. J. W. acknowledges the support from National Natural Science Foundation of China under Grants No. 61632021. P. X. acknowledges the support from National Natural Science Foundation of China under Grants No. 11621091 and 11690031. X. Q. acknowledges the support from National Natural Science Foundation of China under Grants No. 11804389.

%\section{Author contributions}
%J. W.,  X. S. and Y. L. conceived and designed the project. Y. L., J. J., J. Z., G. T. and J. W. provided the theoretical proof and analysis. Y. L., P. Z., D. W., J. D., X. Q., P. X. and  J. W. performed the experiments. Y. L., J. J., A. H., P. X., G. T., X. F., M. D., C. W., X. Y. and J. W. wrote the manuscript. J. W. and X. S. supervised the project.

\clearpage
\onecolumngrid
%\appendix

\setcounter{section}{0}
\setcounter{equation}{0}
\setcounter{figure}{0}
\setcounter{table}{0}
\renewcommand{\theequation}{S\arabic{equation}}
\renewcommand{\thefigure}{S\arabic{figure}}
\renewcommand{\thetable}{S\Roman{table}}
\renewcommand{\bibnumfmt}[1]{[S#1]}
\renewcommand{\citenumfont}[1]{S#1}

\begin{center}\bf\large
    Supplementary Information: \\Sample Caching Markov Chain Monte Carlo Approach to Boson Sampling Simulation
\end{center}

\section{Boson Sampling and Quantum Supremacy}

Boson sampling is a process that by injecting $n$ bosons (typically photons) into the $m$-mode interferometer, and measuring the output distributions of the photons one obtains samples from a boson sampling device. The patterns of the output photons form a complicated probability distribution, as described by Eq.~\ref{EQS:BosonSamplingSolution}
\begin{equation}\label{EQS:BosonSamplingSolution}
{\text {Pr}}(S\rightarrow T)=\frac{|{\text {Per}}(U^{(S,T)})|^2}{\prod_is_i!\prod_jt_j!},
\end{equation}
where $S$ and $T$ are two $m$-dimension vectors describing the input pattern and output pattern of the photons respectively. Specifically, $S=(s_1,s_2,...,s_m)$ with $s_i$ meaning there are $s_i$ photons in the $i^{th}$ input port of the interferometer, and $T$ is defined for the output ports in the same way. In optical implementations, $S$ and $T$ are often described by multi-mode number states. The standard input state is $|\psi_{in}\rangle=|S\rangle=|\underbrace{11...1}_{n}\underbrace{00...0}_{m-n}\rangle$, which is also applied in our simulation. $U^{(S,T)}$ is a sub-matrix by choosing different rows and columns from the transformation matrix $U$ that is decided by the interferometer~\cite{Aaronson2011}. $Per(\cdot)$ is the permanent of the matrix given. For a matrix $A=\{a_{ij}\}_{n\times n}$, the permanent of $A$ is defined as
\begin{equation}\label{EQS:Permanent}
{\text {Per}}(A)=\sum_\sigma\prod_{i=1}^na_{i\sigma_i},
\end{equation}
where $\sigma$ ranges over all the permutations of $\{1,2,...,n\}$.

The exact simulation of boson sampling is hard. While in the case of approximation, the hardness proof of boson sampling is based on a conjecture that the permanent of the matrix with Gaussian elements is hard to approximate unless the polynomial hierarchy collapses. While to ensure the Gaussian feature of the $n\times n$-submatrix, the transformation matrix should be Haar random, and $m$ is in the order of $n^5\log^2n$~\cite{Aaronson2011}. Generally, $m=n^2$ is thought to be sufficient.

Totally, there are up to $\left(_n^{m+n-1}\right)$ output patterns of the photons. However, we care more about the ``collision-free'' cases, where photons don't share the same mode, i.e. no bunching, thus reducing the number of patterns to $\left(_n^m\right)$. The simulation of boson sampling is to draw samples from the distribution described by Eq.~\ref{EQS:BosonSamplingSolution}. It's unavoidable to calculate or approximate permanents, thus the simulation of boson sampling is likely to be hard on classical computers. The two most-efficient permanent-computing algorithms, {\em Ryser}'s algorithm and {\em Balasubramanian-Bax-Franklin-Glynn}'s algorithm are in the time complexity of $O(n^2\cdot 2^n)$ if implemented to pursue the efficiency of parallelism on the target platform with massive parallel processing unit. Though it's relatively easier to achieve the time complexity of $O(n\cdot 2^n)$ when executed serially.

Since boson sampling is hard to simulate on classical computers, it's promising to realize the quantum supremacy through boson sampling, and the benchmark of boson sampling for the supremacy is then raised for discussion. Ref.~\cite{Neville2017} firstly proposed a practical boson sampling simulator through Metropolised independent sampling (MIS), and concluded that the server-level computing hardware is sufficient for 30-photon simulation, and the performance of supercomputer could further afford 50-photon boson sampling~\cite{Wu2018}. With a detailed comparison between the performance of MIS and currently proposed experiments, it's concluded that the supremacy has not been achieved yet, and seems unlikely in the near term.

\section{The Computational Hardness Limit of Classically Simulating Boson Sampling}

In~\cite{Wu2018}, the benchmark on one of the fastest supercomputer, Tianhe-2, suggests that the performance limit of classical computers on boson sampling is to generate a 50-photon sample in about 100 minutes. This sampling rate is obtained on the assumption that the performance of generating a sample could be represented by that of computing a permanent. The computational hardness limit is thus defined in this way as the performance of calculating only one permanent, which asks for that the number of permanents required for one sample should be only one.

The calculation of a permanent is the core problem of classical simulating boson sampling. However, between the calculation of permanent and the generation of a sample, there is a gap caused by sampling algorithms. Current sampling algorithms are not able to reach this computational hardness limit. To generate a sample, often it has to calculate more than one probability, and thus involves calculating more than one permanents, as shown in Fig. 1(b) in main text. For example, the number of permanents required for one sample is a constant (=100) in MIS~\cite{Neville2017}, and for brute force sampling, this number is $\left(_n^{m}\right)$. If an algorithm reaches this limit, the extra cost other than computing one permanent for one effective sample is negligible, and the performance of computing a permanent can represent that of simulating boson sampling. Here we briefly introduce three widely used classical sampling algorithms.

\subsection{Brute force sampling}

Brute force sampling generates samples in a more straightforward way: compute the probabilities of all the output patterns, and then draw samples. Thus it's also called na\"ive sampling. However, even to generate only one sample, brute force sampling has to calculate $\left(_n^m\right)$ permanents in the collision-free regime. Since $m$ is required to be $n^2$, the number of probabilities of the distribution and the corresponding number of permanents required to calculate grows exponentially with $n$ (Fig. 1(b) in main text). It's obvious that the brute force sampling method is no longer feasible when $n$ reaches a certain value. For example, when $n=7$, the quantity of permanents required to calculate is $8.59\times 10^7$, and when $n$ reaches 10, this number grows explosively to $1.73\times 10^{13}$, which seems unacceptable on a classical computer. The other problem of brute force sampling is the storage of the whole distribution. If a single probability is stored in the format of a double precision float number, which occupies 16 bytes on a classical computer, then the storage request for the distribution of a boson sampling instance with 10 photons in 100 modes exceeds 277 Terabytes, and it requires about 156.9 Petabytes for simulating a 30-photon-900-mode instance of boson sampling. Thus brute force sampling is far away from reaching the computational hardness limit of classically simulating boson sampling.

\subsection{Rejection sampling}

\begin{figure}[!t]
  \centering
  \includegraphics[width=0.65\textwidth]{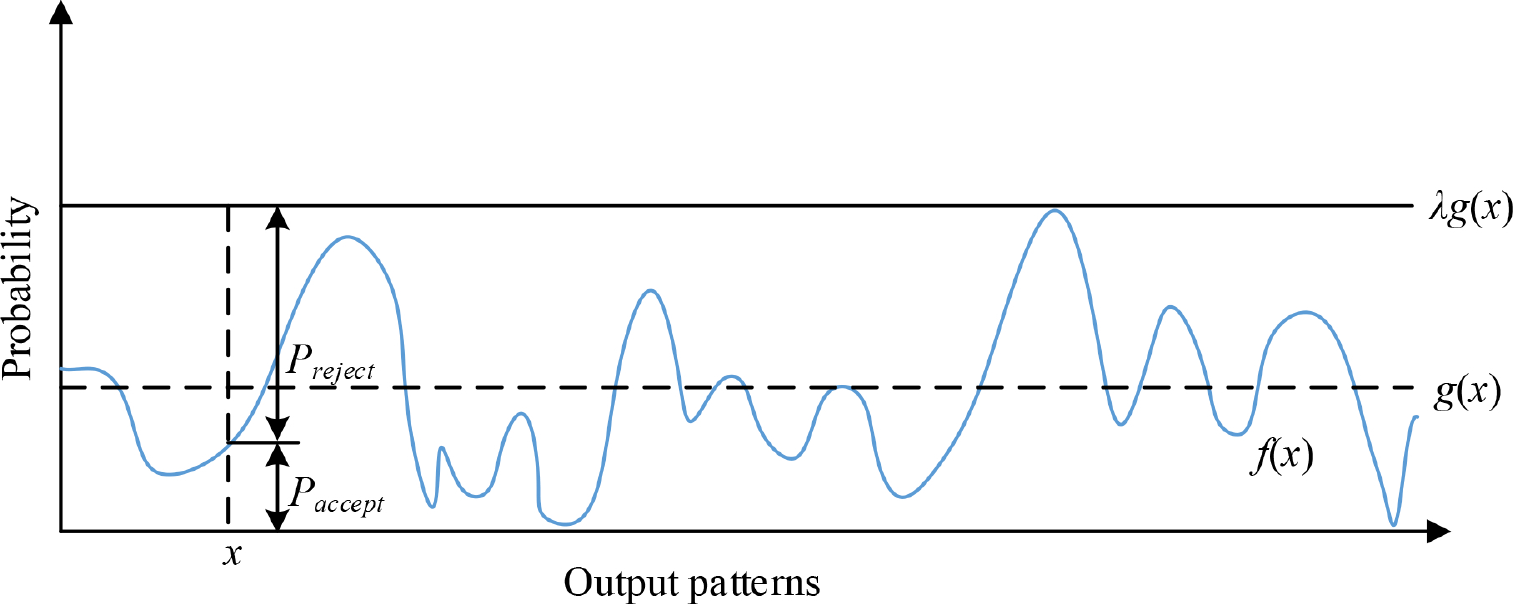}\\
  \caption{\footnotesize The rejection sampling. The sampling is done by first sampling from a easy-to-sample proposal distribution, and judge if output this sample according to a probability decide by ratio between the target distribution and the proposal distribution.}\label{Fig:RejectionSampling}
\end{figure}

Compared to brute force sampling, rejection sampling requires to compute much less permanents for one sample. All what needed is a proposal distribution $g(x)$ that can be efficiently sampled. Before sampling, a parameter $\lambda$ must be decided, so that for any output pattern $x$ of boson sampling, $f(x)<\lambda g(x)$, where $f(x)$ is boson sampling probability for pattern $x$ and $g(x)$ is the corresponding proposal probability. The scheme of rejection sampling is shown in Fig.~\ref{Fig:RejectionSampling}.

The process of rejection sampling repeats the following two steps:
\begin{enumerate}
  \item randomly sample a pattern $x$ from $g(x)$;
  \item calculate $f(x)$, and then output the sample of pattern $x$ with probability $P_{accept}=\frac{f(x)}{\lambda g(x)}$, or discard this sample with probability $P_{reject}=1-P_{accept}$.
\end{enumerate}

Thus for each sample, it's a probabilistic event that the sampler outputs the sample or not, and finally the distribution we draw samples from approaches the target probability distribution. However, it's hard to choose a proposal distribution that overlaps quite well with the boson sampling distribution. A feasible way is to choose the uniform distribution, and let $\lambda$ satisfy that after multiplying $\lambda$, this uniform distribution exactly covers the largest probability of the output patterns in boson sampling, in this case the boson sampling distribution itself has great influence on the success probability, and it brings in another problem: the estimation of the largest probability of boson sampling. Ref.~\cite{Neville2017} discussed about the quantity of permanents used for a sample in rejection sampling. Actually the average probability to accept a sample with uniform proposal is $1/\lambda$, and the computational hardness core problem is repeated for $\lambda$ times for one effective sample. However, though much less than brute force sampling, the number of permanents required to calculate for one effective sample also grows very fast (Fig. 1(b) in main text). In conclusion, rejection sampling is also far from the computational hardness limit of classically simulating boson sampling.

\subsection{Markov chain Monte Carlo}

\begin{figure}[!t]
  \centering
  \includegraphics[width=0.95\textwidth]{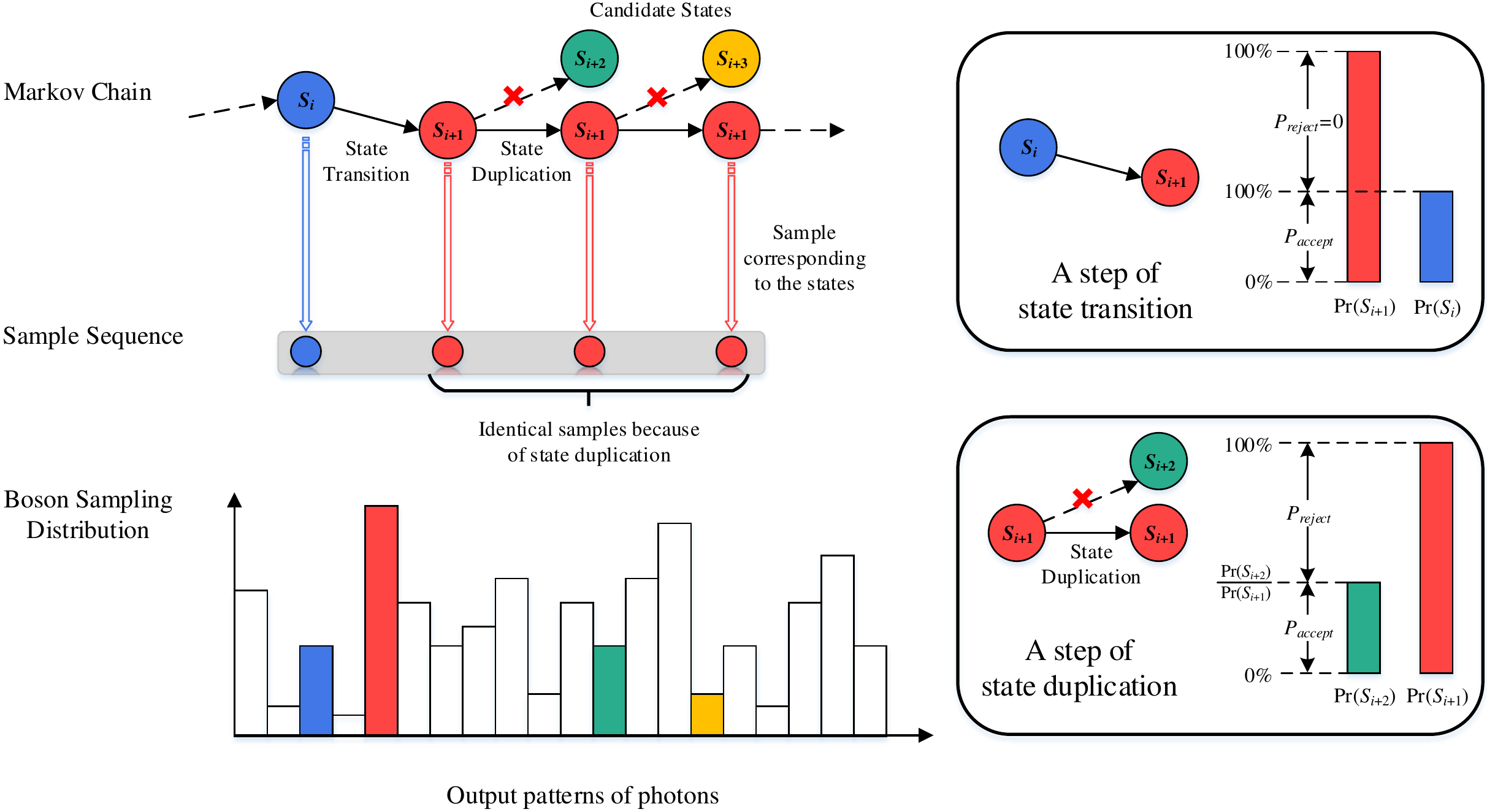}\\
  \caption{\footnotesize The Markov chain Monte Carlo. The sampling is done by constructing a Markov chain. A new node is added on the chain by choosing a candidate state and then probabilistically accept it. If this candidate is rejected, duplicate current state as the new node. Each time a new node is added on the chain, the sample corresponds to the state is outputted. The probability for acceptance/rejection is determined by the ratio of probabilities of the states.}\label{FigS:MCMC}
\end{figure}

In Markov chain Monte Carlo (MCMC), the sampling process is done through constructing a Markov chain with the target distribution as the stationary distribution. The state space, denoted as $S=\{s_1, s_2, ..., s_N\}$, of the Markov chain corresponds to all the $\left(_n^m\right)$ output patterns of boson sampling. Each time the chain expands, a sample could be generated, as shown in Fig.~\ref{FigS:MCMC}. Since we limit the sample space in the collision-free case, the size of the state space is $\left(_n^m\right)$, with each state representing one output pattern of photons.

To be specifically, the algorithm we use is Metropolis-Hastings algorithm with symmetric proposal distribution $g(x)$ which is easy to sample from. The process of the Metropolis-Hastings algorithm is the iteration of the following three steps:
\begin{enumerate}
  \item If current state is $s_i$, choose the next state $s_j$ from distribution $g(s_j|s_i)$, which is the probability to choose $s_j$ as the candidate when current state is $s_i$. The choice of the proposal distribution is arbitrary. For symmetric proposal, which is frequently used, it satisfies $g(s_i|s_j)=g(s_j|s_i)$.
  \item Calculate $p(s_i)$, which is the boson sampling probability for state $s_j$, and accept $s_j$ as the new state with probability $P_{accept}=\min\left(1,\frac{p(s_j)}{p(s_i)}\right)$, or duplicate $s_i$ as the new state with the rest probability;
  \item Output the sample corresponding to the new state.
\end{enumerate}

Note that in step 2, the probability for acceptance would not always be $\min\left(1,\frac{p(s_j)}{p(s_i)}\right)$, because for other Metropolis-Hastings algorithms, the symmetric condition of the proposal density is not necessary, expanding the probability to be $P_{accept}=\min\left(1, \frac{p(s_j)\cdot g(s_i|s_j)}{p(s_i)\cdot g(s_j|s_i)}\right)$. As we will see in the following sections, the choice of proposal density would have influence on the convergence speed of the Markov chain.

Unlike the rejection sampling, the MCMC method will generate a sample even the candidate sample is rejected. Therefore, it seems ideal that for one sample, we just need to calculate one permanent, and thus the computational hardness limit could be reached. However, the sample generated may be erroneous of the correlations between samples. Usually the correlations are eliminated at the cost of impairing the sampling efficiency, and makes it away from the computational hardness limit of classically simulating boson sampling. Here we developed an algorithm to avoid the efficiency impairment while reduce the correlations between samples. We next analyze the autocorrelation problem in MCMC.

\section{The Sample Caching Markov Chain Monte Carlo Method}

We develop an algorithm and try to reach the computational hardness limit by eliminating the autocorrelation of the sequence generated by the MCMC without abandoning samples. To see this, we will introduce the autocorrelation problem in MCMC first.

\subsection{The autocorrelation of the sample sequence generated by Markov chain}

The autocorrelation of the sample sequence is the fact that the samples are correlated, and thus may lead to erroneous samples. The autocorrelation of the sequence at lag $k$ is defined by Eq.~\ref{EQS:AUTOCORRELATION}
\begin{equation}\label{EQS:AUTOCORRELATION}
r_{k}=\frac{\mathds{E}\left[(x_t-\mu)(x_{t+k}-\mu)\right]}{\sigma^2},
\end{equation}
where $x_t$ is value of $t^{th}$ sample, and $\mu$, $\sigma^2$ are the mean and the variance of the distribution. Usually, the true mean and variance are unknown, and are often replaced by the sample mean and variance. The autocorrelation then can be estimated as
\begin{equation}\label{EQS:RK}
r_{k}=\frac{\sum_{t=1}^{n-k}(x_t-\mu_s)(x_{t+k}-\mu_s)}{\sigma_s^2},
\end{equation}
where $\mu_s$ and $\sigma_s^2$ are the mean and variance of the samples. Specially, for first-order autocorrelation, there are other test statistics, such as the Durbin-Watson statistic~\cite{Durbin1971}, as expressed by Eq.~\ref{EQS:DWTest}
\begin{equation}\label{EQS:DWTest}
d=\frac{\sum_{t=1}^T(x_t-x_{t+1})^2}{\sum_{t=1}^Tx_t^2}.
\end{equation}
The relationship between $d$-statistic and the first-order correlation can be given by $r_{1}\approx1-\frac{d}{2}$. We directly estimate the first-order autocorrelation through Eq.~\ref{EQ:1autoc}(main text), which is the special case of Eq.~\ref{EQS:RK} by assigning $k=1$. The value of $r_k$ should be in $[-1,1]$, and it would reflect the autocorrelation in the sequence. The closer $|r_k|$ is to 1, the stronger the samples are correlated, and the sign of $r_k$ indicates if the samples are positively or negatively correlated.

From the definition of autocorrelation, we can analyze the autocorrelation problem. Given a sample sequence $X=\{x_1, x_2, ..., x_N\}$ with $x_i$ as the $i^{th}$ sample taken from a distribution described by $p(\cdot)$ and if the samples are independent, it's easy to observe that
\begin{equation}
\begin{aligned}
&\mathds{E}\left[(x_t-\mu_t)(x_j-\mu_t)\right]/\sigma^2\\
=&\frac{1}{\sigma^2}\sum_{i,j}p(x_i,x_j)(x_i-\mu_t)(x_j-\mu_t) \\
=&\frac{1}{\sigma^2}\sum_{i,j}p(x_i)p(x_j|x_i)(x_i-\mu_t)(x_j-\mu_t) \\
%=&\frac{1}{\sigma^2}\sum_{i,j}p(x_i)p(x_j)(x_i-\mu_t)(x_j-\mu_t) \\
=&\frac{1}{\sigma^2}\sum_{i}p(x_i)(x_i-\mu_t)\sum_{j}p(x_j)(x_j-\mu_t) \\
=&0,
\end{aligned}
\end{equation}
where $p(x_i)$ is the probability for event $x_i$. Note that since the samples are independent, $p(x_j|x_i) = p(x_j)$ for arbitrary $i$ and $j$, and thus result in no autocorrelation for arbitrary $k$.

In one step of state transition in Markov chain with state space $S=\{s_1, s_2, ..., s_N\}$, the probability of the transition from a state to another forms a matrix $P$, with the element at $i^{th}$ row and $j^{th}$ column representing the probability to transit from state $s_i$ to $s_j$, as shown in Eq.~\ref{EQ:MAKOVMATRIX}
\begin{equation}\label{EQ:MAKOVMATRIX}
p(s_j|s_i)=p_{ij}=\left\{\begin{array}{ll}g(s_j|s_i)\cdot\min\left(1,\frac{p(s_j)}{p(s_i)}\right), & i\neq j\\
1-\sum_{l=1}^{i-1}p_{il}-\sum_{l=i+1}^{N}p_{il}, & i=j,
\end{array}\right.
\end{equation}
where $p(s_i)$ is the probability for state $s_i$ and $g(s_j|s_i)$ is the probability from the proposal distribution, which acts as the candidate selection probability. In the Markov chain Monte Carlo, this conditional probability $p(s_j|s_i)$ is just the transition probability $p_{ij}$. However, in most cases, $p_{ij}\neq p(s_j)$, which leads to the correlations between the adjacent samples. For the state-duplication cases, the adjacent samples are exactly identical, which leads to a severe autocorrelation problem and erroneous samples.

\subsection{Jump sampling method}

\begin{figure}[!t]
  \centering
  \begin{minipage}[t]{0.24\linewidth}
  \includegraphics[width=\textwidth]{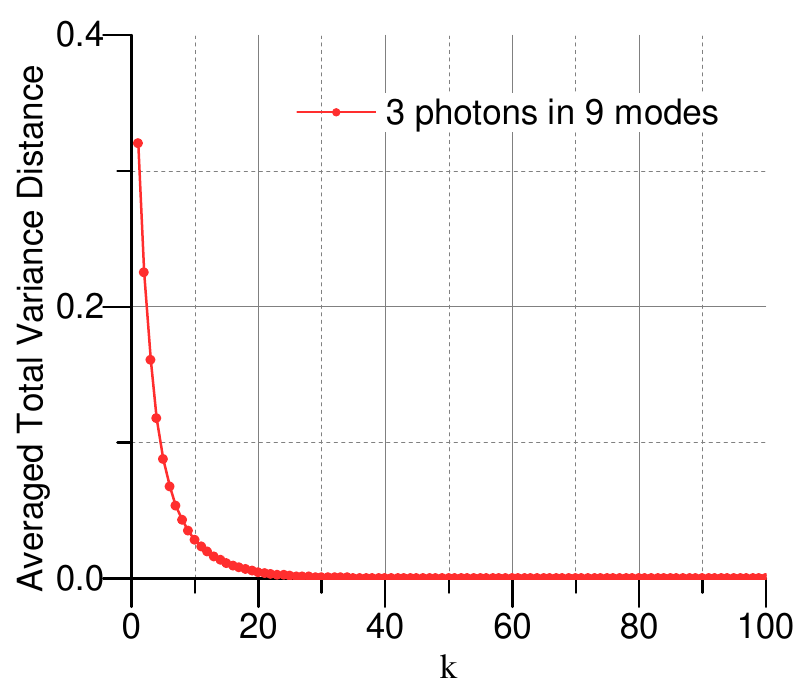}
  \label{fig:side:a}
  \end{minipage}
  \begin{minipage}[t]{0.24\linewidth}
  \includegraphics[width=\textwidth]{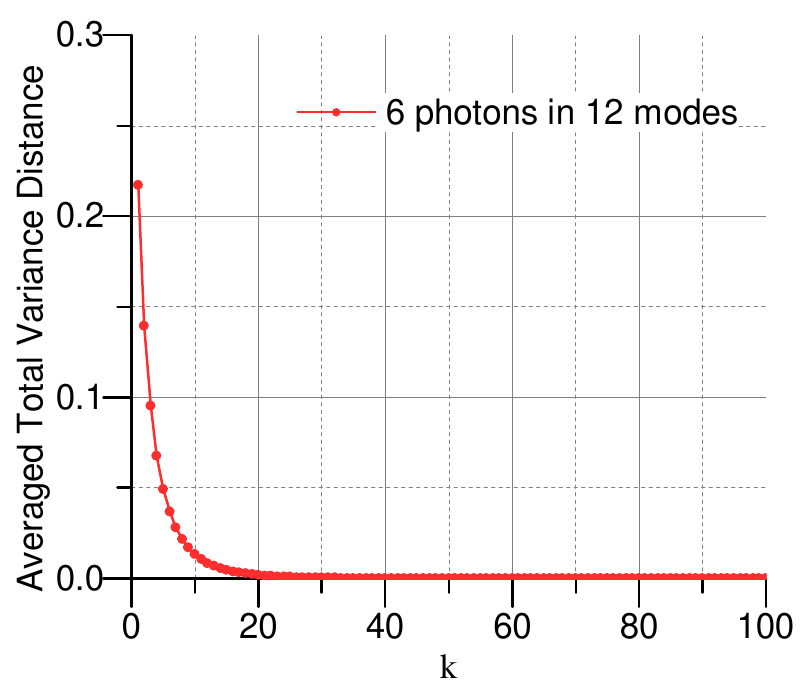}
  \label{fig:side:b}
  \end{minipage}
  \begin{minipage}[t]{0.24\linewidth}
  \includegraphics[width=\textwidth]{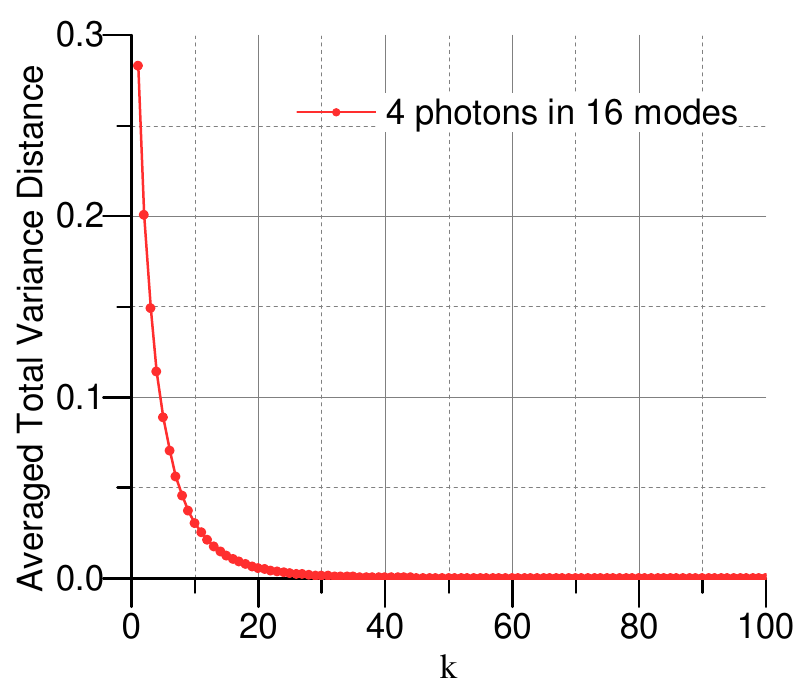}
  \label{fig:side:c}
  \end{minipage}
  \begin{minipage}[t]{0.24\linewidth}
  \includegraphics[width=\textwidth]{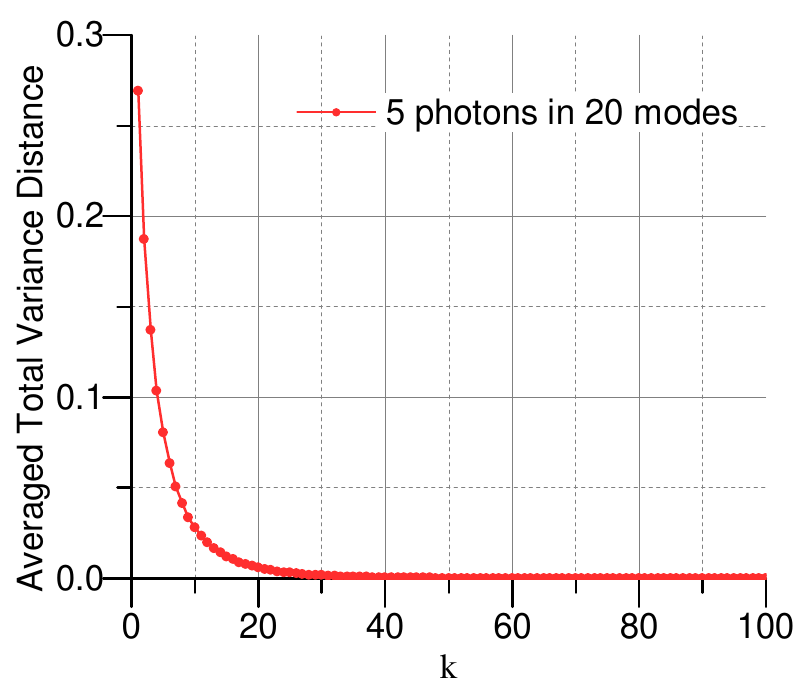}
  \label{fig:side:d}
  \end{minipage}
  \caption{\footnotesize The averaged total variance distance of $P^{(k)}$. Two of the cases do not satisfy the condition that $m=n^2$. The averaged total variance distance reaches blow $10^{-3}$ when the numbers of steps reach 29, 24, 33, 35 respectively.}
  \label{FIGS:CONVERG}
\end{figure}

In Metropolised independence sampling~\cite{Neville2017}, the autocorrelation is eliminated through multi-step transition, i.e. the jump sampling (or thinning procedure). The transition from $s_i$ to $s_j$ in $k$ steps can be described by matrix $P^{(k)}$. It's easy to observe that $P^{(k)}=P^k$. If the Markov chain would converge, then for arbitrary $i$ and $j$, $\lim\limits_{k\to\infty}p^{(k)}_{ij}=p(s_j)$~\cite{Liu1996,Gagniuc2017}. To show this, we define Averaged Total Variance Distance of $P^{(k)}$ between the $k$-step transition probability matrix and the intrinsic probability of the states
\begin{equation}
d_{A}=\sum_{j=1}^N\left(\frac{1}{N}\sum_{i=1}^N\left|p^{(k)}_{ij}-p(s_j)\right|\right),
\end{equation}
where $N$ is the size of the state space. $d_{A}=0$ indicates $p^{(k)}_{ij}=p(s_j)$ for arbitrary $i$ and $j$. In the test, the proposal distribution is $g(s_j|s_i)=\frac{1}{N}$ for arbitrary $i$ and $j$. For example, the results of $d_{A}$ between $P^{(k)}$ for several scales are shown in Fig.~\ref{FIGS:CONVERG}.

Thus as long as the Markov chain would converge, for a certain number $k$, the difference between $p^{(k)}_{ij}$ and $p(s_j)$ is negligible. By throwing away the samples within the $k$ steps, the remained samples are approximately independently. Thus MIS generate an effective sample by calculating $k$ permanents. A proposed value of $k$ is 100, which is sufficient for eliminating the autocorrelation of the sample sequence from more than 30-photon boson sampling scheme, and is with only a constant number from the computational hardness limit of simulating boson sampling. The time estimated for Tianhe-2 supercomputer to generate a 50-photon sample is about 10 days. However, if the samples within the leap is not discarded and is reused, then the efficiency of the algorithm would be greatly enhanced, which is the main idea of our method.

\subsection{Sample caching}

We proposed the Sample Caching Markov Chain Monte Carlo(SC-MCMC) to reduce the autocorrelation without losing any samples. Our protocol mainly contains two parts: a complete MCMC sampler combined with a procedure that we call ``Sample Caching''. The SC-MCMC algorithm is shown as Algorithm~\ref{ALGS:SC-MCMC}.

\begin{algorithm}[H]
  \caption{\small Sample Caching Markov chain Monte Carlo algorithm}
  \label{ALGS:SC-MCMC}
  \begin{algorithmic}[1]
    \Require $Cache$ : Sample Cache; $L$ : Size of the Sample Cache; $SN$: number of samples required
    \Ensure Un-correlated sample sequence
        \For{$i$ = 1:$SN$}
            \State Sample={\bf MCMC}();\Comment{Generate a sample using MCMC}\label{ALGL:MCMC}
            \If{{\bf not} {\bf Full}($Cache$)}
                \State {\bf add}($Cache$, Sample);\Comment{Store the sample in the Cache}
            \Else
                \State u = {\bf UniformRand}([1,$L$])
                \State {\bf Output}($Cache$[u]);\Comment{Pick a sample to output randomly}
                \State {\bf add}($Cache$, Sample, u);\Comment{Store a new sample to the empty slot}
            \EndIf
        \EndFor
        \State Output the samples in the $Cache$ randomly;\Comment{Deal with the samples in the cache when the sampling process is over}
  \end{algorithmic}
\end{algorithm}

\begin{figure}[!htb]
  \centering
    \subfigure[n=3,m=9]{\includegraphics[width=0.22\textwidth]{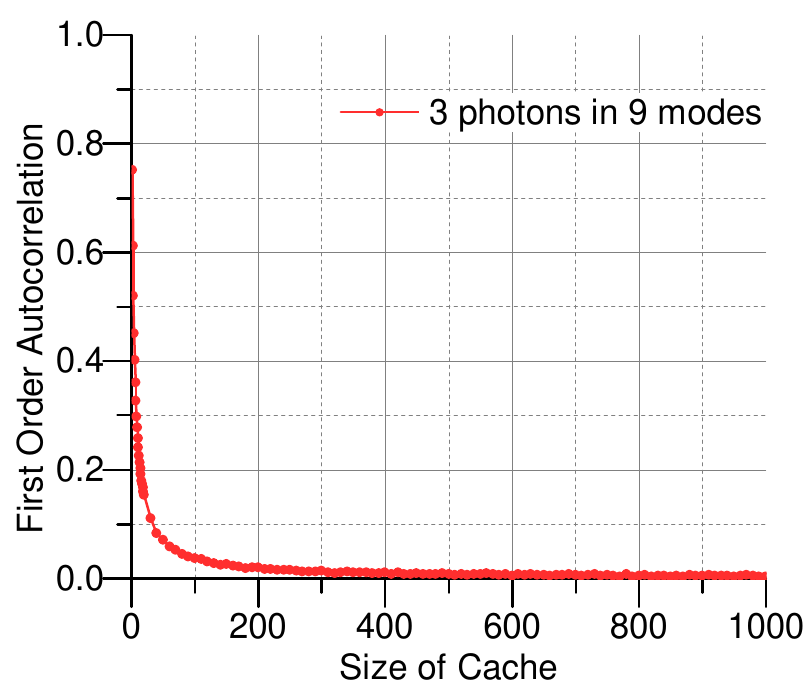}} \hspace{0.1cm}
    \subfigure[n=5,m=25]{\includegraphics[width=0.22\textwidth]{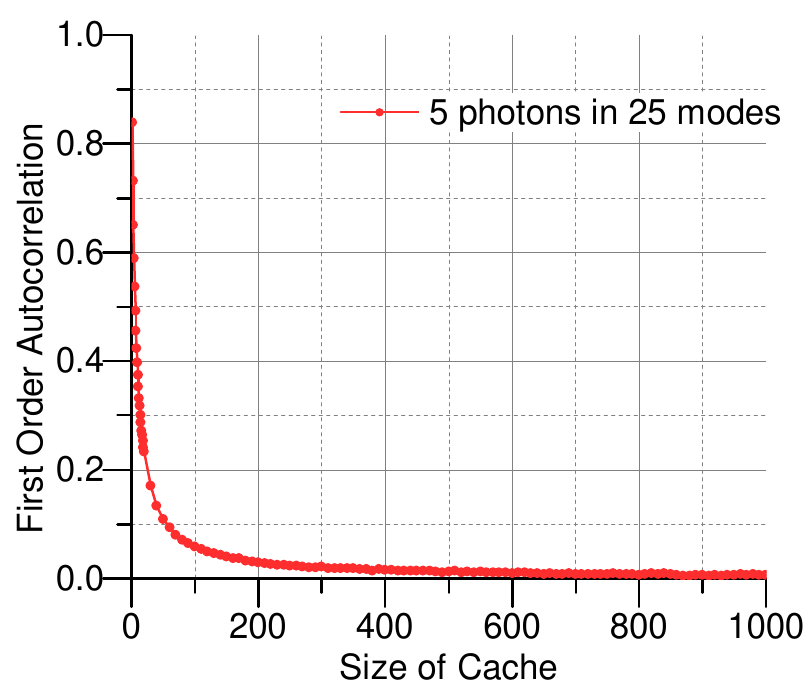}} \hspace{0.1cm}
    \subfigure[n=6,m=36]{\includegraphics[width=0.22\textwidth]{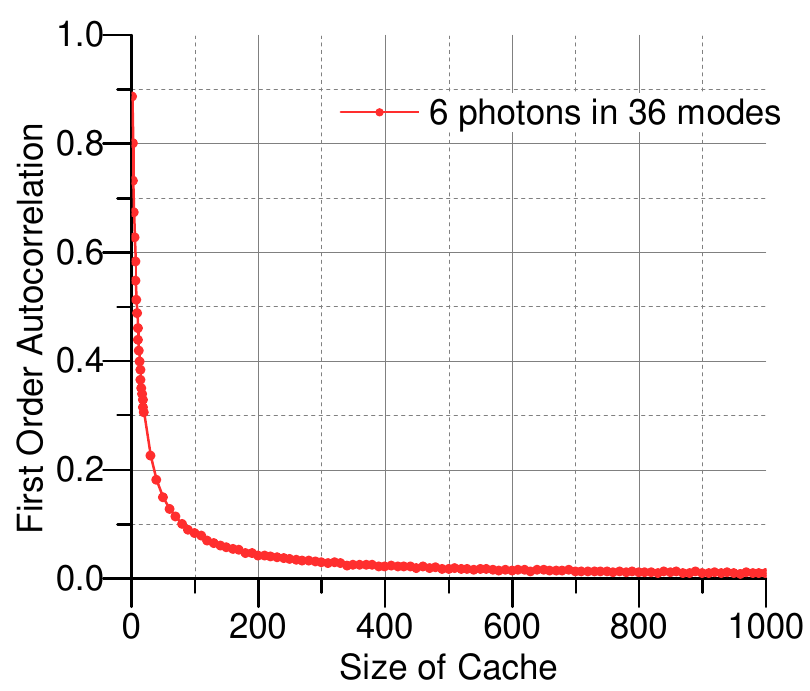}} \hspace{0.1cm}
    \subfigure[n=7,m=49]{\includegraphics[width=0.22\textwidth]{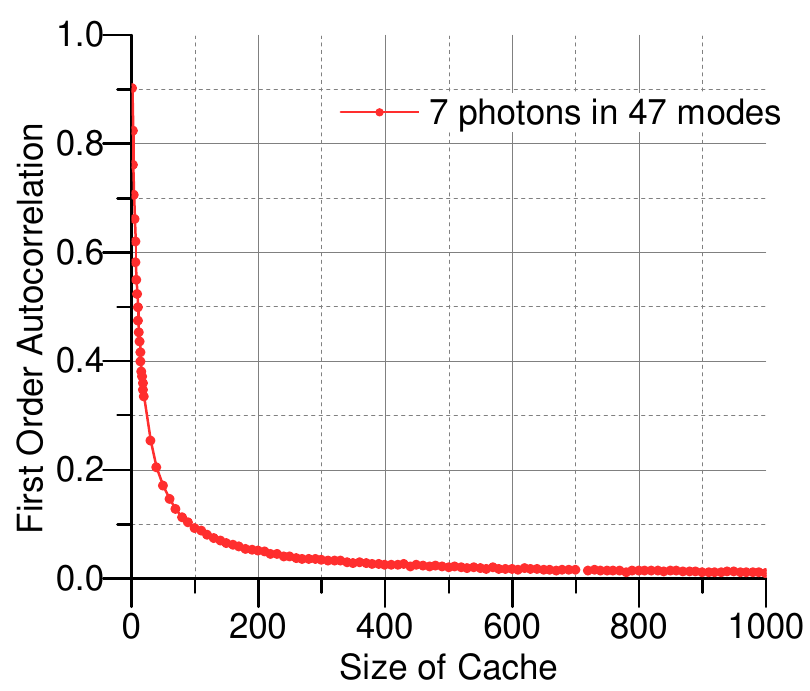}}
    \subfigure[n=8,m=64]{\includegraphics[width=0.22\textwidth]{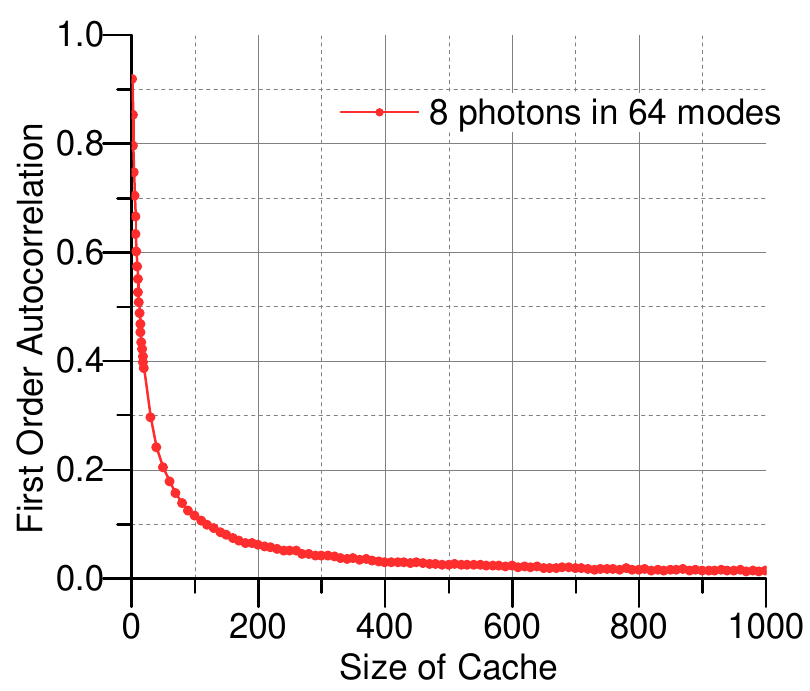}} \hspace{0.1cm}
    \subfigure[n=9,m=81]{\includegraphics[width=0.22\textwidth]{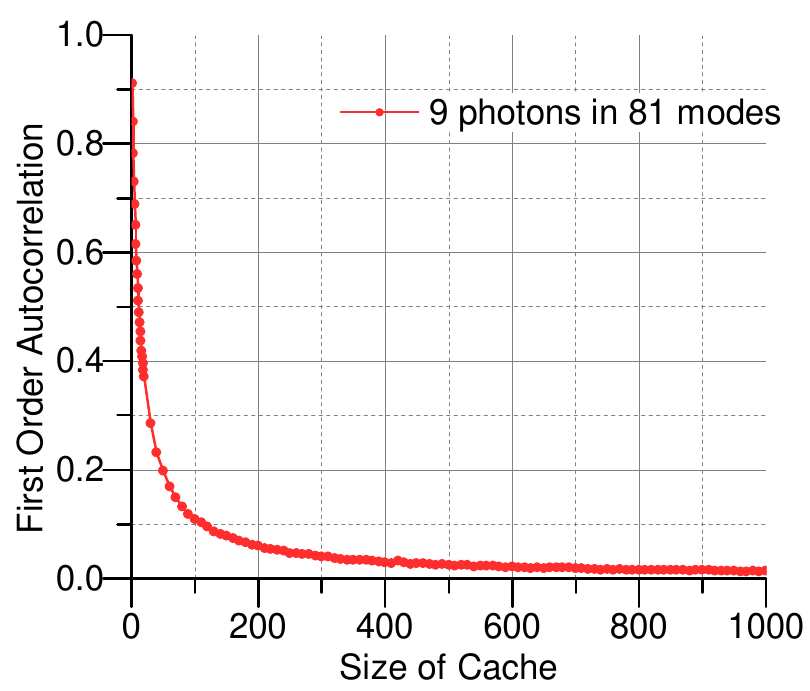}} \hspace{0.1cm}
    \subfigure[n=10,m=100]{\includegraphics[width=0.22\textwidth]{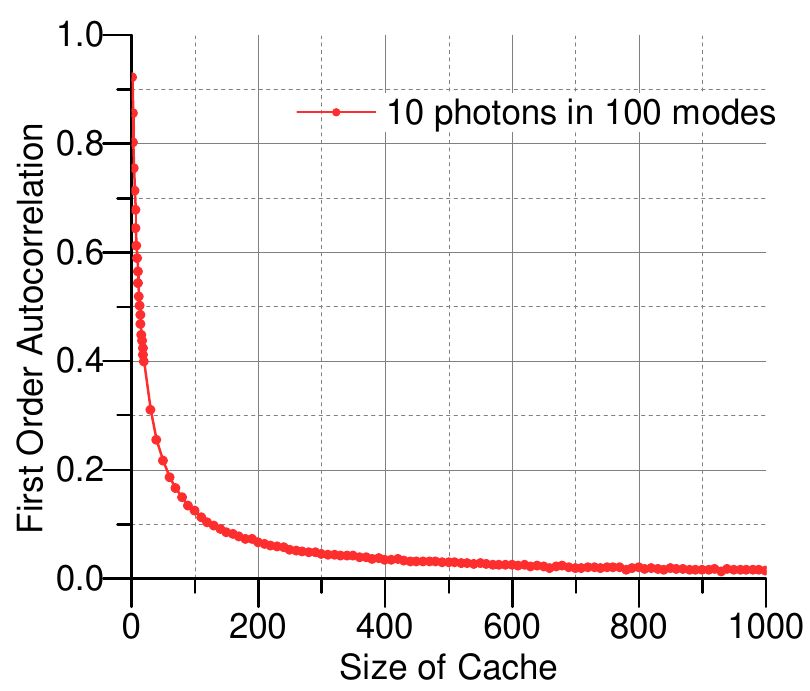}} \hspace{0.1cm}
    \subfigure[n=11,m=121]{\includegraphics[width=0.22\textwidth]{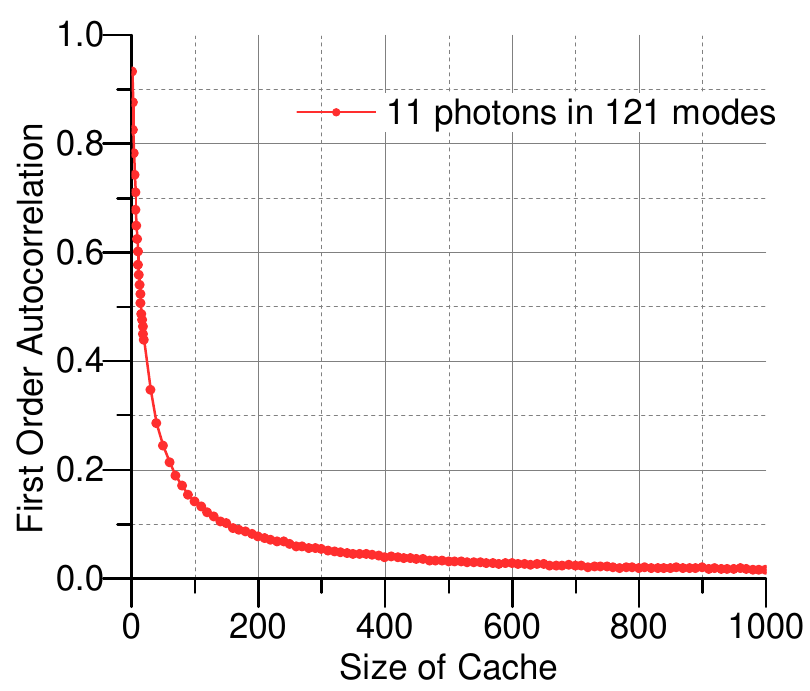}}
    \subfigure[n=12,m=144]{\includegraphics[width=0.22\textwidth]{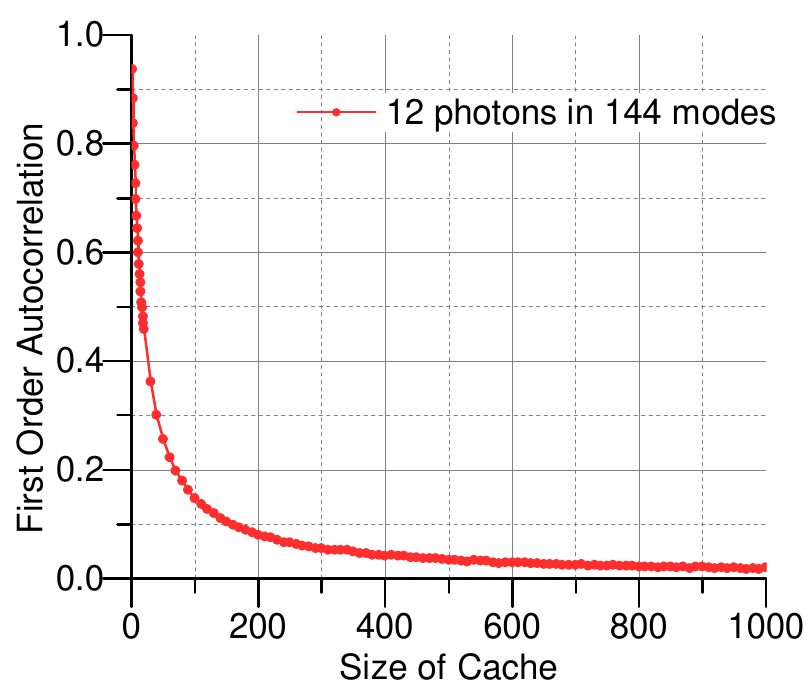}} \hspace{0.1cm}
    \subfigure[n=13,m=169]{\includegraphics[width=0.22\textwidth]{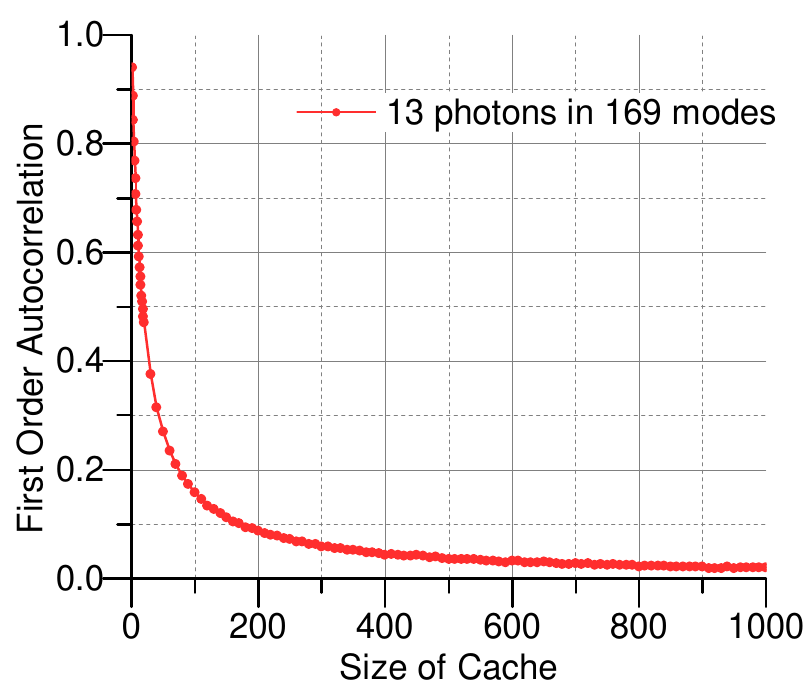}} \hspace{0.1cm}
    \subfigure[n=14,m=196]{\includegraphics[width=0.22\textwidth]{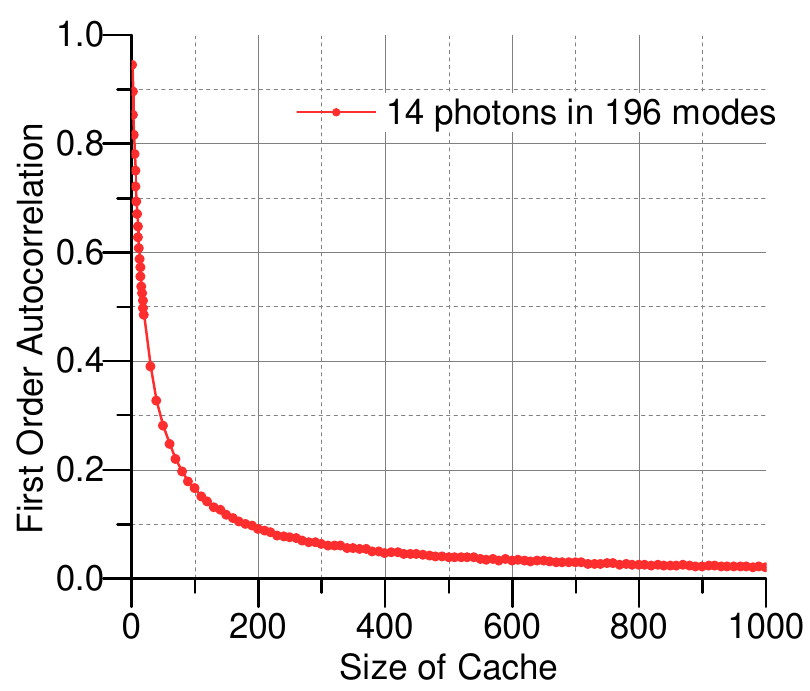}} \hspace{0.1cm}
    \subfigure[n=15,m=225]{\includegraphics[width=0.22\textwidth]{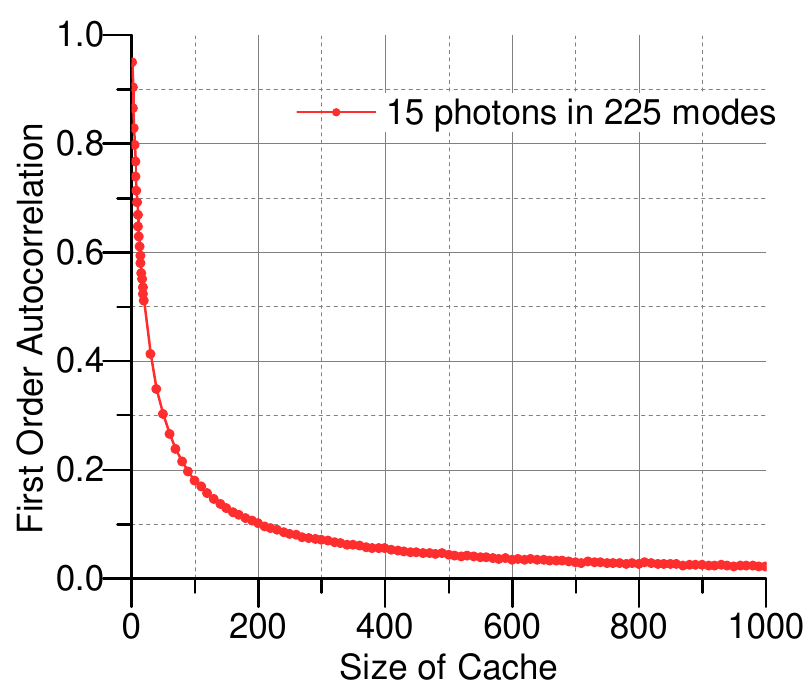}}
    \subfigure[n=16,m=256]{\includegraphics[width=0.22\textwidth]{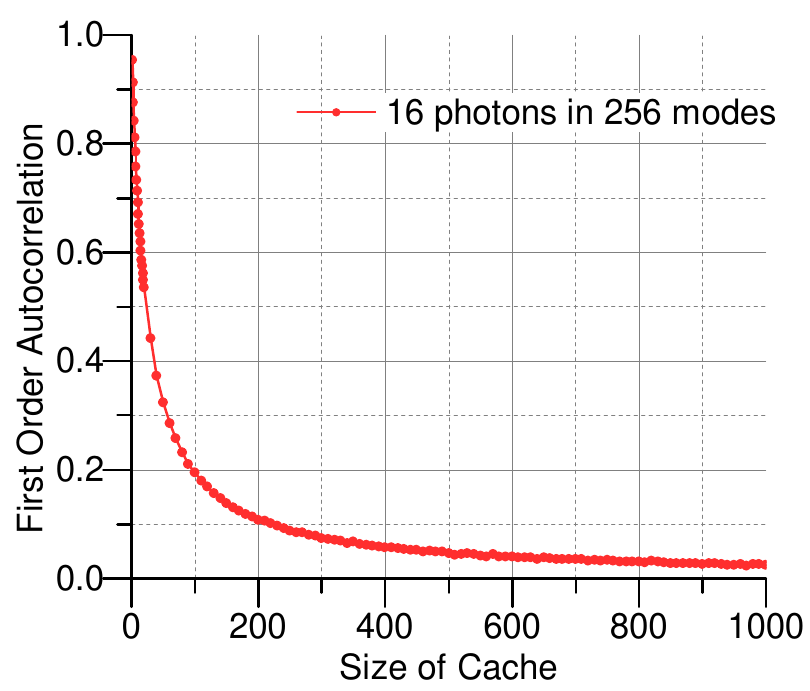}} \hspace{0.1cm}
    \subfigure[n=17,m=289]{\includegraphics[width=0.22\textwidth]{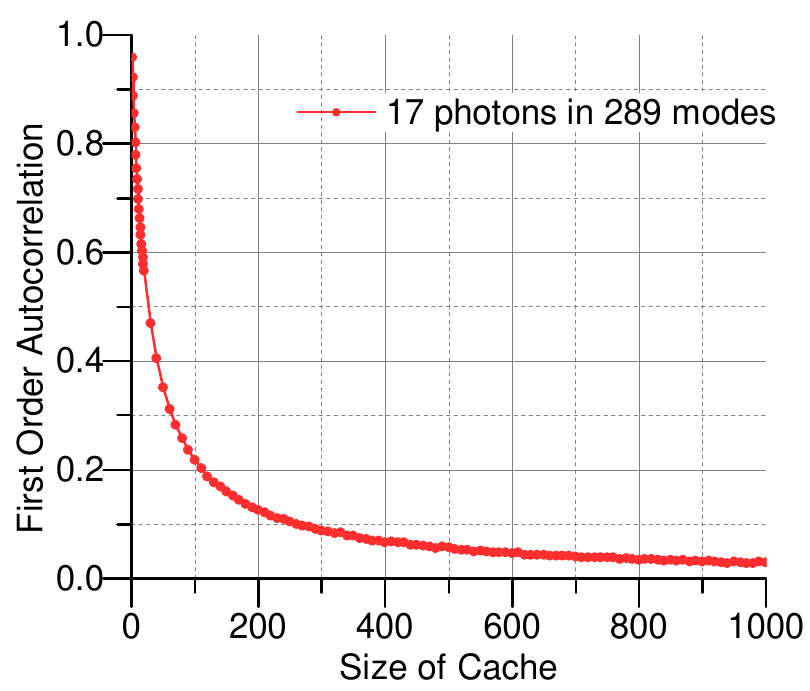}} \hspace{0.1cm}
    \subfigure[n=18,m=324]{\includegraphics[width=0.22\textwidth]{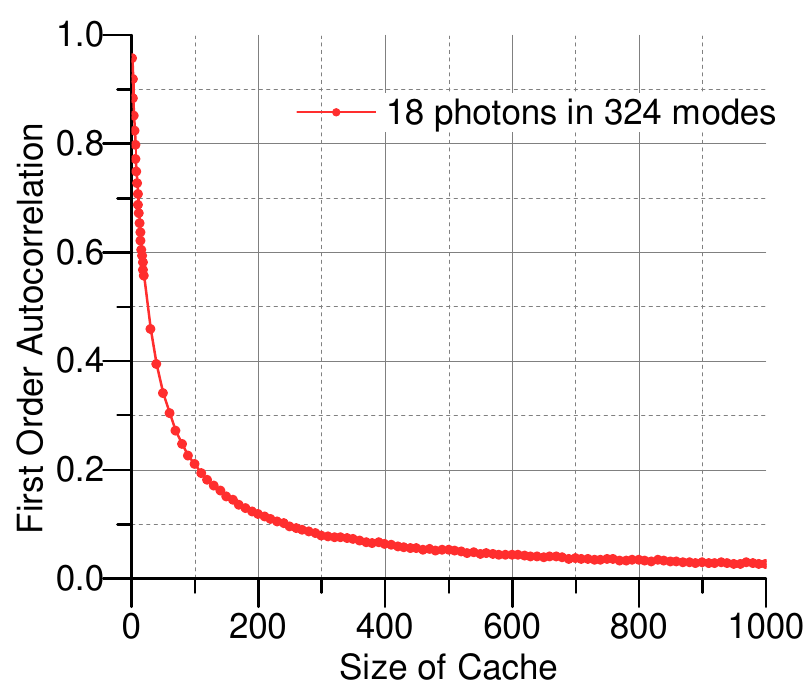}} \hspace{0.1cm}
    \subfigure[n=19,m=361]{\includegraphics[width=0.22\textwidth]{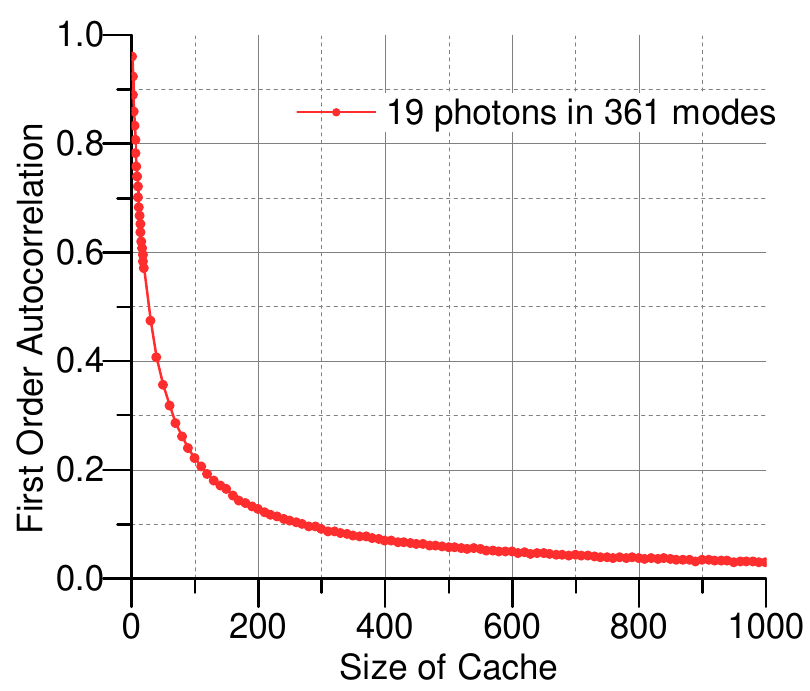}}
    \subfigure[n=20,m=400]{\includegraphics[width=0.22\textwidth]{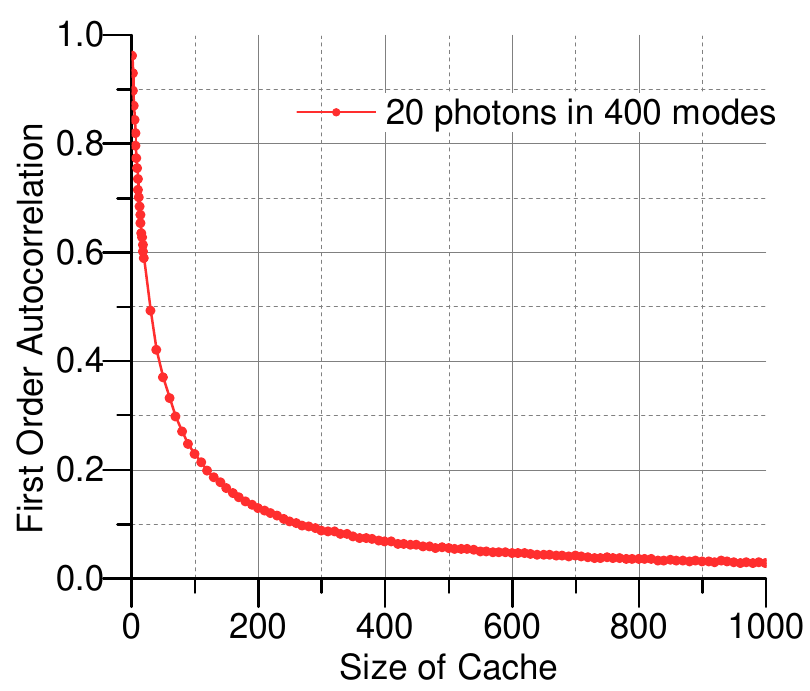}} \hspace{0.1cm}
    \subfigure[n=21,m=441]{\includegraphics[width=0.22\textwidth]{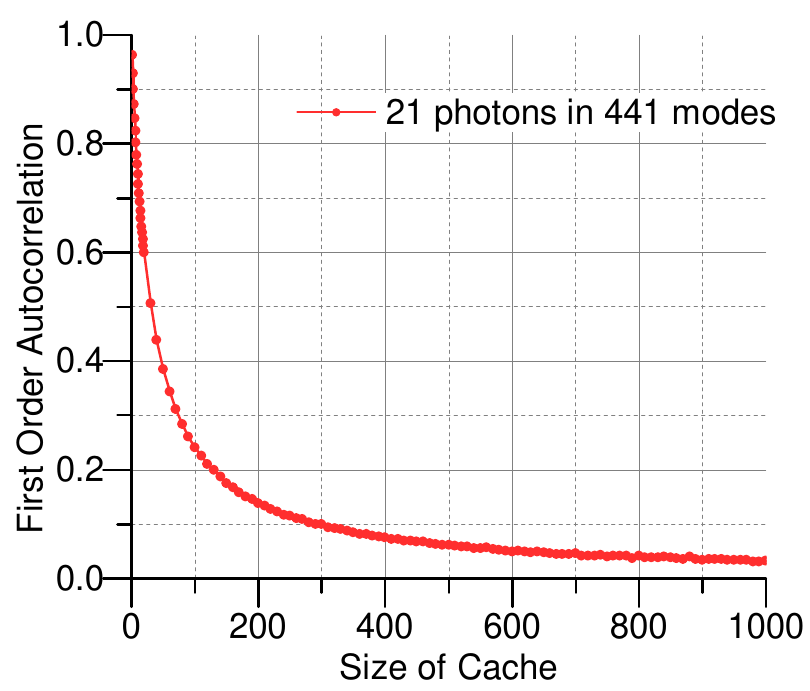}} \hspace{0.1cm}
    \subfigure[n=25,m=625]{\includegraphics[width=0.22\textwidth]{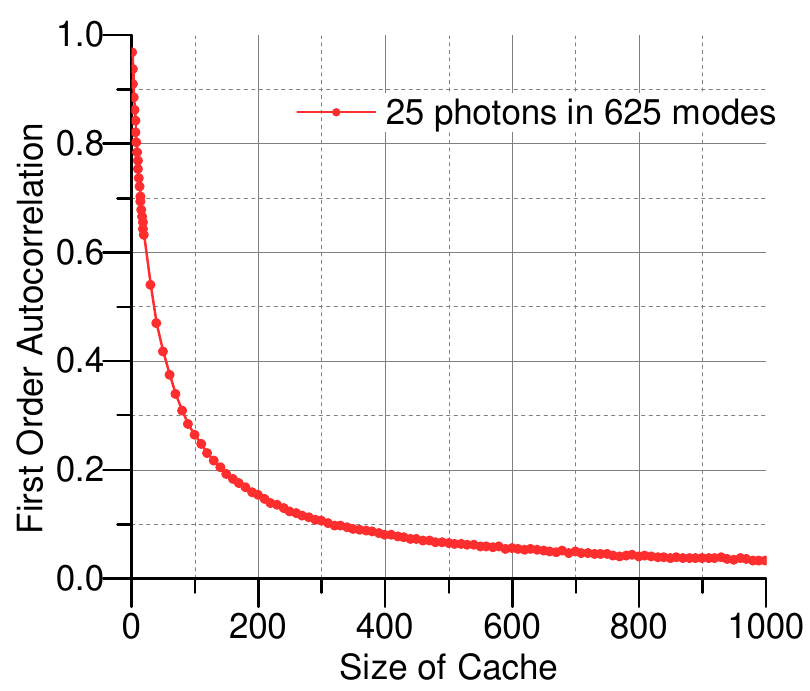}} \hspace{0.1cm}
    \subfigure[n=30,m=900]{\includegraphics[width=0.22\textwidth]{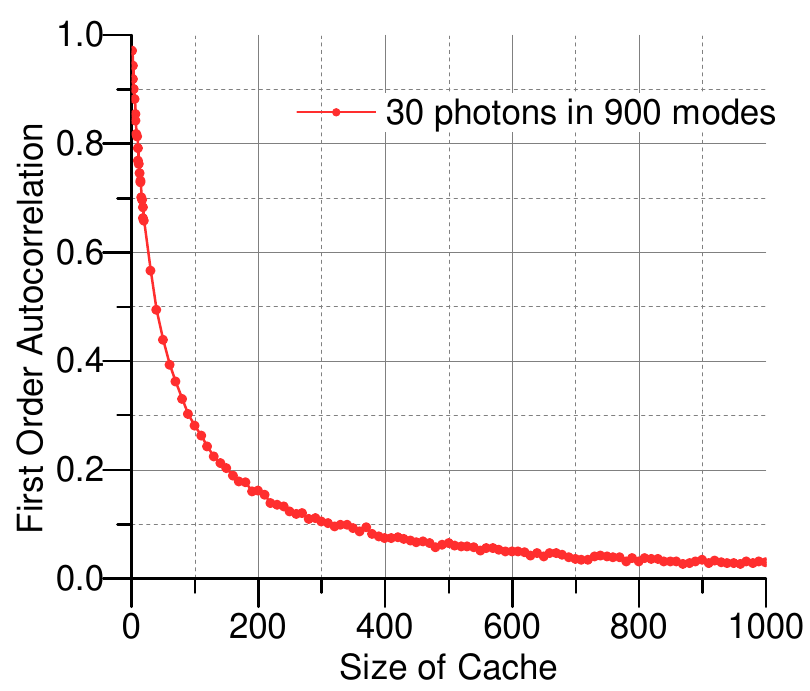}}
  \caption{\footnotesize The autocorrelation is reduced when the size of cache increased. The autocorrelation is negligible when the size of cache arrives at 1,000. Also the reducing speed of autocorrelation is slower in larger scales. For each scale, 1,000,000 samples are taken except the 25-photon-625-mode and 30-photon-900-mode cases where only 200,000 and 20,000 samples are taken respectively.}
  \label{FIGS:AutocorrelationReduceWithIncreaseOfCacheSize}
\end{figure}

\begin{figure}[!htb]
  \centering
  \subfigure[n=5,m=25,MCMC]{\includegraphics[width=0.21\textwidth]{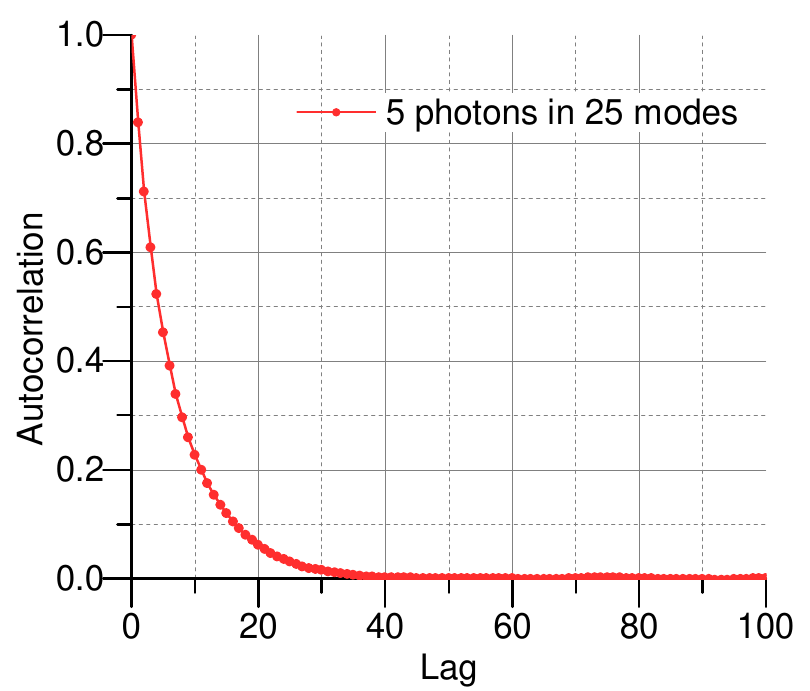}} \hspace{0.2cm}
  \subfigure[n=5,m=25,SC-MCMC]{\includegraphics[width=0.21\textwidth]{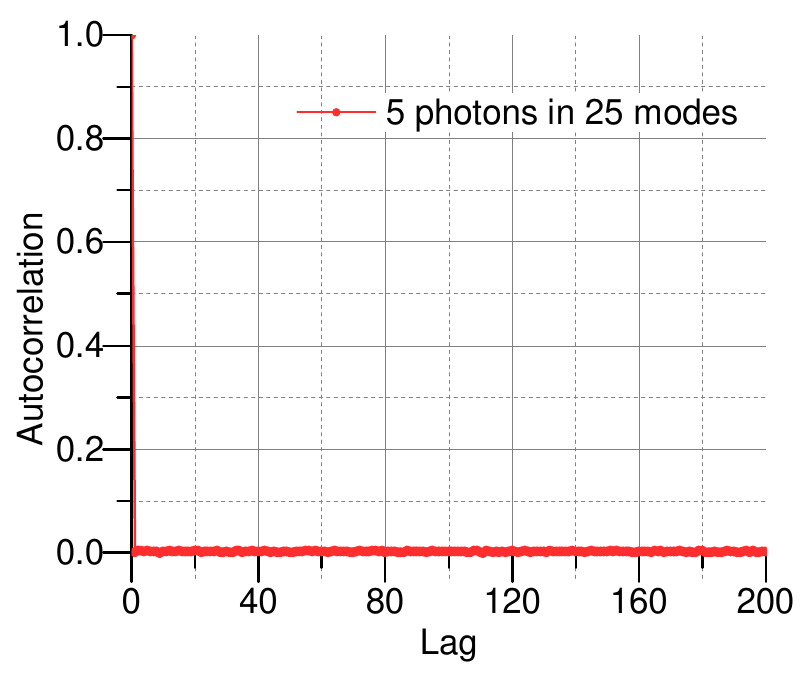}} \hspace{0.2cm}
  \subfigure[n=10,m=100,MCMC]{\includegraphics[width=0.21\textwidth]{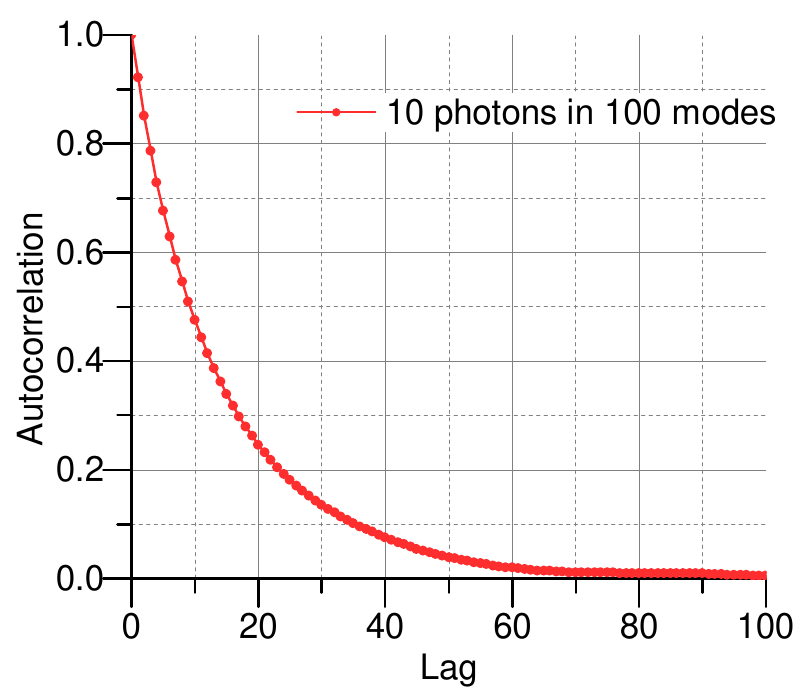}} \hspace{0.2cm}
  \subfigure[n=10,m=100,SC-MCMC]{\includegraphics[width=0.21\textwidth]{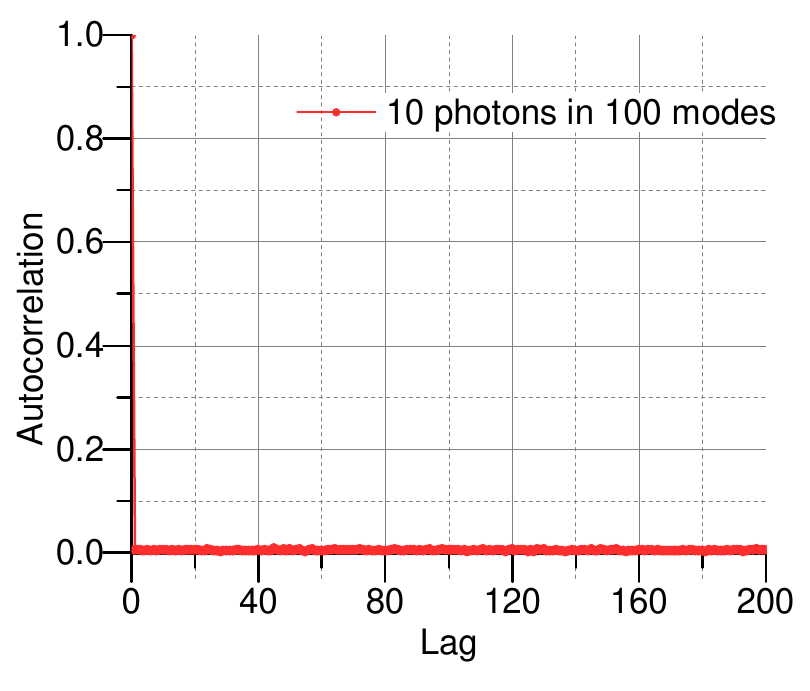}}
  \subfigure[n=12,m=144,MCMC]{\includegraphics[width=0.21\textwidth]{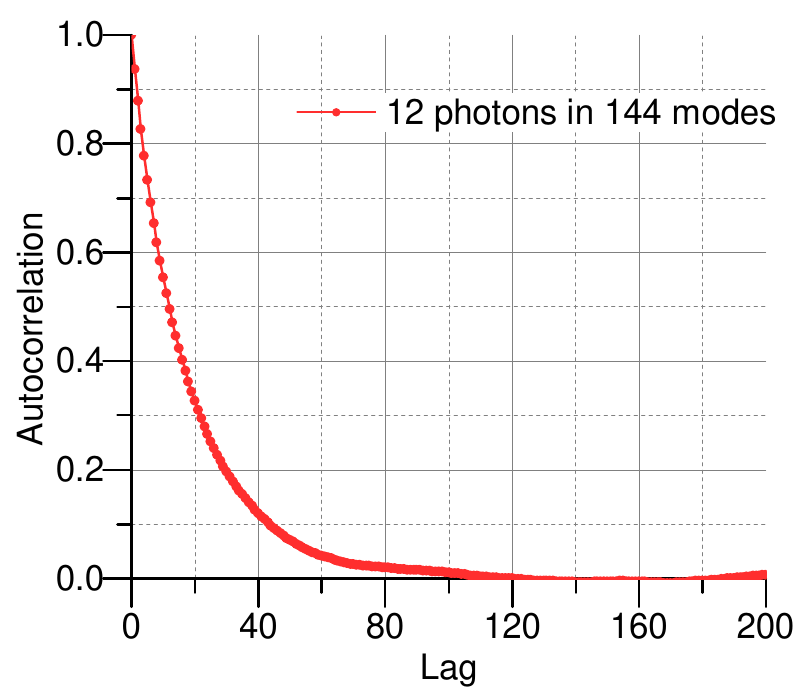}} \hspace{0.2cm}
  \subfigure[n=12,m=144,SC-MCMC]{\includegraphics[width=0.21\textwidth]{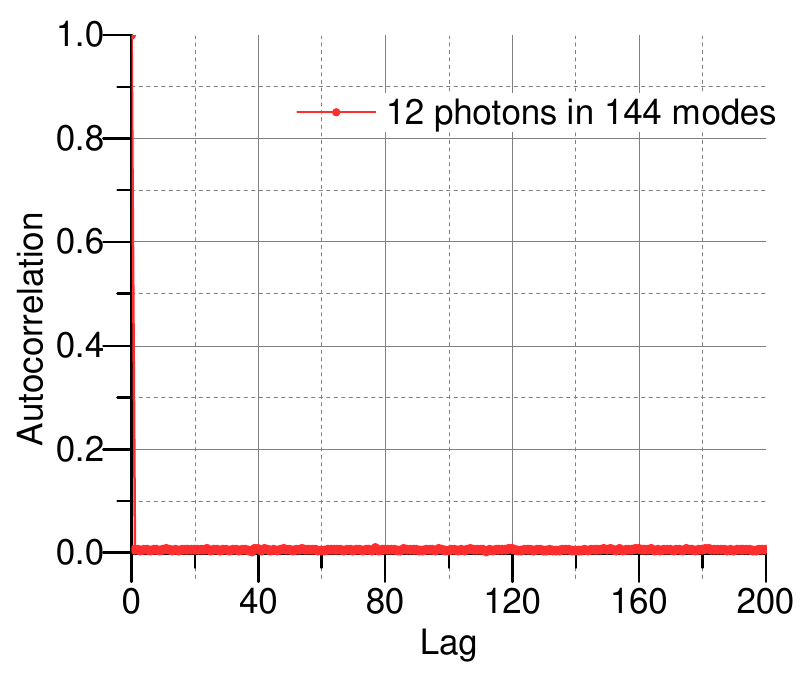}} \hspace{0.2cm}
  \subfigure[n=15,m=225,MCMC]{\includegraphics[width=0.21\textwidth]{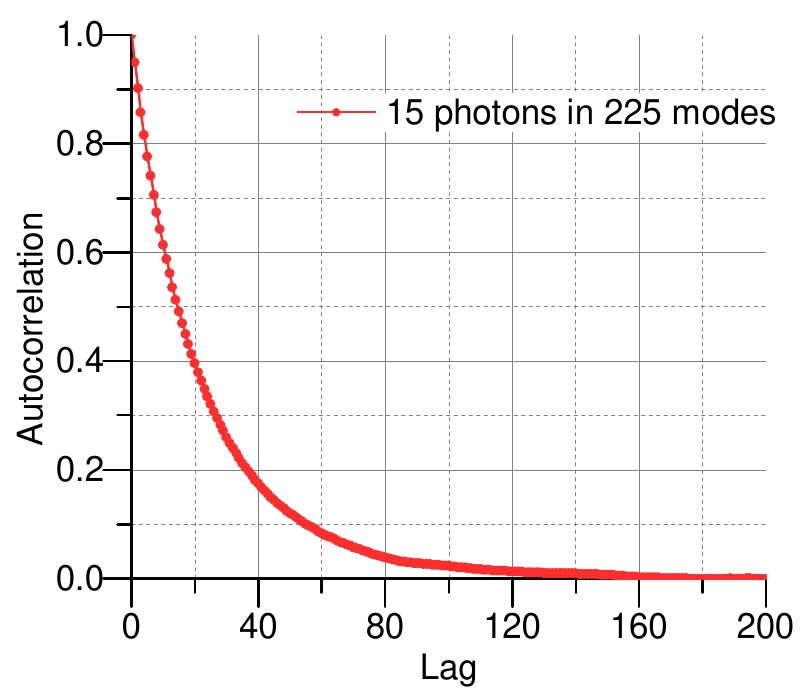}} \hspace{0.2cm}
  \subfigure[n=15,m=225,SC-MCMC]{\includegraphics[width=0.21\textwidth]{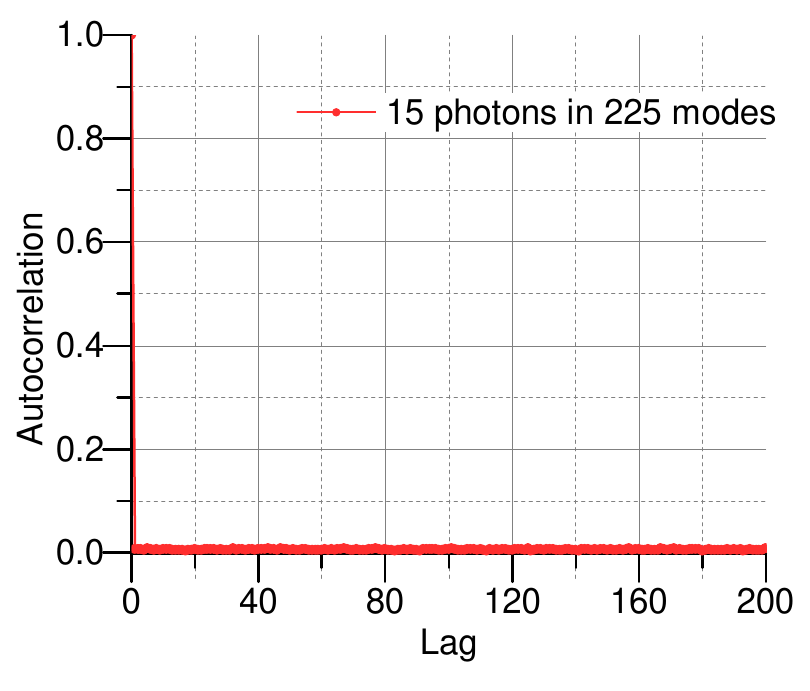}}
  \subfigure[n=17,m=289,MCMC]{\includegraphics[width=0.21\textwidth]{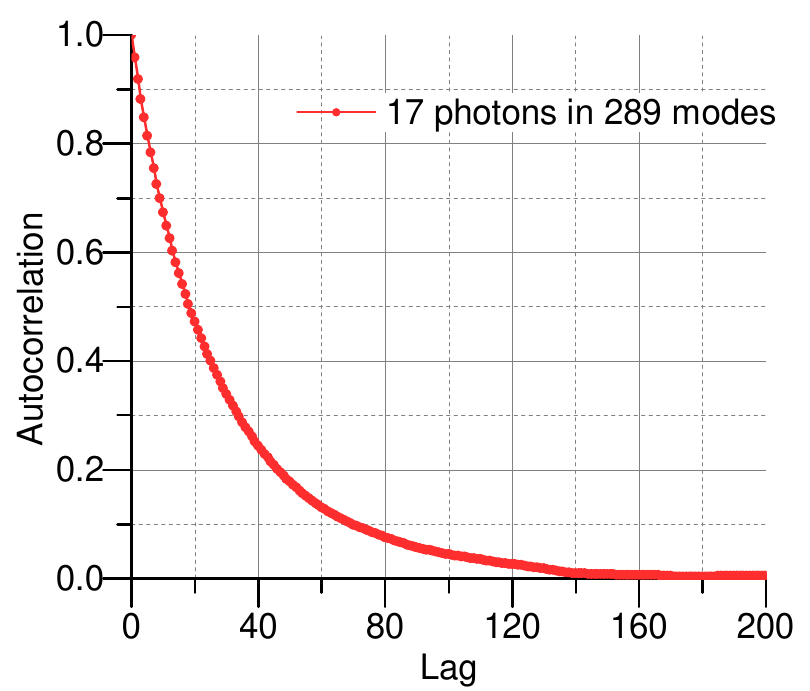}} \hspace{0.2cm}
  \subfigure[n=17,m=289,SC-MCMC]{\includegraphics[width=0.21\textwidth]{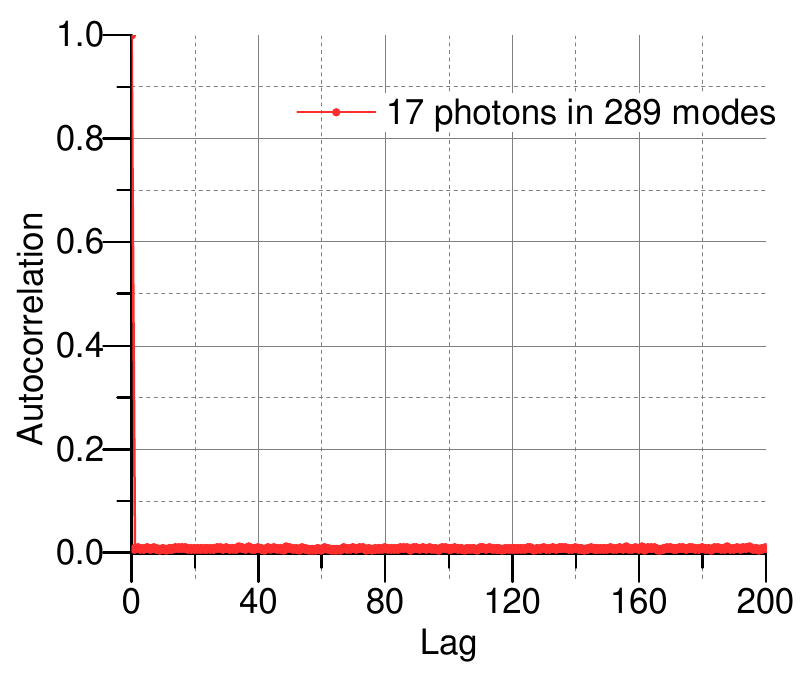}} \hspace{0.2cm}
  \subfigure[n=18,m=324,MCMC]{\includegraphics[width=0.21\textwidth]{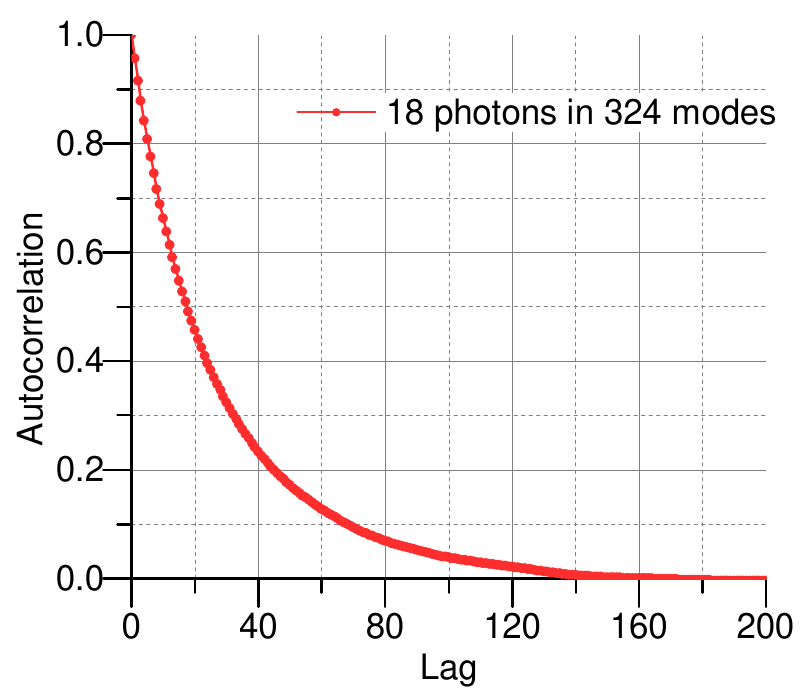}} \hspace{0.2cm}
  \subfigure[n=18,m=324,SC-MCMC]{\includegraphics[width=0.21\textwidth]{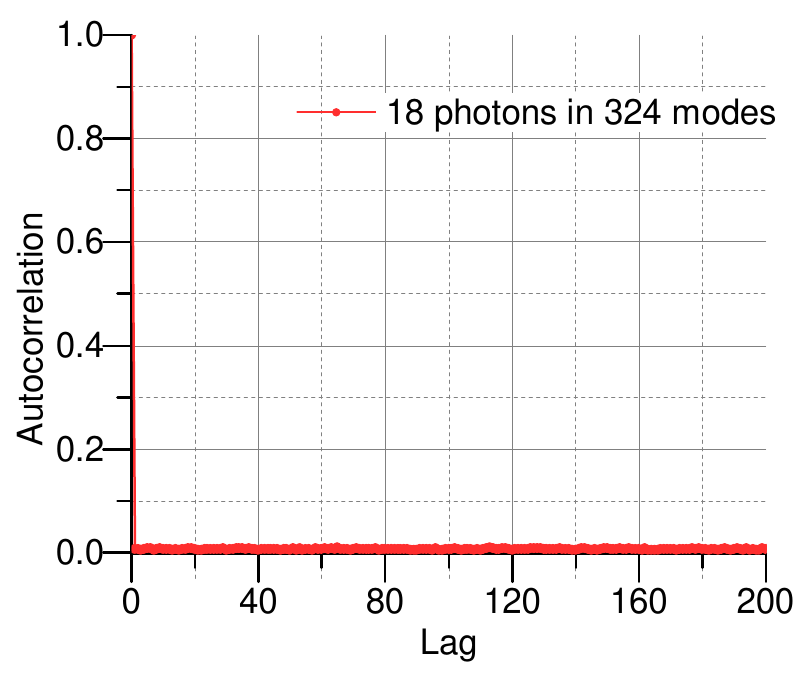}}
  \subfigure[n=19,m=361,MCMC]{\includegraphics[width=0.21\textwidth]{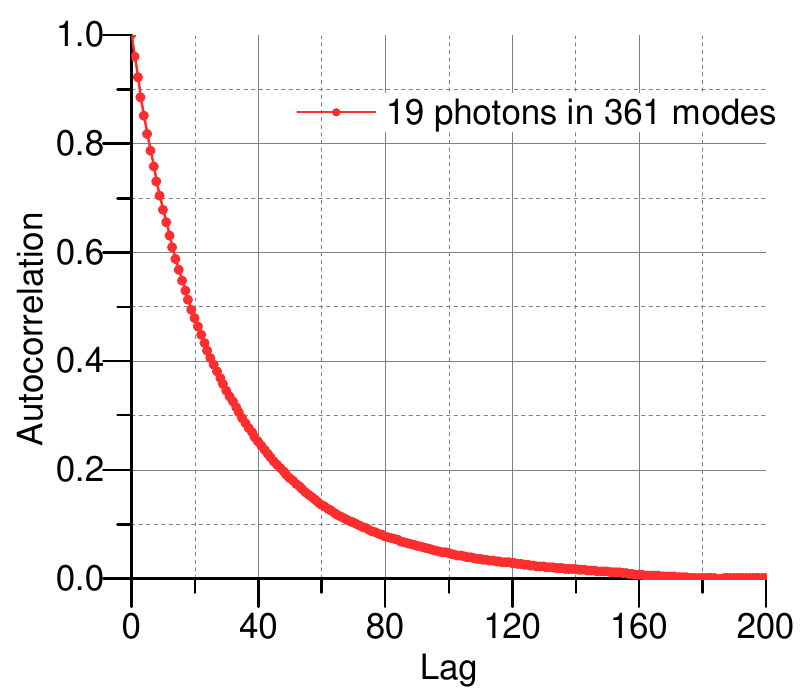}} \hspace{0.2cm}
  \subfigure[n=19,m=361,SC-MCMC]{\includegraphics[width=0.21\textwidth]{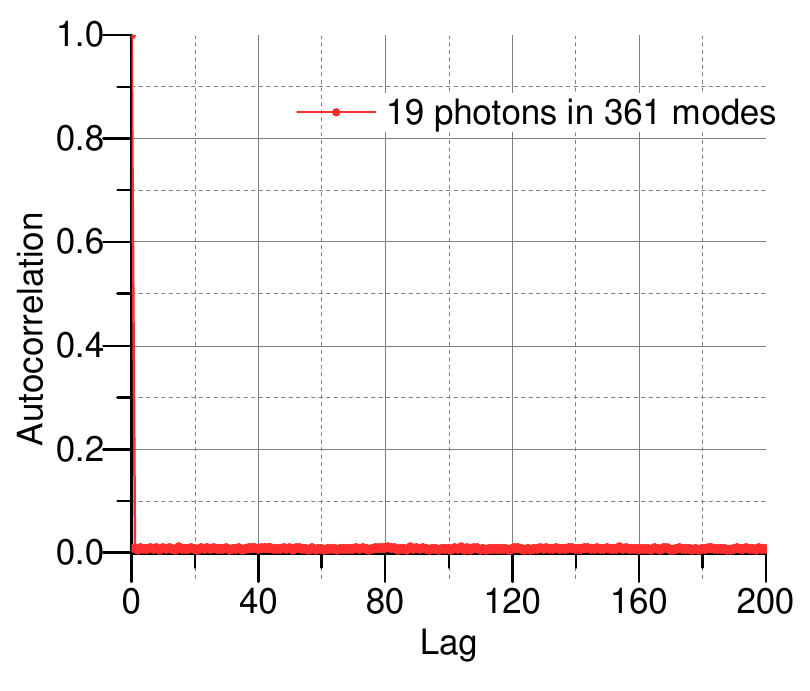}} \hspace{0.2cm}
  \subfigure[n=21,m=441,MCMC]{\includegraphics[width=0.21\textwidth]{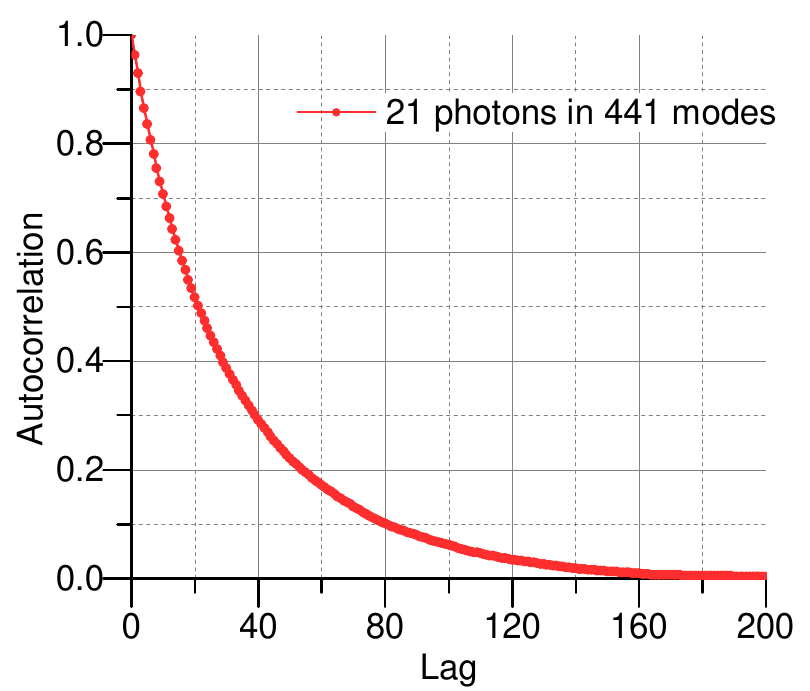}} \hspace{0.2cm}
  \subfigure[n=21,m=441,SC-MCMC]{\includegraphics[width=0.21\textwidth]{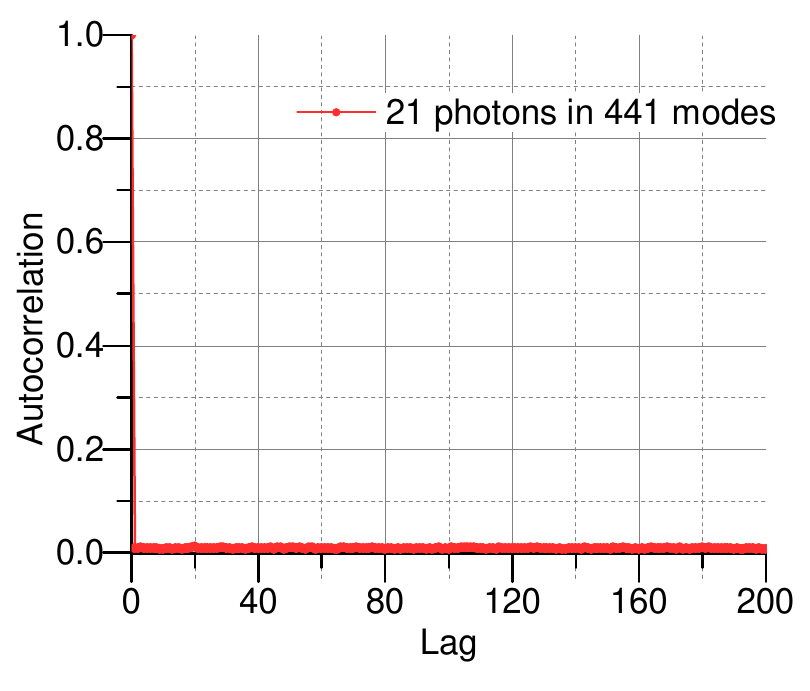}}
  \subfigure[n=25,m=625,MCMC]{\includegraphics[width=0.21\textwidth]{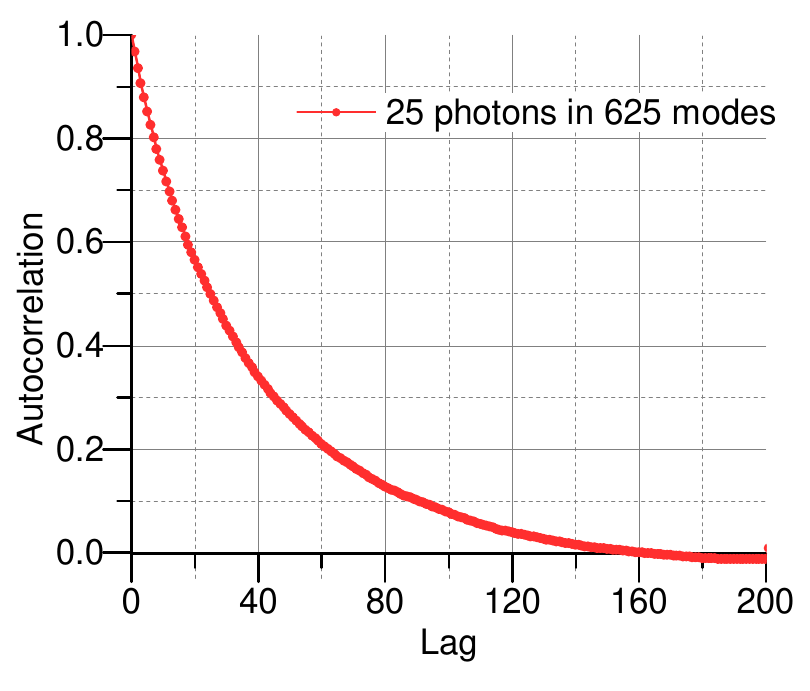}} \hspace{0.2cm}
  \subfigure[n=25,m=625,SC-MCMC]{\includegraphics[width=0.21\textwidth]{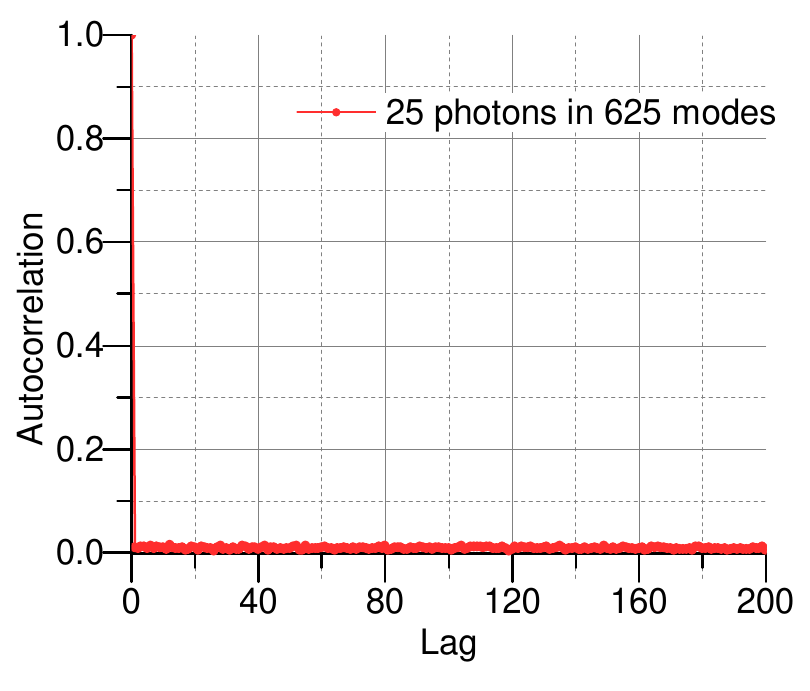}} \hspace{0.2cm}
  \subfigure[n=30,m=900,MCMC]{\includegraphics[width=0.21\textwidth]{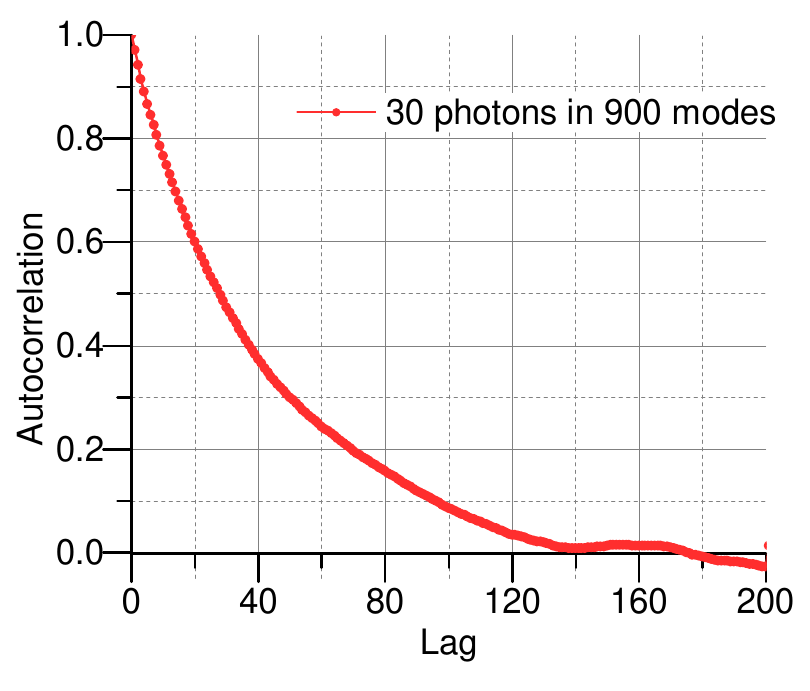}} \hspace{0.2cm}
  \subfigure[n=30,m=900,SC-MCMC]{\includegraphics[width=0.21\textwidth]{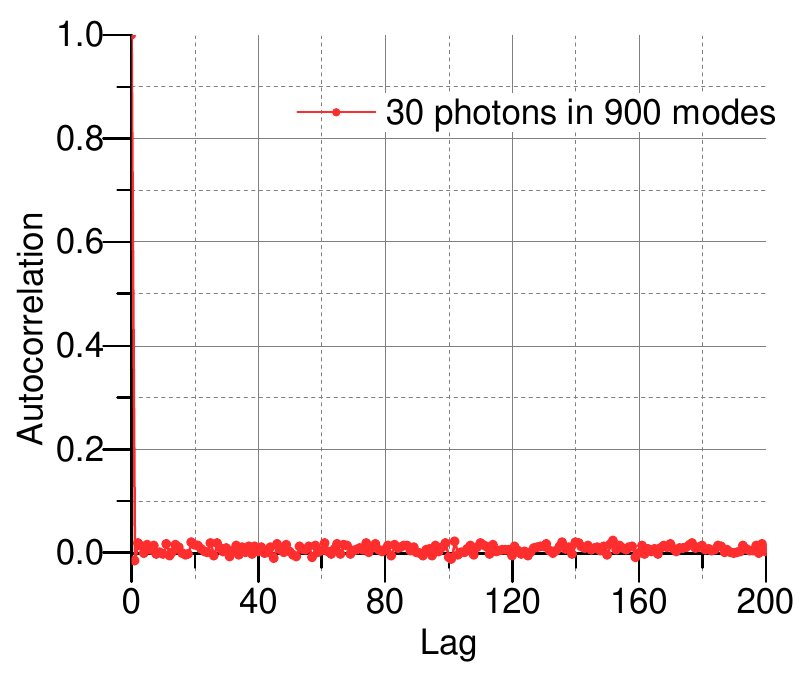}}
  \caption{\footnotesize The autocorrelation of the original sequence and the sequence generated by SC-MCMC. The size of the cache is 4,000. The autocorrelations at arbitrary orders of the SC-MCMC sequences are negligible. The odd columns are the autocorrelation of the original sequences generated by standard MCMC processes, and the even columns are that of the SC-MCMC sequences. The size of the sequence is 1,000,000 for the cases where the number of photons is less than 25, for the 25-photon-625-mode case, the size of the sequence is 200,000, and for the 30-photon-900-mode case, the size of the sequence is 20,000.}
  \label{FIGS:EffectSampleCache}
\end{figure}

By applying sample caching on the sequence, the finally generated samples are nearly independent. Fig.~\ref{FIGS:AutocorrelationReduceWithIncreaseOfCacheSize} shows the influence of sample cache with varied size on the first-order autocorrelation, and Fig.~\ref{FIGS:EffectSampleCache} shows the comparison between the original sequence generated by normal MCMC algorithm and the SC-MCMC generated sequence in autocorrelation at lags up to 200.

The results indicate that, sample caching process reduces the autocorrelation of the sequence to a negligible level, and the samples in the sequence are nearly independent. More important, in this way, all the samples are saved from being discarded, and averagely the generation of one sample only requires the calculation of one permanent. The cache at the size of 4,000 is sufficient for 30-photon cases.

However, the SC-MCMC method requires a longer start-up time, because of the workload to fill the sample cache. If a MIS and a SC-MCMC sampler are working on a same sampling problem, after the same period of warm-up time, the MIS begins to output samples every $k$ steps, while the SC-MCMC has to wait for $L$ steps, and then output samples in every step. We would find that the SC-MCMC sampler would cache up with the MCMC sampler with jump sampling at the sample ordered with $\lceil\frac{L}{K}\rceil$ by the SC-MCMC sampler. Fortunately, we can use a small trick to improve the fall-behind period of SC-MCMC. The solution is more like a combination of the jump sampling and sample caching: when filling the sample cache, we can output samples in every $k$ generations, while the un-chosen samples are stored in the cache till the sample cache is full, and then follow the SC-MCMC protocol. Thus finally, we can develop a improved SC-MCMC algorithm, as shown in Algorithm~\ref{ALG:IMPSC-MCMC}. The warm-up period of MCMC is not focused in our work, because the number of samples for warm-up is constant and limited, while much more samples can be discarded in other sampling algorithms.
%
%\begin{figure}[!htb]
%  \centering
%  \subfigure[MIS vs. SC-MCMC sampler]{
%    \includegraphics[width=0.4\textwidth]{Fig/FigS6aComp2MC.pdf}
%  }
%  \hspace{0.6in}
%  \subfigure[MIS vs. SC-MCMC sampler]{
%    \includegraphics[width=0.4\textwidth]{Fig/FigS6bCompIMPSC-MCMC.pdf}
%  }
%  \caption{\footnotesize The comparison of the performance between MIS and SC-MCMC sampler. (a) The number of permanent required to calculate for generation of different number of samples. The SC-MCMC sampler requires a longer start-up time, and then it can produce samples in every step, and cache up with MIS at the sample ordered with $\lceil\frac{L}{K}\rceil$. In this figure, $k=100$ and $L=4,000$ (b) The improved SC-MCMC sampler is more like a combination of the jump sampling and sample caching: when filling the sample cache, we can output samples in every $k$ generations, while the un-chosen samples are stored in the cache till the sample cache is full, and then follow the SC-MCMC protocol. The data used for simulation is the time required for a $30\times 30$ permanent, which is 0.9912 seconds obtained by averaging the results of 20,000 executions. Note that the common warm-up periods are not shown in the figures.}
%  \label{FIGS:Comp2MC}
%\end{figure}

\begin{algorithm}[H]
  \caption{\small Improved Sample Caching Markov chain Monte Carlo algorithm}
  \label{ALG:IMPSC-MCMC}
  \begin{algorithmic}[1]
    \Require $Cache$ : Sample Cache; $L$ : Size of the Sample Cache; $k$ : jumping step; $SN$: number of samples required
    \Ensure Un-correlated sample sequence
        \For{$i$ = 1:$SN$}
            \State Sample={\bf MCMC}();\Comment{Generate a sample using MCMC}\label{ALGL:MCMC}
            \If{{\bf not} {\bf Full}($Cache$)}
                \If{$(SN - 1)\%k$==$0$}
                    \State {\bf Output}(Sample);    \Comment{Output in the every $k$ samples if the cache is not full}
                \Else
                    \State {\bf add}($Cache$, Sample);\Comment{Store the sample in the Cache}
                \EndIf
            \Else
                \State u = {\bf UniformRand}([1,$L$])
                \State {\bf Output}($Cache$[u]);\Comment{Pick a sample to output randomly}
                \State {\bf add}($Cache$, Sample, u);\Comment{Store a new sample to the empty slot}
            \EndIf
        \EndFor
        \State Output the samples in the $Cache$ randomly;\Comment{Deal with the samples in the cache when the sampling process is over}
  \end{algorithmic}
\end{algorithm}

\section{Theoretical Analysis of Sample Caching Markov Chain Monte Carlo}

The effectiveness of sample cache in the asymptotic condition where the size of the cache is large enough is easy to understand. The standard MCMC process ensures that the samples in the cache follows their own probability. Since we uniformly randomly choose the sample from the cache as the sample to output, each selection is independent. Thus we obtain a correlation-less sequence.

\begin{figure}[!htb]
  \centering
  \includegraphics[width=0.7\textwidth]{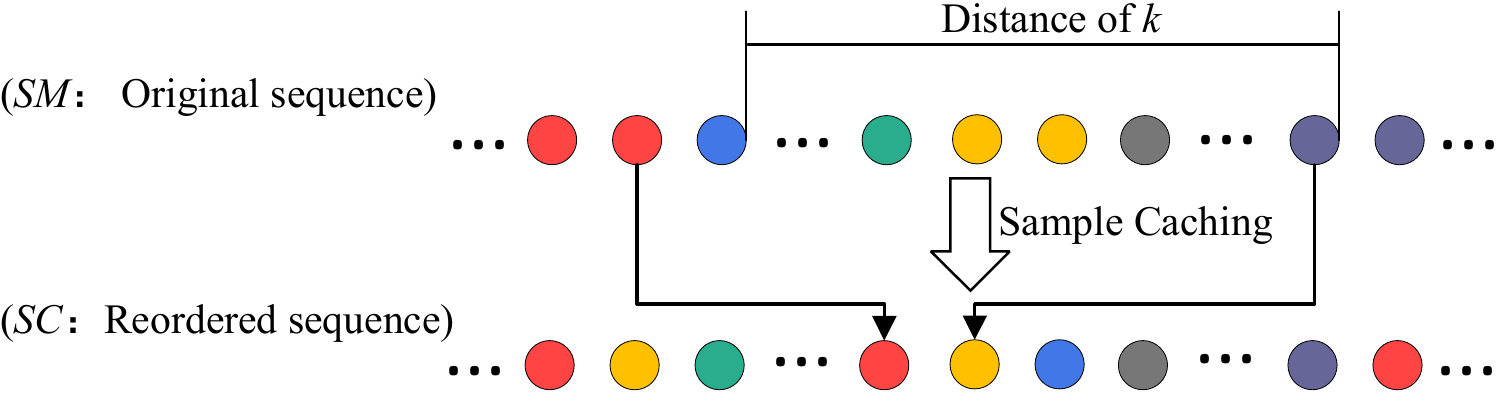}\\
  \caption{\footnotesize Effect of sample caching. The nature of sample caching is the reorder of the samples in the original sample sequence. The adjacent samples in the final sequence may be separated in the original sequence.}\label{Fig:SampleCaching}
\end{figure}

Practically, the size of the cache is limited. In this case, the essence that sample cache works is the reorder of the samples in the sequence, as shown in Fig.~\ref{Fig:SampleCaching}. The samples generated by the MCMC sampler without entering the sample cache forms an original sequence (denoted as $SM$), and the sequence after the sample cache is denoted as $SC$. The adjacent two samples from $SM$ may be separated in $SC$, and the size of cache determines the range that the reorder works.

The reorder process is stochastic. Suppose that two adjacent samples in $SC$ labeled as $x_i$ and $x_j$, and reorder the sequence with a sample cache of size $L$. Now we discuss about the probability that the distance of $x_i$ is in $k$ steps away from $x_j$ in the original sequence before the reorder.

\begin{enumerate}
  \item If $k>0$, it means $x_i$ is $k$ steps ahead of $x_j$ in $SM$.

  In this case, $x_i$ enters the cache first, and in the next $k-1$ steps, $x_i$ is not chosen to output. This probability is $\left(\frac{L-1}{L}\right)^{k-1}$. Then $x_j$ enters the cache, and they are outputted immediately with probability $\frac{1}{L^2}$, or outputted one step later with probability $\frac{L-1}{L}\cdot\frac{1}{L^2}$, or with probability $\left(\frac{L-2}{L}\right)^t\cdot\frac{1}{L^2}$ outputted $t$ steps later, where $t=2,3,...$. Thus in summary, we have
  \begin{equation}
    \begin{aligned}
        p_+(k,L)=&\left(\frac{L-1}{L}\right)^{k-1}\cdot\frac{1}{L^2}\left[1+\frac{L-1}{L}\cdot\sum_{i=0}^\infty\left(\frac{L-2}{L}\right)^i\right]\\
        %=&\left[\frac{1}{L^2}+\frac{L-1}{L}\cdot\frac{1}{L^2}+\frac{L-2}{L}\cdot\frac{L-1}{L}\cdot\frac{1}{L^2}
        %+\left(\frac{L-2}{L}\right)^2\cdot\frac{L-1}{L}\cdot\frac{1}{L^2}+\dots\right]\\
        =&\left(\frac{L-1}{L}\right)^{k-1}\cdot\frac{L+1}{2L^2}.
    \end{aligned}
  \end{equation}

  \item If $k<0$, it means $x_i$ is $k$ steps behind $x_j$ in $SM$.

  In this case, $x_j$ enters the cache first, and in the next $k$ steps $x_j$ is not chosen to output with probability $\left(\frac{L-1}{L}\right)^{k}$. Then these two samples are probable to be outputted in $t$ steps, with probability $\left(\frac{L-2}{L}\right)^t\cdot\frac{1}{L^2}$, where $t=0,1,2,...$. Thus the probability of distance $k$ is
  \begin{equation}
    \begin{aligned}
        p_-(k,L)=&\left(\frac{L-1}{L}\right)^k\frac{1}{L^2}\sum_{i=0}^\infty\left(\frac{L-2}{L}\right)^i \\
                =&\left(\frac{L-1}{L}\right)^k\cdot\frac{1}{2L}.
    \end{aligned}
  \end{equation}
\end{enumerate}

We care more about the absolute value of $k$. The probability that the two adjacent samples are in distance $k$ is $p(k,L)=p_+(k,L)+p_-(k,L)=\left(\frac{L-1}{L}\right)^{k-1}\cdot\frac{1}{L}$. The expected distance of two samples is
\begin{equation}
\begin{aligned}
\mathds{E}(k)&=\sum_{i=1}^\infty i\cdot p(k,L)\\
             &=\sum_{i=1}^\infty i\cdot\left(\frac{L-1}{L}\right)^{i-1}\cdot \frac{1}{L}\\
             &=L.
\end{aligned}
\end{equation}

In this way, the correlated samples are at a lower probability to be at a close distance in the new sequence. This probability is in a inversely proportional relationship with the size of the cache. The distance may directly impact on the correlation between the two samples. From Fig.~\ref{FIGS:CONVERG} we see that if the distance reaches a certain number (denoted as $K$), the correlation of the samples can be ignored. So we can calculate the probability if the correlation of two samples can not be ignored as
\begin{equation}
\begin{aligned}
P_{cr}(L)&=\sum_{i=1}^K\left(\frac{L-1}{L}\right)^{i-1}\cdot \frac{1}{L}\\
%    &=1-\left(\frac{L-1}{L}\right)^K\\
    &=\frac{K}{L}+O\left(\frac{1}{L^2}\right)\\
    &\approx\frac{K}{L}.
\end{aligned}
\end{equation}

We can constraint $P_{cr}$ in a low level to reduce the autocorrelation. That is, let $P_{cr}(L)<\varepsilon$, then $L>\frac{K}{\varepsilon}$. Next we discuss about the practical value of $K$ and $\varepsilon$.

\begin{figure}[!b]
  \centering
    \subfigure[n=7,m=49]{\includegraphics[width=0.23\textwidth]{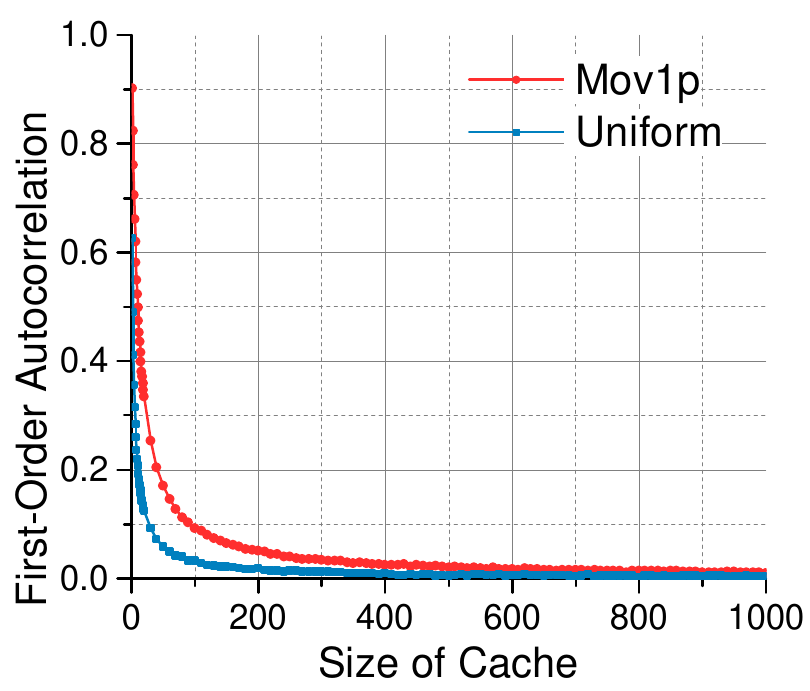}}
    \subfigure[n=10,m=100]{\includegraphics[width=0.23\textwidth]{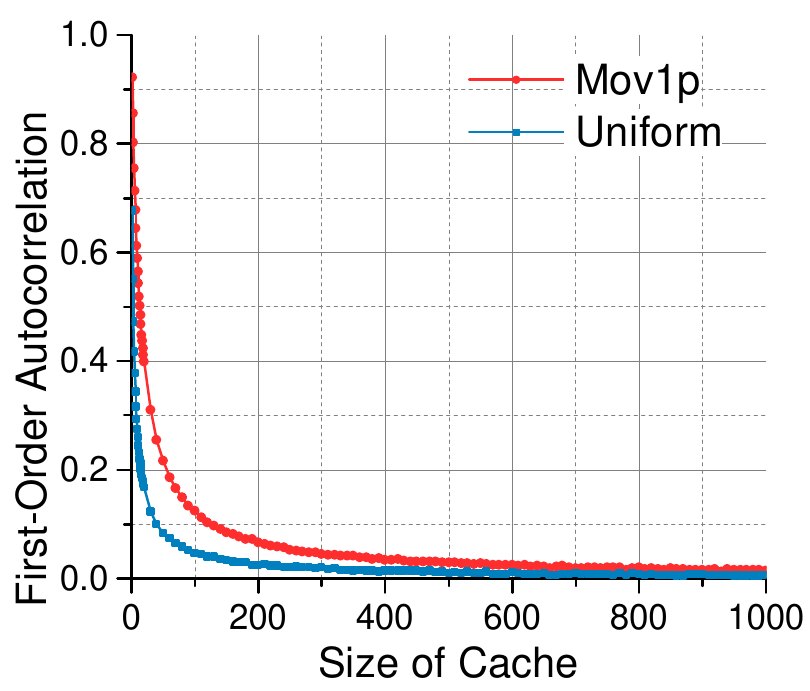}}
    \subfigure[n=13,m=169]{\includegraphics[width=0.23\textwidth]{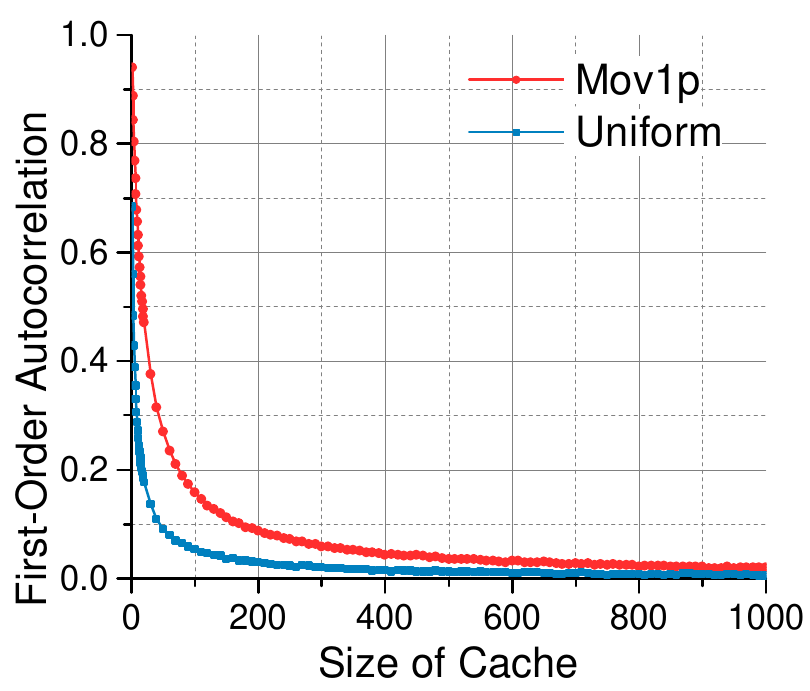}}
    \subfigure[n=16,m=256]{\includegraphics[width=0.23\textwidth]{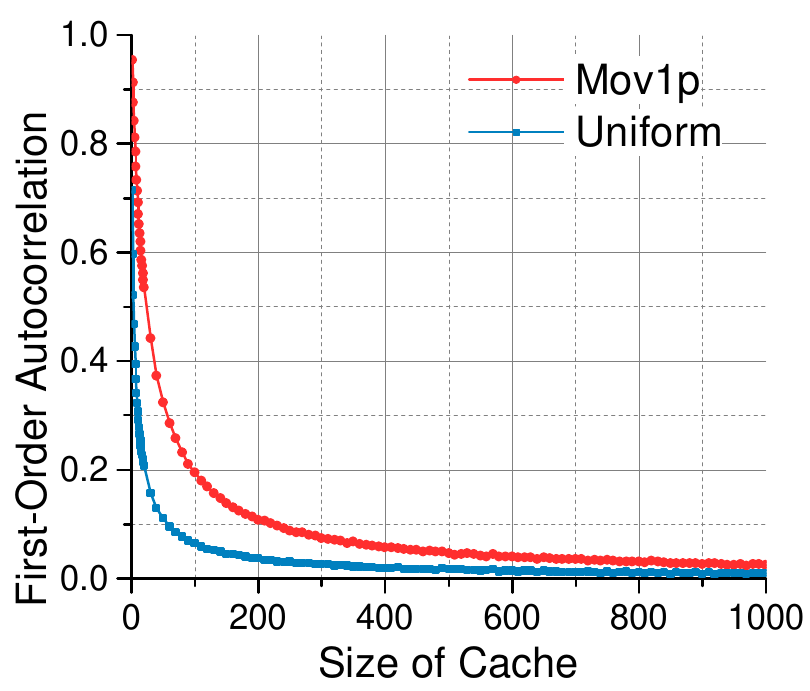}}
  \caption{\footnotesize The autocorrelation of the sequence generated by MCMC with different proposal distribution. ``Mov1p'' is the strategy used in our implementation to expose autocorrelation problem, ``Uniform'' is to choose the candidate uniformly randomly. The uniform choice results in fast convergence speed, and better performance in the autocorrelation than the ``mov1p'' implementation.}
  \label{FIGS:Strategy}
\end{figure}

The value of $K$ should depend on how fast the Markov chain could converge, and can be referred from the jumping step of the MIS. The choice of the easy-to-sample proposal distribution before the reject-accept phase in Markov chain may greatly affect the convergence speed. In our implementation, we tested two proposal distributions. One is the uniform distribution, which indicates that all the states are chosen uniformly randomly as the candidate state no matter what the current state is. The other one generates candidate state by randomly moving one of the photons to another empty mode, as shown in Eq.~\ref{EQS:IMPLG}

\begin{equation}\label{EQS:IMPLG}
g(s_j|s_i)=\left\{
\begin{array}{ll}
\frac{1}{n\cdot(m-n)}, &\text{ patterns correspond to }s_i\text{ differs from that of }s_j\text{ in the position of one photon;}\\
0, &\text{ else},
\end{array}
\right.
\end{equation}
where $n$ is the number of photons and $m$ is the number of modes. For example, in the simulation for $n=2, m=6$, state corresponding to pattern $|010100\rangle$ may transmit to state with pattern $|000101\rangle$ by moving the photon from the second mode to the last mode. This proposal distribution leads to slower convergence of the Markov chain, and thus exposes more severe autocorrelation problem in our test. Practically, the uniform distribution is preferred. For example, as shown in Fig.~\ref{FIGS:Strategy}, the curves are obtained from the sequences obtained using different candidate choosing strategy. ``mov1p'' is the implementation of Eq.~\ref{EQS:IMPLG}; ``Uniform'' means the candidate is chosen uniformly randomly.

\begin{figure}[!htb]
  \centering
    \subfigure[$r_1$ of the sequence with different size of sample cache]{\includegraphics[width=0.24\textwidth]{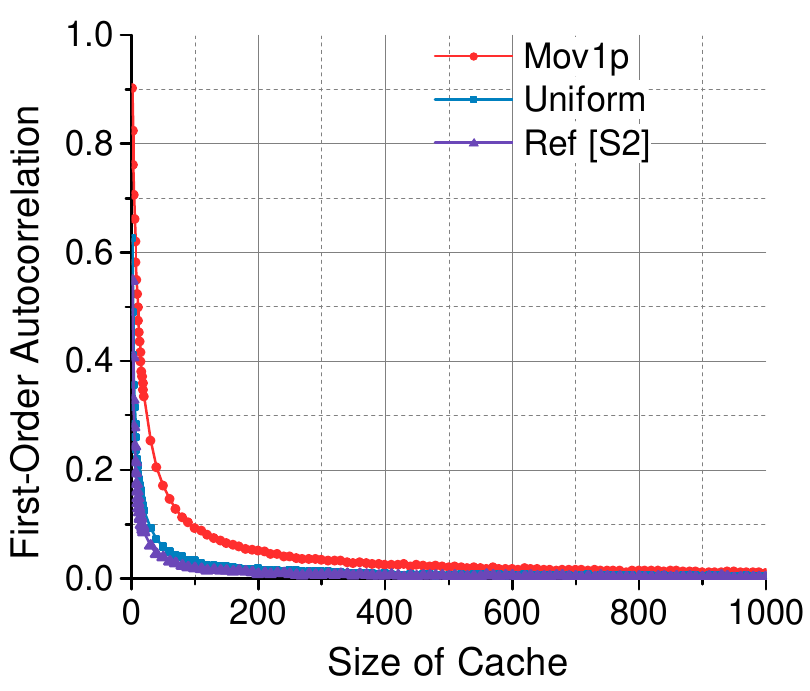}}%
    \subfigure[$L=1$]{\includegraphics[width=0.24\textwidth]{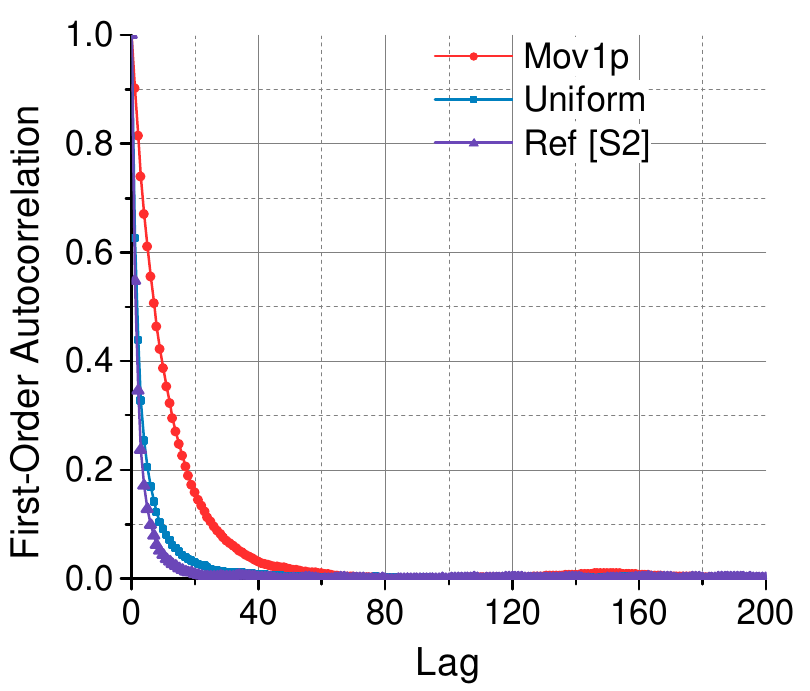}}
    \subfigure[$L=1,000$]{\includegraphics[width=0.24\textwidth]{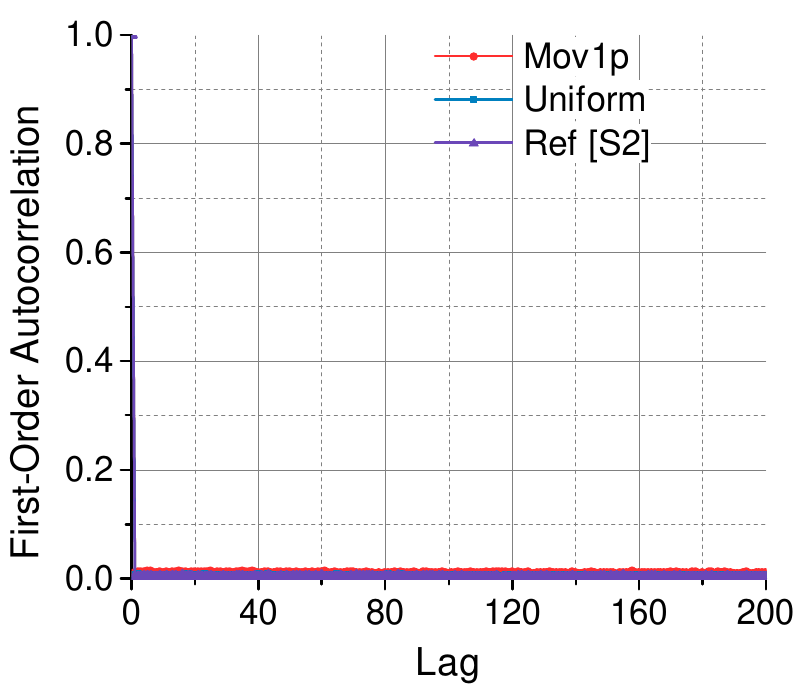}}
    \subfigure[$L=4,000$]{\includegraphics[width=0.24\textwidth]{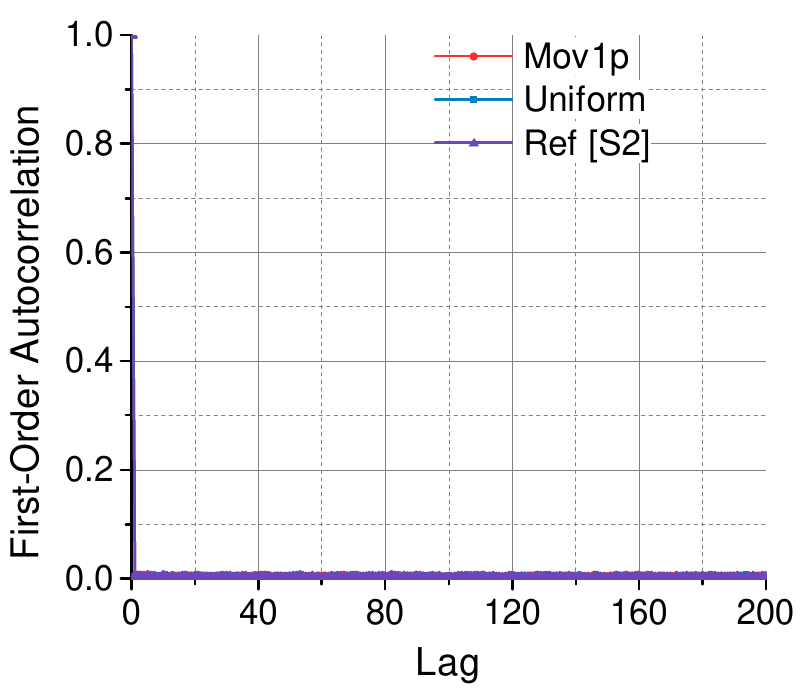}}
  \caption{\footnotesize The comparison between the three proposal distribution. ``Mov1p'' is the strategy used in our implementation to expose autocorrelation problem, ``Uniform'' is to choose the candidate uniformly randomly, and ``Ref [S2]'' is the strategy used in ref.~\cite{Neville2017} where the distribution of distinguishable particles is used as the candidate. The strategy in ref.~\cite{Neville2017} results in fastest convergence speed and further the least correlated samples. (a) The first-order autocorrelation of the sequence produced by SC-MCMC with different sizes of sample cache applied. (b)$\sim$(d) The autocorrelation at different lags (up to 200) of the 3 sequences obtained via varied size of sample cache. $L=1$ reflects the autocorrelation of the original sample sequence. The autocorrelation of the sequence is negligible no matter what proposal density is chosen when the size of sample cache reaches 1,000. When $L$ reaches 4,000, the differences between the methods with different proposal density is negligible. In this test, $n=7$ and $m=49$.}
  \label{FIGS:DistinguishableDistributionCandidate}
\end{figure}

Further, we compared the two strategies with that in ref.~\cite{Neville2017} where the distribution of distinguishable particles is used as the proposal distribution, as shown in Fig.~\ref{FIGS:DistinguishableDistributionCandidate}. The strategy in ref.~\cite{Neville2017} helps produce least correlated samples, and thus a smaller sample cache is sufficient, but introduce the cost of computing an extra $n\times n$ real permanent. With the increase of the size of the sample cache, the autocorrelations of the sequences under different proposal distribution become closer, and are hard to distinguish when the size of the cache reaches a certain level. Thus we claim that no matter what strategy is applied, the sample cache would finally help eliminate the autocorrelation as long as the Markov chain can finally converge, while the difference resides in the size of the sample cache required.

On the other hand, the strategy in ref.~\cite{Neville2017} produces least correlated samples, the sampler with this proposal distribution must be able to sample from the distribution of distinguishable particles efficiently, which involves the calculation of the permanent of a real matrix. In this way, two permanents (one of a complex matrix and one of a real matrix) have to be calculated for one sample. Although it has been proved that the permanent of a positive real matrix is easy to approximate~\cite{Jerrum2004}, the practical calculation on classical computers still requires more extra cost.

With proper proposal distribution, it's claimed that $K=100$ is sufficient for the cases with more than 30 photons. However, unbefitting choice of proposal distribution would lead to higher requirement of the size of the sample cache, while the performance of SC-MCMC would not be affected that much, except a longer start-up time. When the cache is filled, the SC-MCMC sampler still can generate one sample by calculating one permanent. Because of the possibility of bad convergence speed when choosing improper proposal distribution, we set $K=200$.

Next we discuss about the choice of $\varepsilon$, and then we can use a sufficient big cache to eliminate the autocorrelation. \Cref{TABS:SIZECORRERR} gives the examples.

\begin{table}[b]
  \centering
  \caption{\footnotesize $\varepsilon$ and the first-order autocorrelation under different cache size. For the $n=3, m=9$ case (left), $K$ is set to be 29 according to Fig.~\ref{FIGS:CONVERG}. For the $n=4, m=16$ case (right), $K$ is set to be 33. For each sequence 1,000,000 samples are taken. $\bar{k}$ is the average absolute distance in the original sequence of the adjacent samples of the final sequence, $N_{1}$ is the number of adjacent samples in the final sequence those are also adjacent in the original sequence, and $R_1$ is the corresponding ratio over the whole sample sequence, which is of the theoretical value as $\frac{1}{L}$. $F_K$ is the frequency of adjacent samples in the final sequence those are of a distance less than $K$ in the original sequence, and $\varepsilon$ represents the ratio. Finally, $r_1$ is the first order autocorrelation.}\label{TABS:SIZECORRERR}
  \begin{tabular}{cccccccp{1.5cm}ccccccc}
  \cline{1-7}
  \cline{9-15}
    \multicolumn{7}{c}{3p.9m.} & & \multicolumn{7}{c}{4p.16m.}\\
      L  &  $\bar{K}$ & $N_1$  &  $R_1$ & $F_K$  &  $\varepsilon$ & $r_1$  & &  L & $\bar{K}$ & $N_1$  & $R_1$ & $F_K$  &  $\varepsilon$ & $r_1$. \\
  \cline{1-7}
  \cline{9-15}
      10 &    9.9912 & 99592 &  9.96\% & 953263 &    95.33\% & 0.2580 & &  10 &    10.0059 & 100132 & 10.01\% & 968989 &        96.90\% & 0.3215 \\
      20 &   19.9961 & 49948 &  4.99\% & 774238 &    77.42\% & 0.1514 & &  20 &    19.9819 &  50304 &  5.03\% & 816477 &        81.65\% & 0.1949 \\
      30 &   30.0184 & 33599 &  3.36\% & 625093 &    62.51\% & 0.1095 & &  30 &    30.0365 &  33105 &  3.31\% & 673629 &        67.36\% & 0.1386 \\
      40 &   39.9672 & 25031 &  2.50\% & 519809 &    51.98\% & 0.0843 & &  40 &    40.0001 &  24880 &  2.49\% & 566109 &        56.61\% & 0.1070 \\
      50 &   49.9564 & 19909 &  1.99\% & 443412 &    44.34\% & 0.0691 & &  50 &    49.9722 &  19771 &  1.98\% & 486719 &        48.67\% & 0.0854 \\
         &           &       &  \dots  &        &            &        & &     &            &        &  \dots  &        &                &        \\
     100 &   99.9055 &  9913 &  0.99\% & 253312 &    25.33\% & 0.0371 & &  100&   100.0127 &  10081 &  1.01\% & 282310 &        28.23\% & 0.0445 \\
     200 &  199.8007 &  5018 &  0.50\% & 135236 &    13.52\% & 0.0189 & &  200&   199.7608 &   4989 &  0.50\% & 152125 &        15.21\% & 0.0234 \\
         &           &       &  \dots  &        &            &        & &     &            &        &  \dots  &        &                &        \\
     500 &  499.5682 &  1992 &  0.20\% &  56410 &     5.64\% & 0.0062 & &  500&   499.5963 &   1971 &  0.20\% &  64398 &         6.44\% & 0.0096 \\
         &           &       &  \dots  &        &            &        & &     &            &        &  \dots  &        &                &        \\
    1000 &  999.6380 &   994 &  0.10\% &  28471 &     2.85\% & 0.0031 & & 1000&   999.9884 &    967 &  0.10\% &  32528 &         3.25\% & 0.0051 \\
         &           &       &  \dots  &        &            &        & &     &            &        &  \dots  &        &                &        \\
    10000& 9998.9470 &   107 &  0.01\% &   2969 &     0.30\% & 0.0009 & &10000& 10002.9656 &     87 &  0.01\% &   3356 &         0.34\% & 0.0010 \\
  \cline{1-7}
  \cline{9-15}
  \end{tabular}
\end{table}

It meets some challenges to relate $\varepsilon$ directly with the autocorrelation of the sequence. Empirically we found that $\varepsilon=0.1$ may be enough to limit the first-order autocorrelation within a small range (in the order of $10^{-2}$). Thus we claim that $\varepsilon=0.05$ may be sufficient, and then a practical value of $L$ can be obtained by setting $K=200$ and $\varepsilon=0.05$, that is 4,000. As reflected in Fig.~\ref{FIGS:EffectSampleCache}, the SC-MCMC with a cache of size 4,000 can produce nearly independent samples for 30-photon-900-mode boson sampling. Above all, no matter how big the cache is, no sample is discarded.

\section{Numerical Results}

\subsection{The value assigning of the output patterns}

To calculate the autocorrelation within the sequence, each output pattern of photons has to be assigned a value. We tried 3 ways for value assigning.

\begin{enumerate}
  \item Binary to Decimal

      Since the regime that we simulate boson sampling is limited in the collision-free case, the number of photons in each mode would not exceed 1. Thus the pattern can be regarded as a binary number. We can simply transform it into a decimal number as its value. However, this method is not feasible for larger scale because of the limit of the word length of classical computers.
  \item Order of Sort

      We first sort the patterns, and use the numerical order as its value.
  \item $-\log P_i$

      The value is assigned as the logarithm of its probability. This method is also applied in ref. ~\cite{Neville2017}.
\end{enumerate}

\begin{figure}[!htb]
  \centering
    \subfigure[Order of sort]{\includegraphics[width=0.42\textwidth]{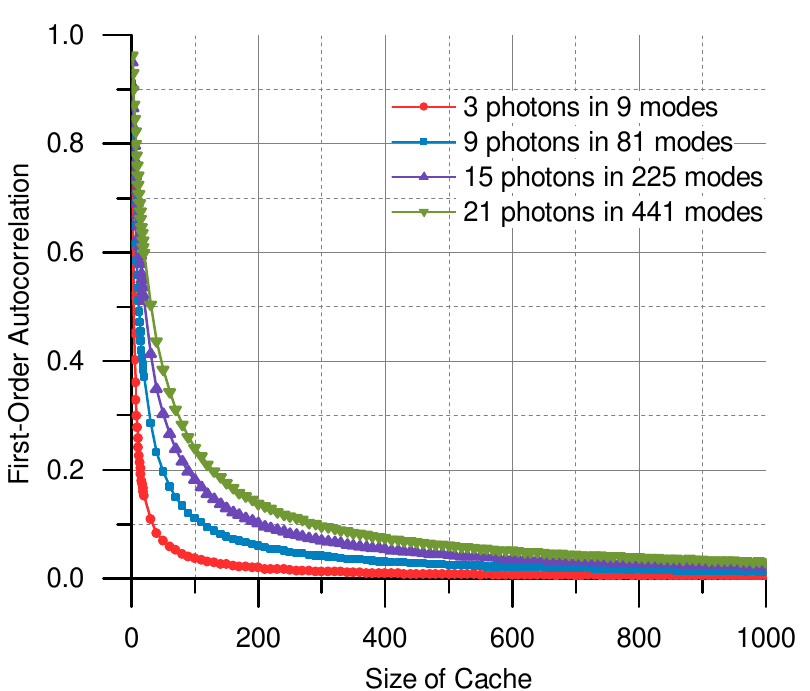}}
    \subfigure[$-\log P_i$]{\includegraphics[width=0.42\textwidth]{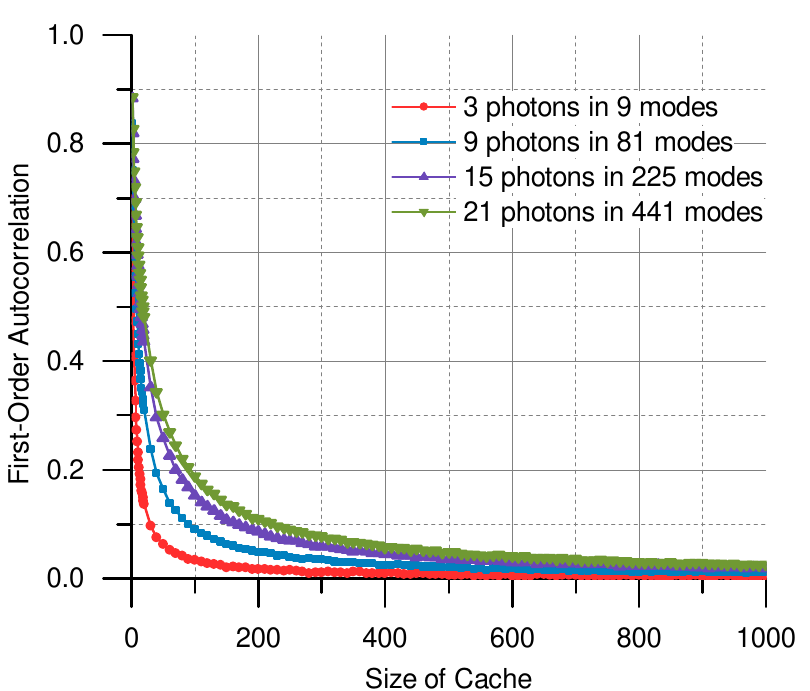}}
  \caption{\footnotesize The value assigning strategy doesn't impact on the effect of the sample cache. The autocorrelation decreases with the increase of the size of cache. Larger scale simulations require bigger caches. (a) The values of patterns are assigned using ``Order of sort'' strategy. (b) The values of patterns are assigned using ``$-\log P_i$'' strategy. The strategies for value assigning do not impact on the effect of sample cache.}
  \label{FIGS:ValueAssigningStrategy}
\end{figure}

For example, when simulating the boson sampling with 3 photons and a 6-mode optical network, in the collision-free regime there are totally 20 output patterns, and the value assigning of the 20 patterns are shown in Tab.~\ref{TABS:ValueAssigningExample}. The assigning strategy doesn't impact on the final result, and the theoretical analysis works no matter what method we use to assign values to the patterns. \Cref{FIGS:ValueAssigningStrategy} shows examples for two different assigning strategies. In our simulation, we use the means of ``Order of sort'' to assign value to each output pattern.

\begin{table}[!h]
  \centering
  \caption{\footnotesize Examples for the value assigning of output patterns. $V_{BD}$ is the vale assigned by transmit the binary values into decimal, $V_{SO}$ is the order of the patterns, and $V_{logP}$ is the logarithm of the probability of the patterns.}
  \label{TABS:ValueAssigningExample}
  \begin{tabular}{ccccccp{1cm}cccccc}
  \cline{1-6}
  \cline{8-13}
    No. & Pattern & Probability & $V_{BD}$ & $V_{SO}$ & $V_{-logP}$&  & No. & Pattern & Probability & $V_{BD}$ & $V_{SO}$ & $V_{-logP}$\\
  \cline{1-6}
  \cline{8-13}
    1 & (0,0,0,1,1,1) & $6.24\times 10^{-2}$ & 7 & 1 & 1.204 & & 11 & (1,0,0,0,1,1) & $8.8\times 10^{-3}$ & 35 & 11 & 2.055\\
    2 & (0,0,1,0,1,1) & $0.71\times 10^{-3}$ & 11 & 2 & 3.144 & & 12 & (1,0,0,1,0,1) & $5.15\times 10^{-2}$ & 37 & 12 & 1.287\\
    3 & (0,0,1,1,0,1) & $1.41\times 10^{-2}$ & 13 & 3 & 1.848 & & 13 & (1,0,0,1,1,0) & $3.82\times 10^{-3}$ & 38 & 13 & 2.417\\
    4 & (0,0,1,1,1,0) & $5.45\times 10^{-3}$ & 14 & 4 & 2.262 & & 14 & (1,0,1,0,0,1) & $1.46\times 10^{-2}$ & 41 & 14 & 1.834\\
    5 & (0,1,0,0,1,1) & $1.12\times 10^{-2}$ & 19 & 5 & 1.949 & & 15 & (1,0,1,0,1,0) & $8.44\times 10^{-3}$ & 42 & 15 & 2.073\\
    6 & (0,1,0,1,0,1) & $2.33\times 10^{-2}$ & 21 & 6 & 1.631 & & 16 & (1,0,1,1,0,0) & $2.67\times 10^{-3}$ & 44 & 16 & 2.572\\
    7 & (0,1,0,1,1,0) & $1.96\times 10^{-2}$ & 22 & 7 & 1.705 & & 17 & (1,1,0,0,0,1) & $6.24\times 10^{-3}$ & 49 & 17 & 2.204\\
    8 & (0,1,1,0,0,1) & $7.05\times 10^{-3}$ & 25 & 8 & 2.151 & & 18 & (1,1,0,0,1,0) & $1.49\times 10^{-2}$ & 50 & 18 & 1.824\\
    9 & (0,1,1,0,1,0) & $2.81\times 10^{-2}$ & 26 & 9 & 1.551 & & 19 & (1,1,0,1,0,0) & $3.16\times 10^{-3}$ & 52 & 19 & 2.499\\
    10 & (0,1,1,1,0,0) & $5.04\times 10^{-3}$ & 28 & 10 & 2.297 & & 20 & (1,1,1,0,0,0) & $2.02\times 10^{-2}$ & 56 & 20 & 1.694\\
  \cline{1-6}
  \cline{8-13}
  \end{tabular}
\end{table}

\subsection{Correctness}

Since SC-MCMC essentially is only the reorder of the sequence, the samples taken are not changed. The standard MCMC process embedded ensures the correctness of the results. Current validation techniques provide the method to distinguish boson sampling distribution from another proposal distribution once. Basically the sampling results in boson sampling should be compared to the uniform distribution and the distribution sampling from distinguishable particles, particularly in the cases where the scales reaches a rather large level. Here we directly compare the frequency of the sampling results to the theoretical probability distribution, and calculate the similarity following Eq.~\ref{EQ:SIMILAR}
\begin{equation}\label{EQ:SIMILAR}
S=\frac{\left(\sum_i\sqrt{P_iQ_i}\right)^2}{\sum_iP_i\sum_iQ_i},
\end{equation}
where $P_i$ and $Q_i$ represent the probability of event labeled as $i$ in distribution $P$ and $Q$ respectively.

\begin{figure}[!t]
  \centering
  \subfigure[n=4,m=16]{\includegraphics[width=0.44\textwidth]{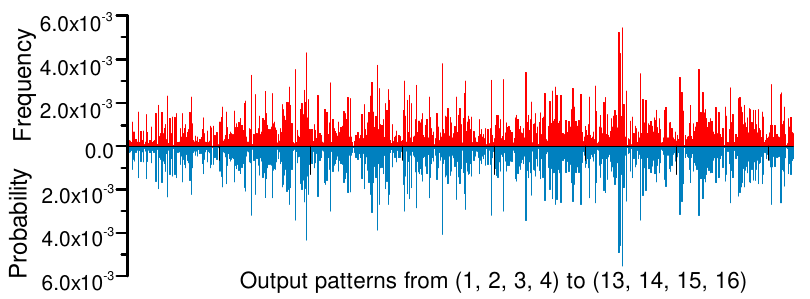}}
  \hspace{0.3cm}
  \subfigure[n=5,m=25]{\includegraphics[width=0.44\textwidth]{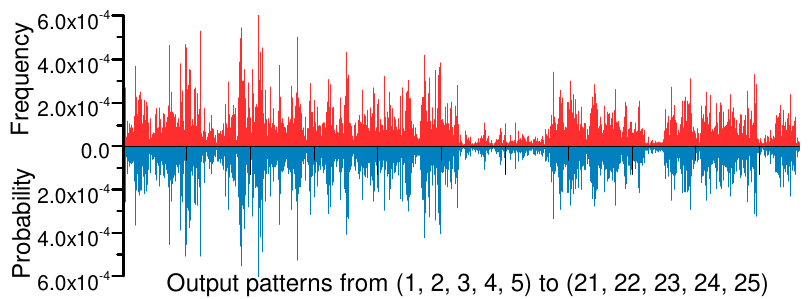}}\\
  \subfigure[n=6,m=12]{\includegraphics[width=0.44\textwidth]{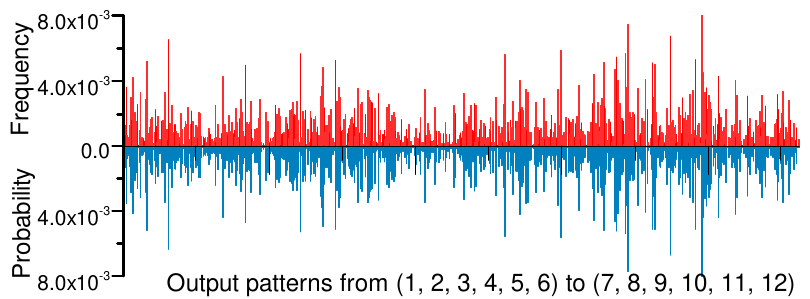}}
  \hspace{0.3cm}
  \subfigure[n=7,m=14]{\includegraphics[width=0.44\textwidth]{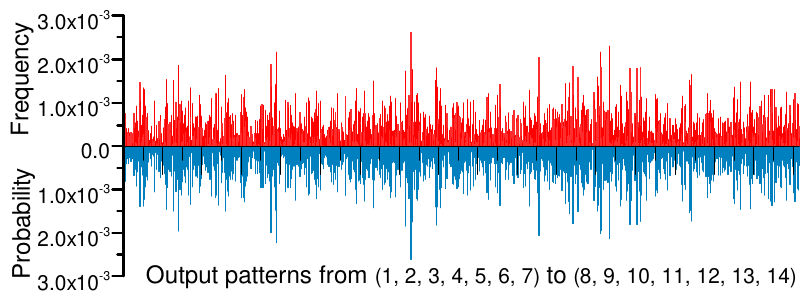}}
  \caption{\footnotesize The correctness of the sampling results. For each scale, 1,000,000 (for 4p.16m.), 6,000,000 (for 5p.25m.), 2,000,000 (for 6p.12m.) and 2,000,000 (for 7p.14m.) samples are taken respectively. The similarities for the two sampling results with the theoretical distribution are 99.90\%, 99.52\%, 99.94\% and 99.80\% respectively.}
  \label{FIGS:SamplingResults}
\end{figure}

Empirically, we found that to reach a high similarity between the frequency of the sampling results and the probability distribution, the number of samples taken should be around~$100\left(_n^m\right)$. Thus it limits the scales of simulation if we only take 1,000,000 samples. As have shown in Fig.~\ref{FIGS:SamplingResults}, the sampling results agree the theoretical probability results well. In small scales we have confirmed the correctness by comparing the frequency graph with the probability distribution, this validation method is not feasible in relatively large scale, which requires corresponding computing resource with the brute force sampler. In large scale simulation, other validation method is required, such as the likelihood ratio test.

If the correctness of the classical sampler is admitted, the classical sampler further provides an approach for the validation of the physical experiments. However, the hardness of boson sampling makes it difficult to validate the experimental results since it requires to calculate exponential permanents to provide theoretical probability distribution, which is exactly what the brute force sampler does. Another method that may help, which is kind of speculative, could be like this: We could obtain two sample sequences, one is from the classical sampler that could be trusted, the other is from the experimental boson sampler, and use some statistic techniques to validate, such as the $K$-$S$ test.

\subsection{The influence of sample cache on high-order autocorrelation}

\begin{figure}[!b]
  \centering
%    \subfigure[n=3,m=9]{\includegraphics[width=0.24\textwidth]{Fig/FigS12apeakmoving9m3p.pdf}}
%    \subfigure[n=5,m=25]{\includegraphics[width=0.24\textwidth]{Fig/FigS12bpeakmoving25m5p.pdf}}
%    \subfigure[n=6,m=36]{\includegraphics[width=0.24\textwidth]{Fig/FigS12cpeakmoving36m6p.pdf}}
%    \subfigure[n=7,m=49]{\includegraphics[width=0.24\textwidth]{Fig/FigS12dpeakmoving49m7p.pdf}}
    \subfigure[n=8,m=64]{\includegraphics[width=0.24\textwidth]{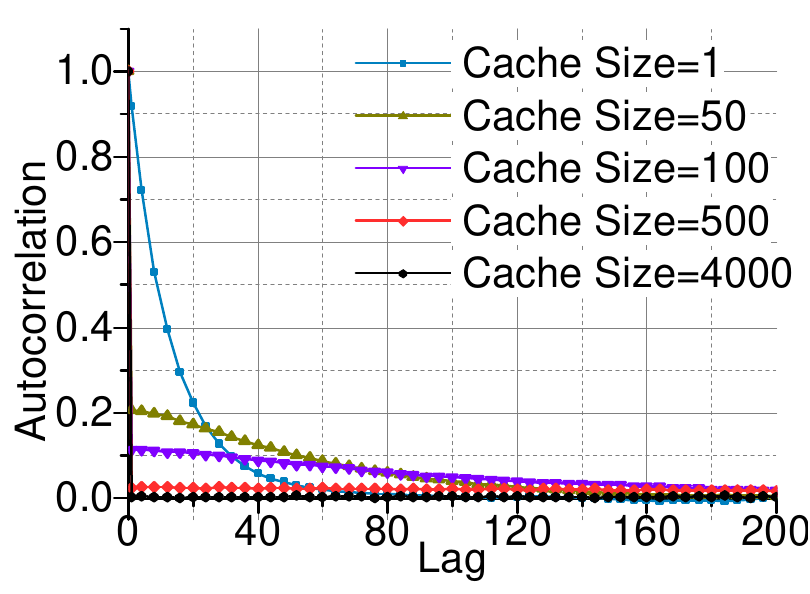}}
    \subfigure[n=9,m=81]{\includegraphics[width=0.24\textwidth]{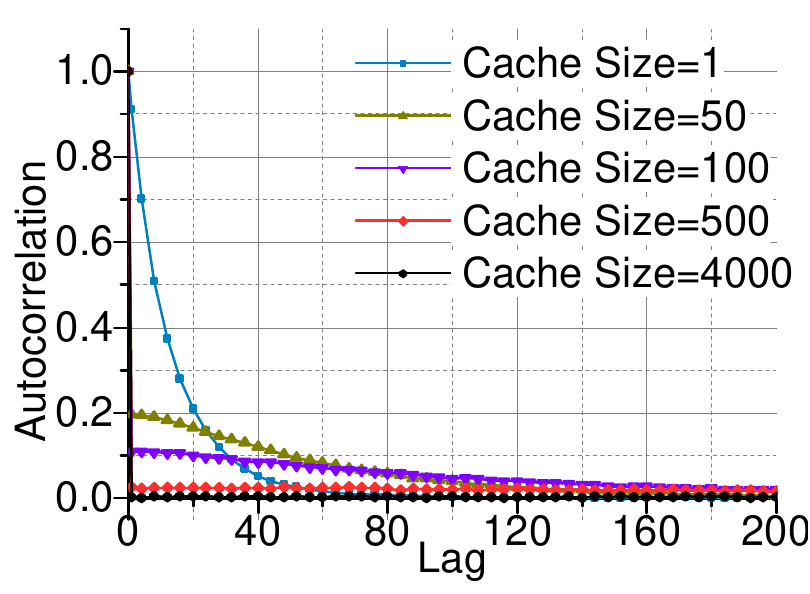}}
    \subfigure[n=10,m=100]{\includegraphics[width=0.24\textwidth]{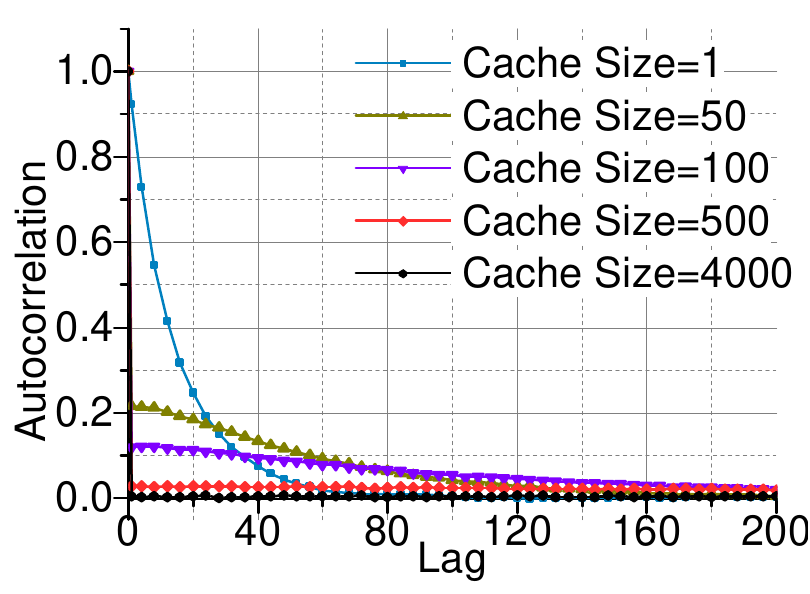}}
    \subfigure[n=11,m=121]{\includegraphics[width=0.24\textwidth]{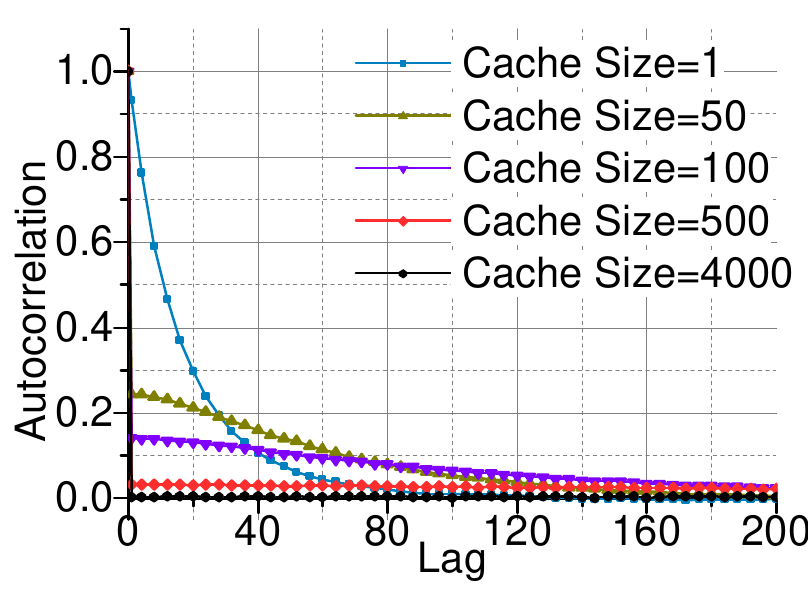}}
    \subfigure[n=12,m=144]{\includegraphics[width=0.24\textwidth]{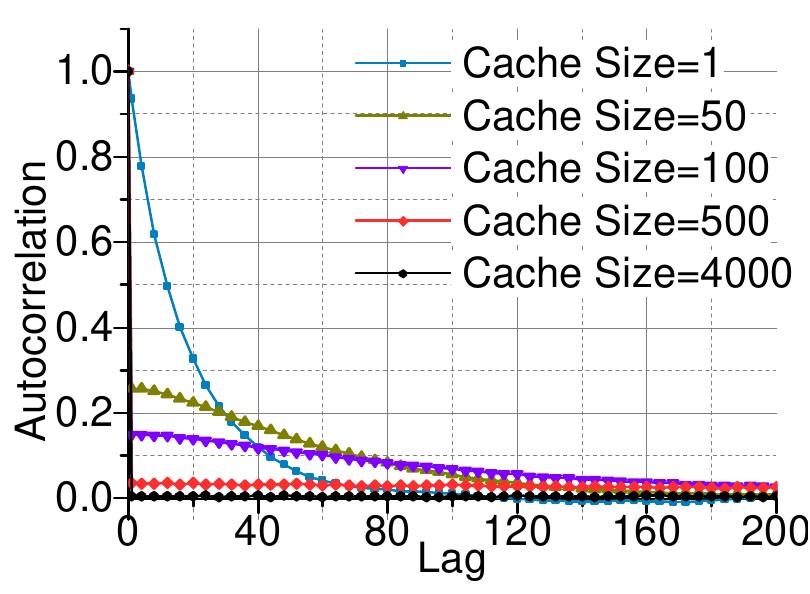}}
    \subfigure[n=13,m=169]{\includegraphics[width=0.24\textwidth]{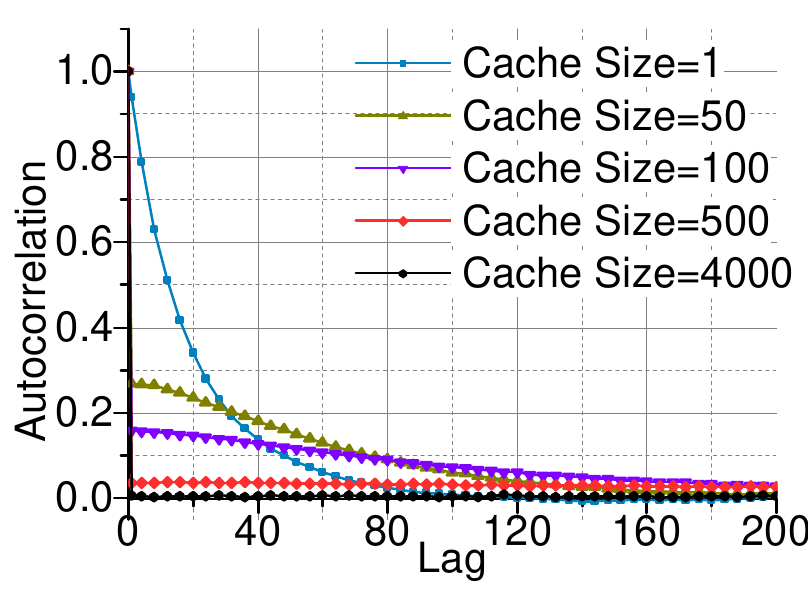}}
    \subfigure[n=14,m=196]{\includegraphics[width=0.24\textwidth]{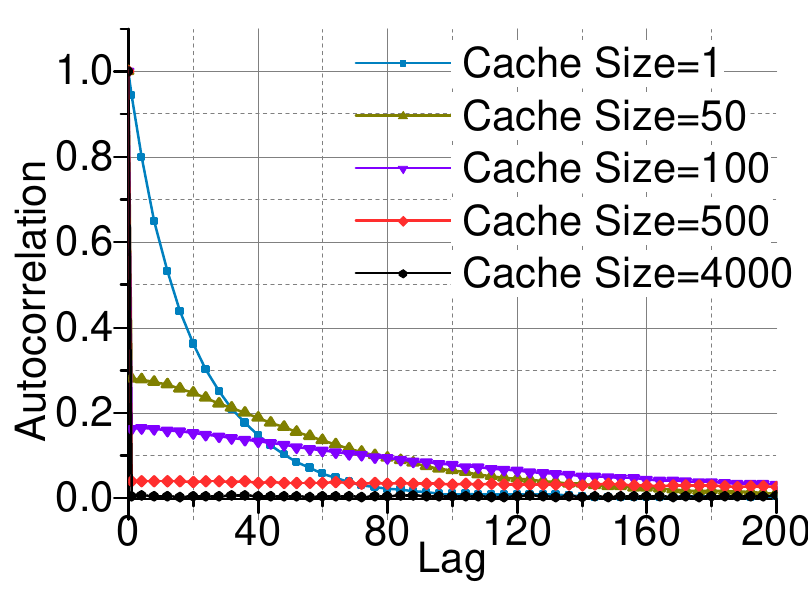}}
    \subfigure[n=15,m=225]{\includegraphics[width=0.24\textwidth]{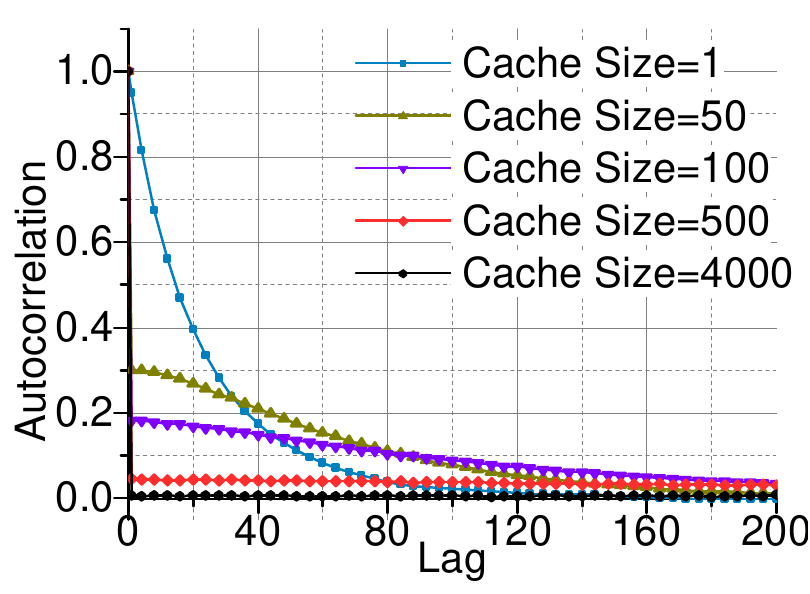}}
    \subfigure[n=16,m=256]{\includegraphics[width=0.24\textwidth]{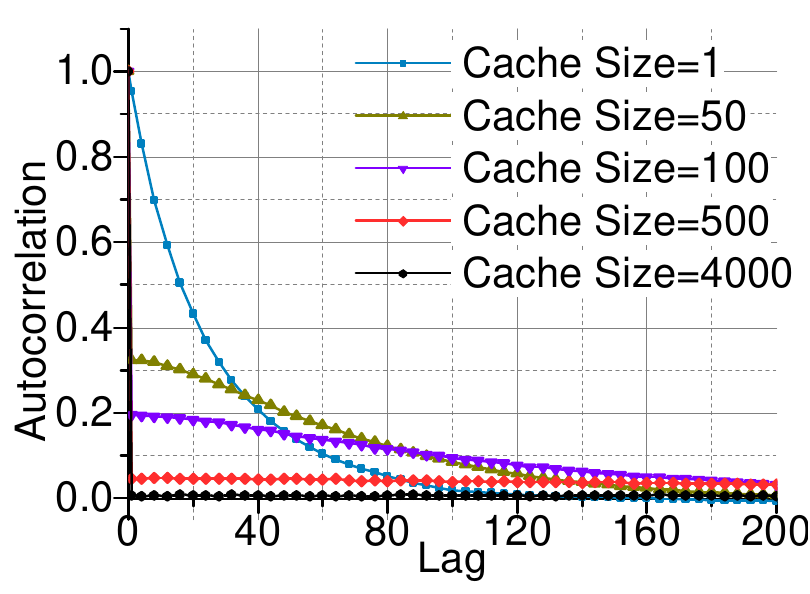}}
    \subfigure[n=17,m=289]{\includegraphics[width=0.24\textwidth]{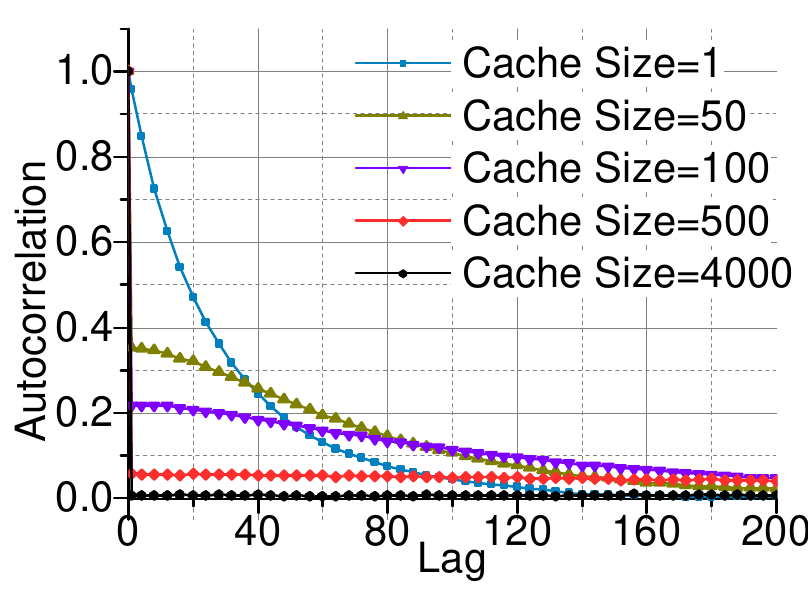}}
    \subfigure[n=18,m=324]{\includegraphics[width=0.24\textwidth]{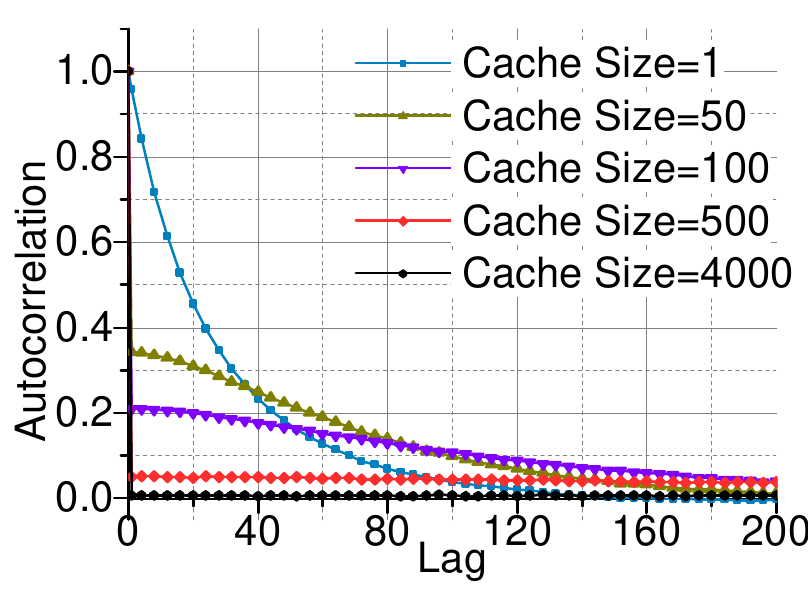}}
    \subfigure[n=19,m=361]{\includegraphics[width=0.24\textwidth]{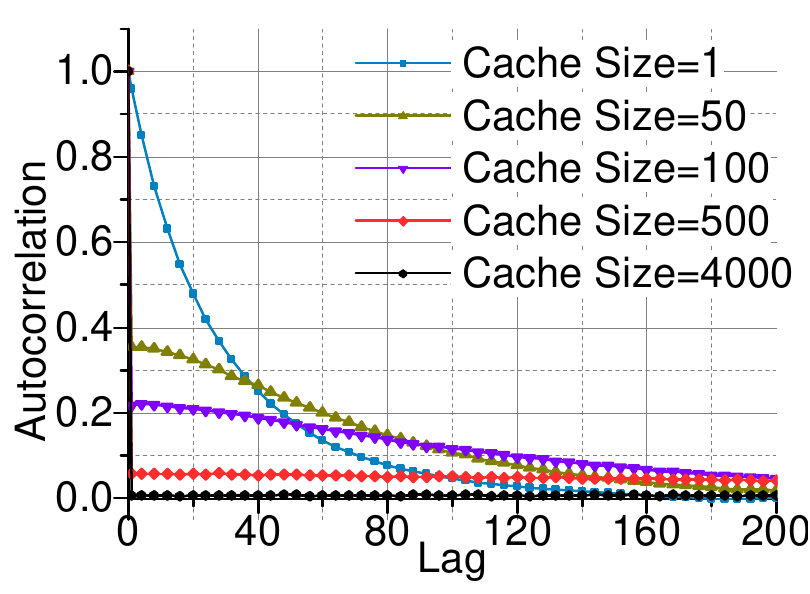}}
    \subfigure[n=20,m=400]{\includegraphics[width=0.24\textwidth]{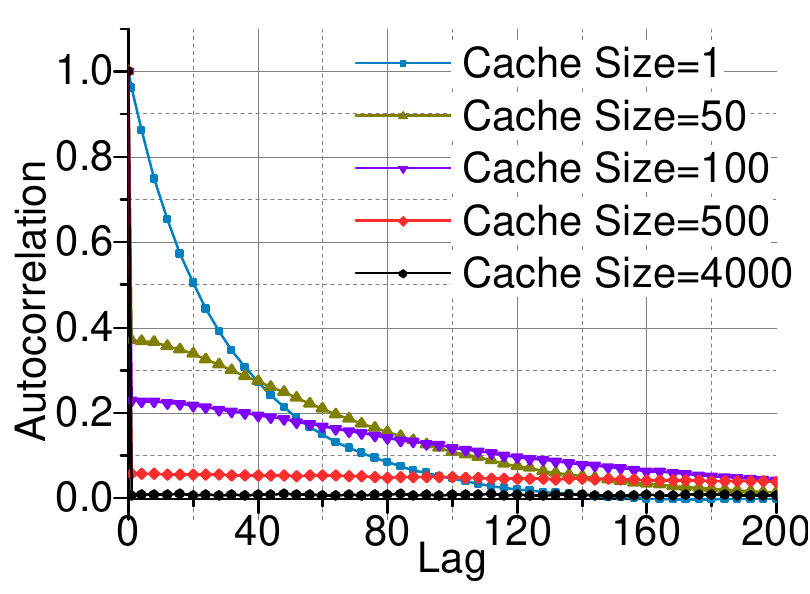}}
    \subfigure[n=21,m=441]{\includegraphics[width=0.24\textwidth]{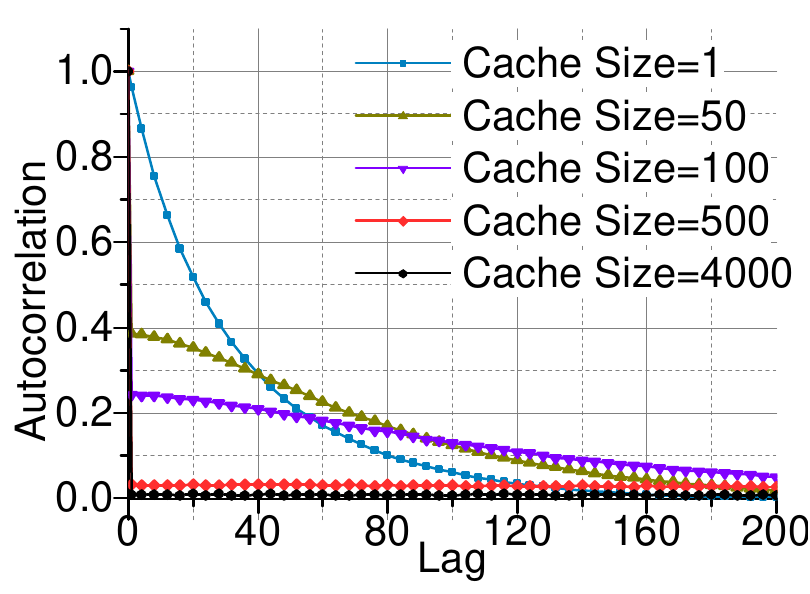}}
    \subfigure[n=25,m=625]{\includegraphics[width=0.24\textwidth]{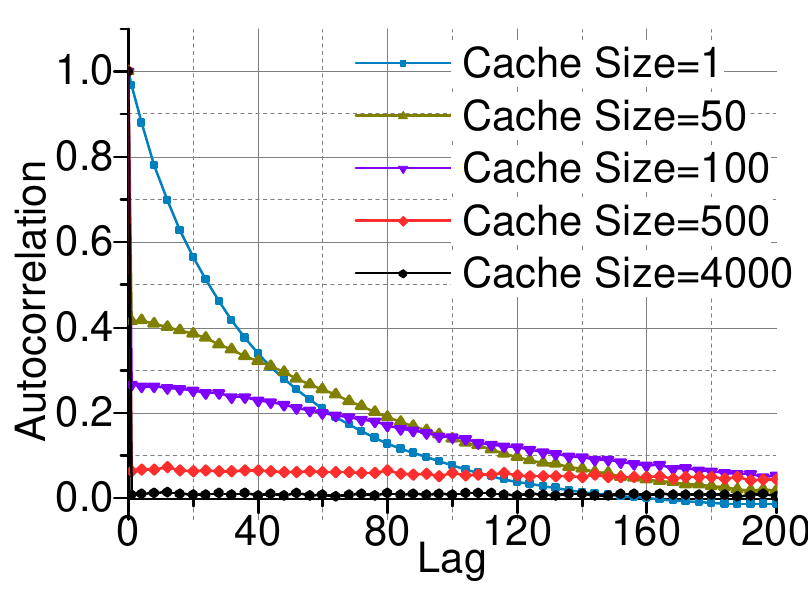}}
    \subfigure[n=30,m=900]{\includegraphics[width=0.24\textwidth]{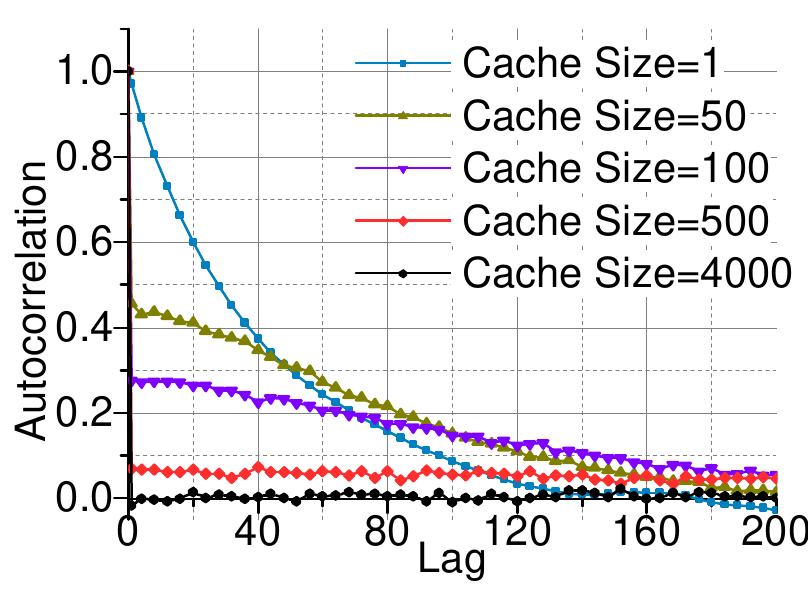}}
  \caption{\footnotesize The autocorrelation at lags up to 200. The autocorrelation spread to the high-order space with the enlarge of the cache. When the size of cache reaches 4,000, the autocorrelations at all lags ($>0$) are negligible, and the samples in the sequence are approximately independent.}
  \label{FIGS:Peakmoving4scales}
\end{figure}

The essence of SC-MCMC is the reorder of samples, so that the correlated samples are scattered to be apart in varied distances. The range of scattering depends on the size of the sample cache. Thus the low-order autocorrelation decreases with the cost that the high-order autocorrelation increases. Until the size of the sample cache reaches a certain degree, the space for the scattering of low-order autocorrelation is large enough, thus the low-order autocorrelation can be reduced with high-order autocorrelation increasing by a negligible quantity. The results of autocorrelation scattering are shown in Fig.~\ref{FIGS:Peakmoving4scales}.

The process of scattering can be observed with the increase of the size of sample cache. The curve indicated by $L=1$ shows the initial autocorrelation at all the lags. When $L=50$, the low-order autocorrelation decreases (e.g. lag from 1 to 20), while the autocorrelations at lags greater than 20 increase. When $L$ reaches 100, the curve tends to be flat, and $L$ reaching 500 makes it flatter. When $L$ arrives at 4,000, the autocorrelation at all the lags is negligible, therefore we obtain a sequence containing nearly independent samples.

\subsection{Performance}

Here we show the performance of our SC-MCMC. The test platforms are Tianhe-2 supercomputer, and another small-scale local clusters. The simulations only take advantages of the CPUs of Tianhe-2 without the usage of accelerators. The performance parameters of each computing nodes of Tianhe-2 (the corresponding parameters of the accelerators are excluded) are listed in Tab.~\ref{TABS:PerfPara}. The sampling results are shown in Tab.~\ref{TABS:Performance}. The largest number of nodes used on Tianhe-2 is 64, and the Intel Xeon Phi accelerators are not used, therefore it is still a server-level cluster.

\begin{table*}[!b]
  \centering
  \caption{\footnotesize Performance parameters of the computing node of Tianhe-2 supercomputer and those of the local cluster.}
  \label{TABS:PerfPara}
  \begin{tabular}{ccp{0.2cm}cp{0.2cm}cp{0.2cm}c}
  \hline
  \hline
    \multicolumn{2}{c}{\multirow{3}{*}{{\bf Item}}} &&                                  \multicolumn{5}{c}{{\bf Parameters of a node}}                                        \\
  \cline{4-8}
                                                   &&&                    \multicolumn{3}{c}{{\bf Tianhe-2}}                           && \multirow{2}{*}{\bf Local Cluster}  \\
  \cline{4-6}
                                                   &&&            With Accelerators           &&         Without Accelerators          &&                                     \\
  \hline
  \multicolumn{2}{c}{Peak Performance}              &&              3.43Teraflops             &&             422.4Gigaflops            &&             201.6Gigaflops          \\
  \multirow{2}{*}{Processors:}  &        CPU        &&  Intel Xeon E5 $\times$ 2 (24 cores)   &&  Intel Xeon E5 $\times$ 2 (24 cores)  &&  Intel Xeon E5 $\times$ 2 (12 cores)\\
                                &    Accelerators   &&  Intel Xeon Phi $\times$ 3 (171 cores) &&               $\setminus$             &&               $\setminus$           \\
  \multicolumn{2}{c}{Memory Storage Capacity}       &&                  72GB                  &&                   64GB                &&                   16GB              \\
  \multicolumn{2}{c}{Interconnect Network}          &&              TH Express-2              &&               TH Express-2            &&                InfiniBand           \\
  \hline
  \hline
  \end{tabular}
\end{table*}

\begin{table}[!htb]
  \centering
  \caption{\footnotesize Simulation results using SC-MCMC. The scales reflect the number of photons (p.) and modes (m.) in the boson sampling scheme. $T_{total}$  is the time used for the whole sampling process, $T_{1Sample}$ is the average time for one sample, $T_{per}$ is the time used on the calculation of permanents, and $T_{1Per}$ is the average time for one permanent. $Rate$ is the sampling rate when using $N$ nodes on the specified platform. $r_1$ is the first-order autocorrelation of the sequence when using a sampling cache with size of 4,000. The execution on Tianhe-2 only uses the CPUs, while the Intel Xeon Phi accelerators are not applied. When the number of photons are less than 17, the main cost of the calculation is the start-up of the calculation (note that the time for one permanent is in the order of $10^{-4}$ seconds), rather than the permanents. After that ($n\geq 17$), the time used in the simulation well confirm to the rule that the execution time doubles when the number of photons increases by 1, which means the quantity of computation of permanent becomes the main part of the simulation.}\label{TABS:Performance}
  \begin{tabular}{rccrrrcrccc}
    \hline
    \hline
        Scale       & Platform   & $N$ &    $S$\hspace{0.4cm}    & \hspace{0.3cm}Rate(Hz) & \hspace{0.3cm}$T_{total}$(s) & \hspace{0.2cm}$T_{1Sample}$(s) & \hspace{0.3cm}$T_{per}$(s) & \hspace{0.2cm}$T_{1Per}$(s)  & $\%_{Per}$& $r_1$ \\
    \hline
        20p.400m.   &   Tianhe-2 &  4  &   500,000 &  152.03 &  3288.88 & 0.00658 &  3185.82 & 0.00637 & 96.87\%  &  0.0097 \\
        25p.625m.   &   Tianhe-2 & 32  &   200,000 &   22.51 &  8884.56 & 0.04442 &  8804.99 & 0.04402 & 99.10\%  &  0.0089 \\
        30p.900m.   &   Tianhe-2 & 64  &    20,000 &    1.01 & 19836.72 & 0.99184 & 19825.83 & 0.99129 & 99.95\%  & -0.0162 \\
    \hline
           3p.9m.   &   Cluster  &  1  & 1,000,000 & 4263.29 &   234.56 & 0.00023 &   226.51 & 0.00023 & 96.57\%  &  0.0018 \\
          4p.16m.   &   Cluster  &  1  & 1,000,000 & 4163.90 &   240.16 & 0.00024 &   229.83 & 0.00023 & 95.70\%  & -0.0005 \\
          5p.25m.   &   Cluster  &  1  & 1,000,000 & 4082.18 &   244.97 & 0.00024 &   231.99 & 0.00023 & 94.70\%  &  0.0005 \\
          6p.36m.   &   Cluster  &  1  & 1,000,000 & 3919.72 &   255.12 & 0.00026 &   238.02 & 0.00024 & 93.30\%  &  0.0036 \\
          7p.49m.   &   Cluster  &  1  & 1,000,000 & 3795.73 &   263.45 & 0.00026 &   242.39 & 0.00024 & 92.01\%  &  0.0014 \\
          8p.64m.   &   Cluster  &  8  & 1,000,000 & 3491.04 &   286.45 & 0.00029 &   257.62 & 0.00026 & 89.94\%  &  0.0023 \\
          9p.81m.   &   Cluster  &  8  & 1,000,000 & 3429.28 &   291.61 & 0.00029 &   272.66 & 0.00027 & 93.50\%  &  0.0035 \\
        10p.100m.   &   Cluster  &  8  & 1,000,000 & 3195.79 &   312.91 & 0.00031 &   290.86 & 0.00029 & 92.95\%  &  0.0045 \\
        11p.121m.   &   Cluster  & 32  & 1,000,000 & 2780.55 &   347.64 & 0.00035 &   296.36 & 0.00030 & 85.25\%  &  0.0035 \\
        12p.144m.   &   Cluster  & 32  & 1,000,000 & 2897.39 &   357.68 & 0.00036 &   326.69 & 0.00033 & 91.34\%  &  0.0045 \\
        13p.169m.   &   Cluster  & 32  & 1,000,000 & 2818.12 &   413.30 & 0.00041 &   371.23 & 0.00037 & 89.82\%  &  0.0058 \\
        14p.196m.   &   Cluster  & 32  & 1,000,000 & 2813.13 &   537.99 & 0.00054 &   489.10 & 0.00049 & 90.91\%  &  0.0041 \\
        15p.225m.   &   Cluster  & 32  & 1,000,000 & 2133.86 &   789.08 & 0.00079 &   731.82 & 0.00073 & 92.74\%  &  0.0057 \\
        16p.256m.   &   Cluster  & 32  & 1,000,000 &  800.30 &  1252.05 & 0.00125 &  1183.67 & 0.00118 & 94.54\%  &  0.0064 \\
        17p.289m.   &   Cluster  & 32  & 1,000,000 &  317.46 &  3156.84 & 0.00316 &  3074.90 & 0.00307 & 97.40\%  &  0.0063 \\
        18p.324m.   &   Cluster  & 32  & 1,000,000 &  159.97 &  6251.32 & 0.00625 &  6156.73 & 0.00616 & 98.49\%  &  0.0072 \\
        19p.361m.   &   Cluster  & 32  & 1,000,000 &   78.94 & 12667.81 & 0.01267 & 12558.42 & 0.01256 & 99.14\%  &  0.0049 \\
        20p.400m.   &   Cluster  & 32  & 1,000,000 &   37.98 & 26326.21 & 0.02633 & 26199.45 & 0.02620 & 99.52\%  &  0.0067 \\
        21p.441m.   &   Cluster  & 32  & 1,000,000 &   18.09 & 55293.32 & 0.05529 & 55146.15 & 0.05515 & 99.73\%  &  0.0081 \\
    \hline
    \hline
  \end{tabular}
\end{table}

The percentage of time spent on permanents approaches 100\% with the scale grows. Though with the increase of number of computing nodes, the extra cost for the initialization of the computation also increases. However, compared to the computation of permanents, this is a negligible cost when the number of photons reaches a certain number. For example, when using 32 nodes, the percentage of time on spent on permanents reaches nearly 100\% when the number of photons reaches 21, while the average time on computing one permanent is only 0.055 seconds. Thus we conclude that SC-MCMC can reach the computational hardness limit of classically simulating boson sampling.

\section{Numerical simulation}
\label{SEC:NumericalSimulation}
\begin{figure}[!htb]
  \centering
    \subfigure[Fixing $p=16,000$]{\includegraphics[width=0.35\textwidth]{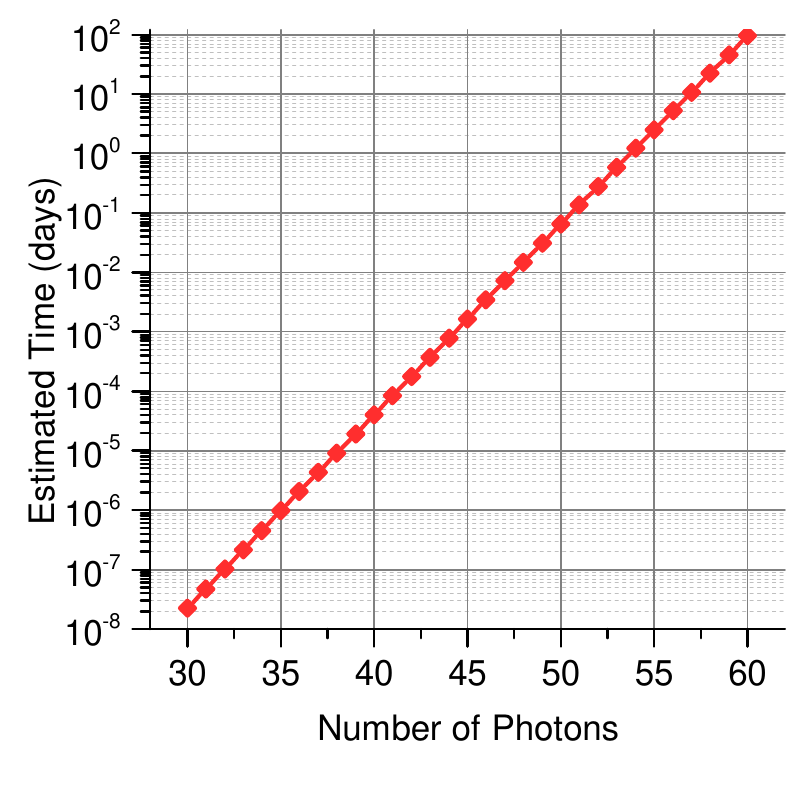}}
    \hspace{2cm}
    \subfigure[Fixing $n=59$]{\includegraphics[width=0.35\textwidth]{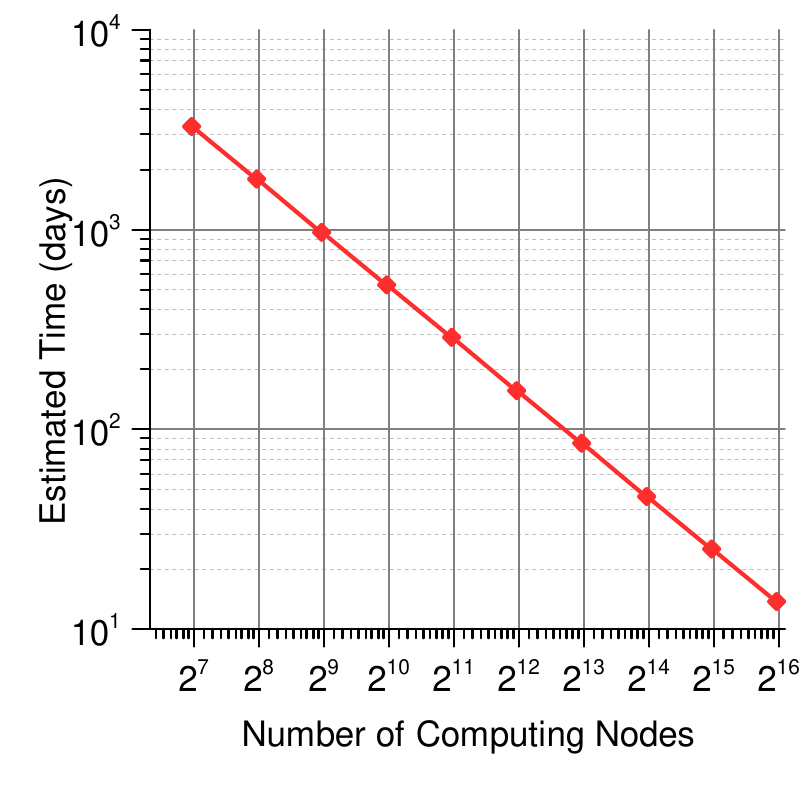}}
  \caption{\footnotesize The Scalability of the algorithm in terms of $n$ and $p$ respectively. (1) The time grows exponentially with the increase of $n$. Specifically, it grows by 1.95 times when $n$ increases by 1. (b) The time nearly halves when the number of nodes used doubles, which indicates that if the computing capability doubles, the time used would reduce by half.}
  \label{FIGS:Scalability}
\end{figure}

The precise estimation must base on some test first. In~\cite{Wu2018}, they have measured the scalability curve of calculating permanents on Tianhe-2 supercomputer. Our algorithm makes it feasible to regard the performance of calculating permanents as the performance of simulating boson sampling, which reveals the scaling in terms of $n$ (the number of photons) and $p$ (the number of computing nodes) respectively, based on the assumption that we are allowed to take a lot of samples. By fixing $p=16,000$, the scaling of $n$ is described by Eq.~\ref{eq:scalabilityn}, and by fixing $n=59$, the scaling of $p$ is shown in Eq.~\ref{eq:scalabilityp}
\begin{equation}
\label{eq:scalabilityn}
T(n)=1.9925\cdot n^22^n\times 10^{-15},\\
\end{equation}
\begin{equation}
\label{eq:scalabilityp}
T(p)=\frac{1.9675\times 10^{10}}{p^{0.8782}}.
\end{equation}
The scalability curve of simulating boson sampling on Tianhe-2 are shown in Fig.~\ref{FIGS:Scalability}.

As reported in~\cite{Neville2017}, the MIS could generate a 50-photon sample on the supercomputer within under 10 days, while in a corresponding time (within about 11 days), our method could be able to generate a 57-photon sample.

\begin{figure}[h]
  \centering
  \includegraphics[width=0.5\textwidth]{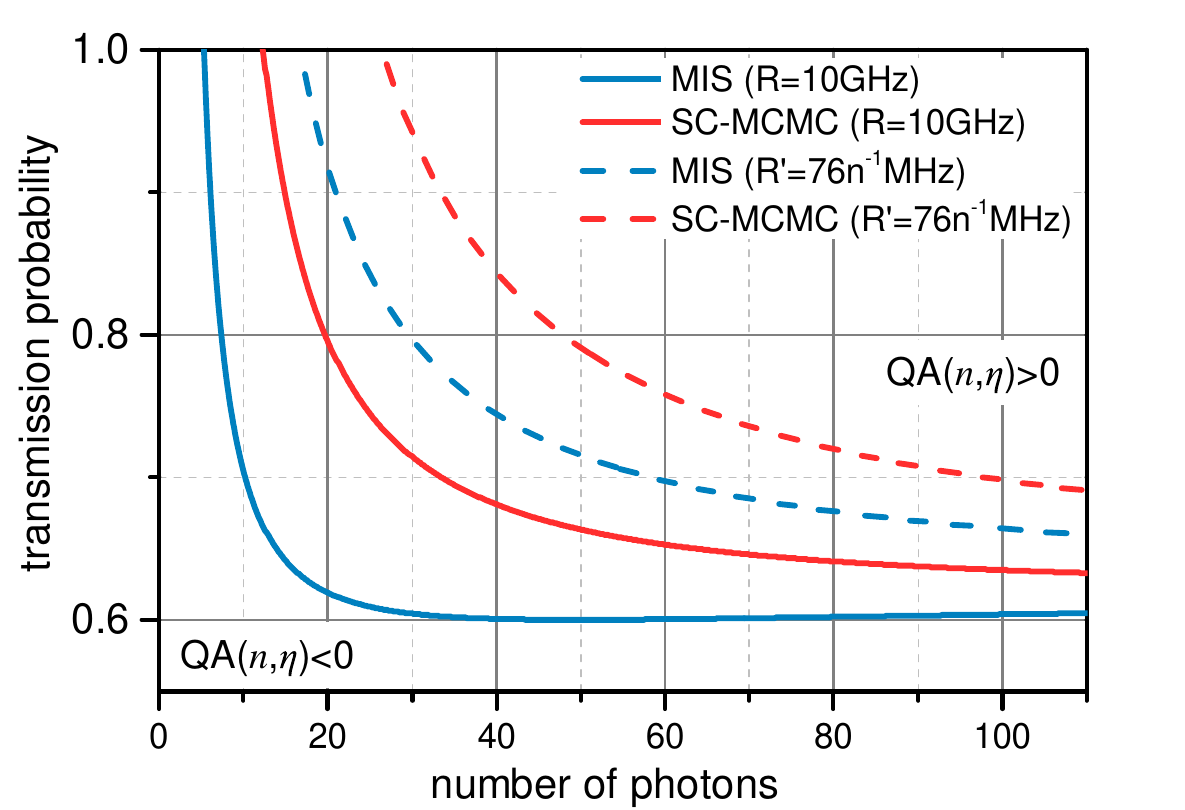}\\
  \caption{\footnotesize The performance comparison between the quantum Boson sampling device and the classical computer. The blue solid (dashed) line represents $QA(n,\eta)=0$ with the classical runtime estimated by MIS as $t_c=3\cdot n^22^n\times10^{-13}$ with the $n$-photon repetition rate of the photon source $R_q = 10{\text {GHz}}$ ($R'_q=76n^{-1} {\text {MHz}}$, the leading parameter of proposed photon source~\cite{Wang2019}, which is obtain from a $76$MHz quantum dot source), and the red lines are the results obtained from SC-MCMC.}
  \label{FIGS:NewThreshold4n}
\end{figure}

To make a more detailed comparison with the physical setups, we rewrite the function of quantum advantage defined in~\cite{Neville2017} as
\begin{equation}
QA(n, \eta)=\log\left(\frac{t_c}{t_q}\right),
\end{equation}
where $n$ is the number of photons, $\eta$ is the transmission probability of a photon, which is the key to the sampling efficiency of a physical realization. $t_c$, $t_q$ represent the estimated time for the simulation of boson sampling instance on classical computers and quantum computers respectively. The value of function $QA$ represents the competition between classical computers and quantum computers. The quantum advantage exists in the situations with $QA>0$ where quantum devices are faster than classical computers, and the border line of $QA=0$ is the balance between the classical device and quantum physical setup. The estimated quantum run time is
\begin{equation}\label{EQS:QT}
t_q(n, \eta)=(R_q\cdot \eta^n\cdot P_{CF})^{-1},
\end{equation}
where $P_{CF}$ is the probability for a collision-free event, which depends on the size of the network. It can be further refined for a square network with $m=n^2$ or a linear network with $m=4n$ which both have been experimentally realized, as Eq.~\ref{EQS:QTREFINED} shows.
\begin{equation}\label{EQS:QTREFINED}
\begin{aligned}
t_q^{m=n^2}&=\frac{e}{R_q\eta^n},   \\
t_q^{m=4n}&=\frac{1}{R_q}\left(\frac{5}{4\eta}\right)^n,
\end{aligned}
\end{equation}
where $e$ in $t_q^{m=n^2}$ and $1.25^n$ in $t_q^{m=4n}$ are the approximation to $P_{CF}^{-1}$ in corresponding cases. $R_q$ takes the value of $10$GHz, which is beyond the reach of the experimentally demonstrated photon sources. The leading parameter of proposed photon source is $R'_q=76n^{-1}{\text {MHz}}$~\cite{Wang2019}. Here we update the $t_c$ as
\begin{equation}
t_c=T(n)=1.9925\cdot n^22^n\times10^{-15},
\end{equation}
where $t_c$ is the estimated classical run time for an instance of size $n$ bosons in $n^2$ modes on Tianhe-2 supercomputer. Here, Fig.~\ref{FIGS:NewThreshold4n} shows the performance comparison quantum runtime as $t_q^{m=4n}$.

With the same experimental techniques, which decides the value of $\eta$, the number of photons required for positive quantum advantage is increased, as shown in Tab.~\ref{TABS:NumberIncreaseN2} for squared network and in Tab.~\ref{TABS:NumberIncrease4N} for the linear network. Currently $\eta$ is less than 0.4~\cite{Spring2013,Broome2013,Tillmann2013,Spagnolo2014,Bentivegna2015,Zhong2018,Wang2019}, and a small increment of $\eta$ could greatly reduce the number of photons required for positive quantum advantage.

\begin{table}[!htb]
  \centering
  \caption{\footnotesize The increment of required photon number for $QA(n,\eta)=0$ for a network with $m=n^2$. The number of photons required for $QA(n,\eta)=0$ increases according to the value of $\eta$. $N_{MIS}$ is the least number of required photons for $QA(n,\eta)=0$ according to the result obtained by MIS if the transmission probability realized in physical experiment is $\eta$. $N_{c_{t}}$ is correspond number via SC-MCMC.}\label{TABS:NumberIncreaseN2}
  \begin{tabular}{c|cccccccccc|ccccccccc}
    \hline
    \hline
      $R_q$          & \multicolumn{10}{c|}{$10$GHz} & \multicolumn{9}{c}{$76n^{-1}$MHz}\\
    \hline
      $\eta$       & 0.55 & 0.60 & 0.65 & 0.70 & 0.75 & 0.80 & 0.85 & 0.90 & 0.95 & 1 & 0.60 & 0.65 & 0.70 & 0.75 & 0.80 & 0.85 & 0.90 & 0.95 & 1\\
    \hline
      $N_{MIS}$    & 15 & 12 & 10 &  8 &  8 &  7 &  7 &  6 &  6 &  6 & 44 & 32 & 26 & 22 & 19 & 17 & 16 & 15 & 14\\
      $N_{t_c}$ & 45 & 29 & 22 & 18 & 16 & 14 & 13 & 12 & 11 & 11 & 69 & 49 & 39 & 33 & 29 & 26 & 24 & 22 & 20\\
      $N_{t_c}-N_{MIS}$  & 30 & 17 & 12 & 10 &  8 &  7 &  6 &  6 &  5 &  5 & 25 & 17 & 13 & 11 & 10 &  9 &  8 &  7 &  6\\
    \hline
    \hline
  \end{tabular}
\end{table}

\begin{table}[!htb]
  \centering
  \caption{\footnotesize The increment of required photon number for $QA(n,\eta)=0$ for a network with $m=4n$. The number of photons required for $QA(n,\eta)=0$ increases according to the value of $\eta$. $N_{MIS}$ is the least number of required photons for $QA(n,\eta)=0$ according to the result obtained by MIS if the transmission probability realized in physical experiment is $\eta$. $N_{c_{t}}$ is correspond number via SC-MCMC.}\label{TABS:NumberIncrease4N}
  \begin{tabular}{c|ccccccc|ccccccc}
    \hline
    \hline
      $R_q$          & \multicolumn{7}{c|}{$10$GHz} & \multicolumn{7}{c}{$76n^{-1}$MHz}\\
    \hline
      $\eta$       & 0.70 & 0.75 & 0.80 & 0.85 & 0.90 & 0.95 & 1 & 0.70 & 0.75 & 0.80 & 0.85 & 0.90 & 0.95 & 1\\
    \hline
      $N_{MIS}$    & 11 &  9 &  8 &  7 &  7 &  6 &  6 & 59 & 39 & 30 & 25 & 21 & 19 & 17\\
      $N_{t_c}$ & 34 & 25 & 20 & 17 & 15 & 14 & 13 & 99 & 64 & 48 & 40 & 34 & 30 & 27\\
      $N_{t_c}-N_{MIS}$  & 23 & 16 & 12 & 10 &  8 &  8 &  7 & 40 & 25 & 18 & 15 & 13 & 11 & 10\\
    \hline
    \hline
  \end{tabular}
\end{table}

The minimum value of $\eta$ is also raised if the number of photons is restricted within a reasonable value, say 100 photons. Tab.~\ref{TABS:ETAIncrement} gives the increment of $\eta$ under different experimental parameters.

\begin{table}[!htb]
  \centering
  \caption{\footnotesize The increment of the minimum value of $\eta$, the transmission probability of a single photon, when the number of photons is limited under 100. For the curves, the minimum value is reached when the photon number is 100. The increment varies according to the shape of the network and the repetition rate of the photon source. $\eta_{MIS}$ is the minimum value of $\eta$ according to the curves obtained by MIS, and $\eta_{SC-MCMC}$ is the correspond value via SC-MCMC.}\label{TABS:ETAIncrement}
  \begin{tabular}{c|cc|cc}
    \hline
    \hline
        Network            &   \multicolumn{2}{c|}{$m=n^2$}    &   \multicolumn{2}{c}{$m=4n$}    \\
    \hline
        Repetition Rate   &\hspace{0.4cm}$R_q=10$GHz\hspace{0.4cm}&\hspace{0.4cm}$R'_q=76n^{-1}$MHz\hspace{0.4cm}&\hspace{0.4cm}$R_q=10$GHz\hspace{0.4cm}&\hspace{0.4cm} $R'_q=76n^{-1}$MHz\hspace{0.4cm} \\
    \hline
        $\eta_{MIS}$                    &       48.81\% &       53.67\%     &       60.41\% &       66.42\%    \\
        $\eta_{SC-MCMC}$                &       51.32\% &       56.43\%     &       63.52\% &       69.83\%    \\
        $\eta_{SC-MCMC}$-$\eta_{MIS}$   &        2.51\% &        2.76\%     &        3.11\% &        3.41\%    \\
    \hline
    \hline
  \end{tabular}
\end{table}

If the number of photons is not limited, the minimum value of $\eta$ will still not approach 0. To see this, we write the classical runtime as
\begin{equation}
t_{cg}=an^b\cdot 2^n,
\end{equation}
where $a$ is the scaling coefficient of the algorithm, and $1\leq b\leq 2$ because an efficient implementation of the permanent calculating algorithm has a time complexity of $O(n2^n)$. The quantum runtime is referred from Eq.~\ref{EQS:QT}. Letting $t_{cg}=t_c$ will give
\begin{equation}
\eta=\sqrt[n]{\frac{1}{R\cdot P_{CF}\cdot an^b\cdot 2^n}}.
\end{equation}
It converges when $n\to \infty$. For the square network, we have
\begin{equation}
\lim_{n\to \infty}\eta=\lim_{n\to \infty}\frac{1}{2}\sqrt[n]{\frac{e}{R\cdot an^b}}=0.5,
\end{equation}
and for the linear network with $m=4n$, we have
\begin{equation}
\lim_{n\to \infty}\eta=\lim_{n\to \infty}\frac{1}{2}\sqrt[n]{\frac{5^n}{R\cdot an^b\cdot4^n}}=0.625.
\end{equation}

The possibly existence of $n^{-1}$ in $R_q$ will not affect the convergence. This convergence indicates that with the proposed experimental parameters, the transmission probability of a single photon is likely to have to be improved to above a threshold value, or the quantum advantage may never be demonstrated by boson sampling.

\bibliographystyle{unsrt}

\end{document}